\documentclass[bj,numbers]{imsart}

\usepackage{algorithm,algpseudocode}
\usepackage{multirow}
	\usepackage{thmtools}
\usepackage{xr}
\usepackage{booktabs}

\startlocaldefs
\newcommand{\R}{\mathbb{R}}

\newcommand{\p}{\mathbb{P}}

\newcommand{\E}{\mathbb{E}}
\newcommand{\pp}{\mathcal{P}}
\newcommand{\FF}{\mathcal{F}}

\def\argmax{\mathop{\mbox{argmax}}}
\def\argmin{\mathop{\mbox{argmin}}}
\def\lf{\lfloor}
\def\rf{\rfloor}

\theoremstyle{plain}

\newtheorem{theorem}{Theorem}[section]
\newtheorem{lemma}[theorem]{Lemma}
\newtheorem{corol}[theorem]{Corollary}
\newtheorem{proposition}{Proposition}
\newtheorem{remark}{Remark}[section]
\newtheorem{definition}[theorem]{Definition}
\newtheorem{example}{Example}[section]

\endlocaldefs
\makeatletter
\def\cleartheorem#1{%
	\expandafter\let\csname#1\endcsname\relax
	\expandafter\let\csname c@#1\endcsname\relax
}
\makeatother
\begin{document}
	
	\begin{frontmatter}
		\title{Multiscale jump testing and estimation  under complex temporal dynamics}
		\runtitle{Multiscale jump detection}
		
		\begin{aug}
			\author[A]{\inits{}\fnms{Weichi}~\snm{Wu}\ead[label=e1]{wuweichi@mail.tsinghua.edu}}
			\author[B]{\inits{}\fnms{Zhou}~\snm{Zhou}\ead[label=e2]{zhou@utstat.toronto.edu}}
			\address[A]{Center for Statistical Science, Department of Industrial Engineering, Tsinghua University, China\printead[presep={,\ }]{e1}}
			
			\address[B]{Department of Statistical Science, University of Toronto, Canada\printead[presep={,\ }]{e2}}
		\end{aug}
		
			\begin{abstract}	We consider the problem of detecting jumps in an otherwise smoothly evolving trend whilst the covariance and higher-order structures of the system can experience both smooth and abrupt changes over time. The number of jump points is allowed to diverge to infinity with the jump sizes possibly shrinking to zero. The method is based on a multiscale application of an optimal jump-pass filter to the time series, where the scales are dense between admissible lower and upper bounds.  For a wide class of non-stationary time series models and  trend functions, the proposed method is shown to be able to detect all jump points within a nearly optimal range with a prescribed probability asymptotically under mild conditions.  For a time series of length $n$, the computational complexity of the proposed method is $O(n)$ for each scale and $O(n\log^{1+\epsilon} n)$ overall, where $\epsilon$ is an arbitrarily small positive constant. Numerical studies show that the proposed jump testing and estimation method performs robustly and accurately under complex temporal dynamics.
		
	\end{abstract}
		
		
	\end{frontmatter}
	



	
	\section{Introduction}
	Time series data with complexly evolving distributional properties are frequently collected in many  applications. One prominent feature of such series is that the trend, covariance, and higher-order cumulants may simultaneously experience both abrupt and smooth changes over time. 
	Many examples of such complex temporal dynamics can be found, for instance, in signal processing where both abrupt and smooth changes are frequently observed in the oscillatory patterns of a signal (\cite{huang1998empirical}, \cite{daubechies2011synchrosqueezed}); and in financial econometrics where the volatility of time series can be both smoothly and abruptly evolving over periods of stable and risky markets (\cite{dahlhaus2006statistical}, \cite{stuaricua2005nonstationarities}).  
	The purpose of this paper is to perform jump testing and estimation under the aforementioned complex temporal dynamics; in particular,  we aim to accurately and efficiently test and estimate all jumps in a piece-wise smooth trend when the covariance and higher-order structures of the series are smoothly and abruptly evolving under mild conditions.

	\subsection{The multiscale jump testing and estimation method}  \label{Sec-MJPD}
	Assume that we observe a time series $\{y_{i,n}\}_{i=1}^n$ which follows the model
	\begin{align}
	y_{i,n}=\beta_n(i/n)+\varepsilon_{i,n}, \label{model 2}
	\end{align}
	where $\beta_n(t)$ is a piece-wise smooth function with $m_n$ jump points $0<d_1<d_2<...<d_{m_n}<1$,  and $\{\varepsilon_{i,n}\}_{i=1}^n$ is a centered error sequence whose convariance and higher-order structures may experience both smooth and jumps over time. Here the number of jumps $m_n$ is allowed to diverge to infinity. We denote $d_0=0$ and $d_{m_n+1}=1$ for convenience and the formal definition of the class of piece-wise smooth functions $\beta_n(t)$ belongs to will be given in Section \ref{class-psmooth}. The key component of our multiscale jump point detection (MJPD) method is an optimal smooth filter 
	$W(x)$, $x\in\R$, such that the discrete application of $W$ at time $t$ and scale $s$
	is approximately zero for all sufficiently small $s$ when $t$ is a smooth point of $\beta_n(\cdot)$; and the latter application is approximately equal to $\Delta_{t,n}$ for all sufficiently small $s$ when $t$ is a jump point, where $ \Delta_{t,n}=\lim_{a\downarrow t}\beta_n(a)-\lim_{a\uparrow t}\beta_n(a)$ is the (signed) jump size. Here the optimality of a filter refers to the best sensitivity in detecting jumps among all filters in a large functional class whose detailed definition is deferred to Section \ref{Filter-Construct}. Throughout this article we shall call such $W$ optimal jump-pass filters and we refer the readers to Figure \ref{Filter} for a graph of $W$ used in our simulations and data analysis.
	
	The MJPD method builds upon normalized applications of $W$ to the time series $y_{i,n}$ at time $t$ and scale $s$,
	\begin{align}
	H(t,s):=\frac{1}{\sqrt{ns}}\sum_{j=1}^ny_{j,n} W\left(\frac{j/n-t}{s}\right).
	\end{align} 
	Observe that in principle $H$ will be large at jump points of $\beta_n(\cdot)$. However, due to non-stationary trend and covariance, scales that are appropriate in detecting the jumps vary constantly over time and are difficult to estimate. For this reason, multiscale methods are particularly important for jump detection under complex temporal dynamics.  In this paper, for each time point $t$, we consider a statistic $G(t, \underline {s}_n, \bar {s}_n)$ which is the maximum of a self-normalized version of $H(t,s)$ at all scales $s$ between admissible lower bound $\underline{s}_n$ and upper bound $\bar {s}_n$ with $ \underline {s}_n\ll \bar {s}_n\ll 1$. 
	Notice that $G(t, \underline {s}_n, \bar {s}_n)$ demonstrates the strongest evidence supporting that $t$ is a jump point among all scales from $\underline {s}_n$ to $\bar {s}_n$. Under complex temporal dynamics where appropriate scales for jump detection evolve constantly over time and are elusive, the multiscale statistic $G(t, \underline {s}_n, \bar {s}_n)$ is expected to be adaptive as it summarizes the strongest evidence of a jump over a wide range of scales. In particular, the difficult task of scale selection is alleviated. Naturally, our method tests the existence and estimates the locations of the jump points according to the magnitudes of $G$ across time. Details are shown in Algorithm \ref{Algorithm1} in Section \ref{JumpDec}.  Sparse versions of the latter maximum-over-multiple-scales idea have been used in, among others, \cite{horowitz2001adaptive}, \cite{zhang2003adaptive}, \cite{gao2008nonparametric} for nonparametric adaptive testing where $O(\log n)$ or less scales were considered. Finally, our method consists of a second-stage local cumulative-sum-based procedure to further improve the accuracy of the estimated jump locations.

	\subsection{Multiscale asymptotics and estimation accuracy}
	The most important step for the implementation of MJPD lies in the theoretical investigation of $\sup_{t\in T_n^d}G(t, \underline {s}_n, \bar {s}_n)$, the maximum deviation of $G(t, \underline {s}_n, \bar {s}_n)$ over time. Here $ T_n^d$ is the collection of all time points except radius $\bar {s}_n$ neighborhoods around the jump points.  To the best of our knowledge, deriving the asymptotic limiting distributions of various multiscale procedures has been an open and challenging problem, even for data sets that are independent and scales that are relatively sparse (see for instance the discussions in 
	\cite{frick2014multiscale}). As a result, upper and lower probability bounds were typically used in such procedures which oftentimes leaded to conservative inference. Alternatively, computationally intensive bootstrap or simulation-based methods can be used for the inference; see for instance 
	\cite{gao2008nonparametric}, \cite{schmidt2013multiscale} and \cite{khismatullina2018multiscale}. However, the bootstrap implementation typically results in long computation time for longer time series which may be undesirable in some jump detection situations. Section \ref{sec:ce} contains a more detailed discussion in this aspect. 
	
	As one main contribution of the paper, we derive the limiting distribution of $\sup_{t\in T_n^d}G(t, \underline {s}_n, \bar {s}_n)$ under complex temporal dynamics which enables MJPD to detect all jump points with a prescribed probability asymptotically. The latter distribution is pivotal and it involves $\underline {s}_n$, $\bar {s}_n $ and $W$ in a complicated way; see Theorem \ref{Thm1} for the details. Tail probabilities of the distribution can be accurately and efficiently calculated based on the closed form formula of its CDF. The derivation of the limiting distribution requires a delicate Gaussian approximation step which establishes that, under complex temporal dynamics, the maximum deviation of  $G(t, \underline {s}_n, \bar {s}_n)$ can be well approximated by that of a Gaussian multiscale statistic. Then we utilize Weyl's formula for the volume of tubes (\cite{weyl1939volume}, \cite{sun1993tail}, \cite{sun1994simultaneous}) to derive the limiting law of the maximum deviation of the latter Gaussian multiscale statistic. Since its asymptotics are established under complex temporal dynamics, MJPD is robust to a large class of smooth changes in the trend as well as smooth and abrupt changes in the second and higher order structures. 


	As a second main contribution of the paper, we establish that the estimation accuracy of MJPD is nearly optimal, where the near optimality refers to the fact that the jump point estimation rate of MJPD is identical to that of parametric jump estimation except a factor of logarithm.  Here the number of jumps is allowed to diverge to infinity with the jump sizes shrinking to 0 at sufficiently slow rates. In particular, the estimation rate for MJPD is nearly the same as the parametric jump detection rate when the trend is piece-wise constant and the errors are i.i.d. (cf. e.g. \cite{siegmund1988},  \cite{dumbgen1991asymptotic} and \cite{muller1997two}). In other words, jump detection in the trend under complex temporal dynamics can be performed with nearly the same order of accuracy as in the independent case when MJPD is used. Technically, the above optimality results require careful manipulations of empirical processes of non-stationary time series for which probabilistic and moment bounds for non-stationary partial sums in \cite{liu2013probability} and large deviation results for heavy tailed sums in \cite{mikosch1998large} are useful.   On the other hand, note that the optimality here is not in the sense of decision theoretical minimax risk over a large class of change point detection procedures (cf. e.g. \cite{ritov1990decision}, \cite{beibel1996note}).

	\subsection{Computational efficiency}\label{sec:ce}
	As increasingly longer time series are being collected,  the issue of efficient computation becomes more and more important for jump detection. In our implementation of MJPD, three major efforts are made towards fast and accurate estimation of the jump points. The first effort is in fact the aforementioned derivation of the limiting distribution of $\sup_{t\in T_n^d}G(t, \underline {s}_n, \bar {s}_n)$ which enables one to obtain the critical values of MJPD almost instantly without resorting to computationally intensive resampling or simulation methods. For comparison purposes, in the simulation studies of Section \ref{Sec::Simu} 
	and Section \ref{5000Sample-Result} of the online supplement, we perform another jump estimation method called SIM which estimates critical values of $\sup_{t\in T_n^d}G(t, \underline {s}_n, \bar {s}_n)$ via the multiplier bootstrap. For a time series of length 5000, it is reported that it takes approximately 28 minutes for SIM to finish calculating 5000 bootstrap replicates on a fast desktop computer equipped with intel i7-8700 CPU. Consequently bootstrap-based methods, at least in their ordinary forms, will take a long time to detect jumps in time series data with large amount of observations.
	
	Our second effort is the use of the fast sum updating algorithm (\cite{seifert1994fast}, \cite{fan1994fast}, \cite{langrene2019fast}) to evaluate $\{H(\frac{i}{n},s)\}_{i=1}^n$ which reduces the computational cost of the latter quantity from $O(n^2s)$ to $O(n)$ for each scale $s$. Specifically, our implementation of the fast sum updating algorithm makes use of the piece-wise low-order-polynomial form of the optimal filter $W$ and calculates $H(\frac{i+1}{n},s)$ from $H(\frac{i}{n},s)$ with an $O(1)$ computational cost. Readers are referred to Section \ref{Compu-Issue} for the details.
	
	The third effort we made is an efficient sparsification of the scales. Observe that the theory of MJPD is established over all scales from $\underline {s}_n$ to $\bar {s}_n$. In practice, we recommend evaluating $G(t, \underline {s}_n, \bar {s}_n)$ on a sparse set of scales ${\cal G}_n$, where ${\cal G}_n$ is a sequence of $O([\log n]^{1+\epsilon})$ scales starting from $\underline {s}_n$ and ending in $\bar {s}_n$. Here $\epsilon$ is an arbitrarily small positive constant. An important justification for the latter sparsification is the theoretical result established in this paper that the temporal maximum deviations of $G(t, \underline {s}_n, \bar {s}_n)$ evaluated on $[\underline s_n, \bar s_n]$ and ${\cal G}_n$ coincide asymptotically; see Theorem \ref{Save_b} in Section \ref{Compu-Issue}. Combining with our second effort, we conclude that the total computational cost for MJPD is $O(n[\log n]^{1+\epsilon})$ in view of the fact that the second-stage local cumulative-sum-based estimating procedure only costs $o(n)$ computational time.  Finally, though not implemented at the time of writing, MJPD is ideal for parallel computing as the calculations of $\{H(\frac{i}{n},s)\}_{i=1}^n$ across different scales are totally independent.
	
	
	\subsection{Literature review}
	For i.i.d. or stationary errors, the problem of detecting jumps in a piece-wise smooth signal was considered in, among others, \cite{muller1992change},  \cite{eubank1994nonparametric}, 
	\cite{loader1996change},  \cite{gijbels1999estimation}, \cite{qiu2003jump}, \cite{chen2012testing} and \cite{chen2022inference}  via single scale methods.   Exceptions include \cite{zhang2016testing} who considered testing smooth trend versus an jump alternative when the errors are locally stationary using a single scale kernel-based method. To our knowledge, heteroscedasticity and autocorrelation robust multiple jump detection algorithms for non-stationary time series models have not been studied in the literature. MJPD can be viewed as a multiscale extension of \cite{muller1992change}  to the case of complex temporal dynamics.  Meanwhile, second stage refinement in jump detection was studied in, for instance, \cite{muller1997two} and \cite{gijbels1999estimation} where optimality of the two-stage methods was theoretically proven when the number of jumps is bounded and known. Our optimality results on the second-stage refinement generalize those of \cite{muller1997two} and \cite{gijbels1999estimation} to the case of unknown number of jump points and non-stationary and dependent errors.

	Here we would like to discuss the distinction between jump detection considered in this paper and change point detection in statistics. There is a huge literature in change point detection. See for instance \cite{bai1997estimating}, \cite{bai1998estimating}, \cite{qu2008testing}, \cite{shao2010}, \cite{killick2012optimal}, \cite{frick2014multiscale}, and \cite{dette2016detecting},  among many others. 
	Most of the aforementioned papers rely on the assumption that the parameter of interest is a {\it piece-wise constant} function of time without smooth changes in order to segment the sequence into stable sections. However, in some real data applications it may be more appropriate to distinguish between smooth and abrupt changes and characterise the parameter of interest by {\it piece-wise smooth} functions since smooth or slow changes in the underlying data generating mechanism are widely observed in many physical, social, and economic systems over time. \textcolor{black}{Examples include but are not limited to temperature data (\cite{zhang2016testing}), hydrology data (\cite{muller1992change}), and macroeconomic data (\cite{chen2022inference})}. When algorithms based on the piece-wise constant assumption are applied to a smoothly varying system, typically many spurious jump points will be flagged and it could be difficult for the user to discover various features of the underlying smooth curve such as linearity or convexity. \textcolor{black}{See our data analysis for a detailed discussion.} The difference between jump detection and change point detection is characterized by the aforementioned difference in modelling the dynamics of the parameter of interest. 
	
	
	The rest of the paper is organized as follows. Introductions to and definitions of  the class of piecewise smooth functions considered in this paper, jump-pass filters and piece-wise locally stationary processes are presented in Section \ref{Sec:Preliminaries}. In Section \ref{JumpDec} we discuss MJPD in detail. MJPD asymptotics and asymptotic optimality of the two-stage procedure are established, and the associated algorithms are given there. In Section \ref{Sec:implement} we investigate implementation issues, including choices of filters, efficient computation tuning parameter selection.  Simulation results are provided in Section \ref{Sec::Simu}. We analyze a SP500 daily return dataset in Section \ref{sp500]} and identify important jump dates. We conclude the paper and provide some discussions in Section \ref{Sec:discussion}. 
	Finally,  more simulation results, 
	and proofs of the theoretical results of the paper are gathered in the supplemental material. 
	
	\section{Preliminaries} \label{Sec:Preliminaries}

\subsection{ A Class of piece-wise smooth functions} \label{class-psmooth}In this section we shall rigorously define the class of piece-wise smooth trend functions $\beta_n(\cdot)$ considered in the main article. To this end, we shall henceforth consider the class $\mathcal M(m_n, \Delta_n, \gamma_n, k)$ which consists of all functions $f:[0,1]\rightarrow \mathbb R$ such that for some  constants $C_{Lip}$ and $\bar C$,
\begin{description}
	\item (F1) 
	$f(\cdot)$ has $m_n$ discontinuous points $0=d^f_0<d^f_1<d^f_2<...<d^f_{m_n}<d^f_{m_n+1}=1$, and $f(x)\in \mathcal C^k((d^f_i,d^f_{i+1}),C_{Lip})$ for $x\in (d^f_i,d^f_{i+1}) $, $0\leq i\leq m_n$. In addition, $f^{}$ is either right or left continuous at $\{d_i^f,0\leq i\leq m_n+1\}$. 
	\item (F2) 
	$|f(d_i^f+)-f(d^f_i-)|\geq \Delta_n>0$ for $1\leq i\leq m_n$, where for any function $g$ and $a\in\mathbb R$, $g(a+)=\lim_{s\downarrow a} g(s)$ and $g(a-)=\lim_{s\uparrow a} g(s)$. 
	\item (F3) 
	$\min_{0\leq i\leq m_n} |d_{i+1}^f-d_i^f|\geq \gamma_n>0$.
	\item (F4) $\sup_{a\in [0,1)}|f^{(u)}(a+)|\leq \bar C,\sup_{a\in (0,1]}|f^{(u)}(a-)|\leq \bar C \  \forall~ 0\leq u\leq k$.
\end{description}
Condition (F1) means that $f(\cdot)$ is piece-wise smooth with Lipschitz continuous $k_{th}$ order derivative. 
(F2) puts a lower bound  $\Delta_n$ on the minimum jump size.  (F3) restricts that the minimum space among $\{d^f_i\}_{i=0}^{m_n+1}$ is at least $\gamma_n$. Condition (F4) controls the overall smoothness of derivatives of $f(\cdot)$, which is important when the number of jump points $m_n$ diverges.  Notice that before and after a jump point, the derivatives of $f(\cdot)$ are allowed to be different under our setting. Furthermore, note that our trend model includes the following special case 
\begin{align}\label{f1}
 f(\cdot)=f_0(\cdot)+\sum_{s=1}^{m_n}a_{s}I_{A_s}(\cdot),
\end{align}
where $f_0(\cdot)$ is a smooth function at $[0,1]$, $I_{A_s}(\cdot)$ are indicator functions and $A_s$ are disjoint intervals such that $\cup_s A_s=[0,1].$ Notice that in \eqref{f1}, the left and right derivatives of $f(\cdot)$ at a jump point are equal. A high dimensional version of \eqref{f1} has been considered by, among others, \cite{chen2022inference} with bounded number of jump points and high dimensional, strictly stationary noise. 
In this paper we consider the model
\begin{align}
y_{i,n}=\beta_n(i/n)+\varepsilon_{i,n}, 
\end{align}
where $\beta_n(\cdot)\in \mathcal M(m_n, \Delta_n, \gamma_n, k)$ and $(\varepsilon_{i,n})_{i=1}^n$
are non-stationary errors satisfying $\E(\varepsilon_{i,n})=0$.
The goal of the paper is to estimate the number and the locations of the jump points $\{d^{\beta_n}_i, 1\leq i\leq m_n\}$. 
In both the main article and the supplemental material, we  omit the superscript $\beta_n$ in $\{d_i^{\beta_n}, 1\leq i \leq m_n\}$ for the sake of brevity.  
\textcolor{black}{We shall  compare our method with \cite{chen2022inference} in more detail at the end of Section \ref{JumpDec}.}

	\subsection{Piece-wise locally stationary time series} This subsection is devoted to the modelling of $\{\varepsilon_{i,n}\}_{i=1}^n$. As pointed out in the  introduction, many real-world time series are non-stationary. Often the data generating mechanism of such series can evolve both smoothly and abruptly over time. In this paper we adopt a flexible nonparametric device to model this complex temporal dynamics, which is the piece-wise locally stationary (PLS) time series framework \cite{zhou2013heteroscedasticity}. In the following, for any $d$ dimensional vector $\mathbf v=(v_1,...,v_d)^T$, denote by $|\mathbf v|=(\sum_{i=1}^dv_i^2)^{\frac{1}{2}}$ its Euclidean norm. For a random vector $\mathbf x$, write $\|\mathbf x\|_q=\left(\E |\mathbf x|^q\right)^{\frac{1}{q}}$ for its $\mathcal L^q$ norm. Denoted by $\mathcal C^k(I, C_{Lip})$ the collection of continuous functions that has $k$ times Lipschitz continuous derivatives on interval $I$ with Lipschitz constant $C_{Lip}$.
	
	\begin{definition}\label{def1}(Piece-wise locally stationary processes) Let $\eta=(\eta_i)_{i\in \mathbb Z}$ be a sequence of $i.i.d.$ random variables, and $\FF_i=(\eta_s,s\leq i).$ The sequence $(\varepsilon_{i,n})_{i=1}^n$ is called PLS with $l$ break points if there exist constants $0=c_0<c_1<...<c_l<c_{l+1}=1$ and possibly nonlinear filters $L_s, 0\leq s\leq l$ such that
		\begin{align}\label{eq:PLS} 
		\varepsilon_{i,n}=L_j(i/n,\FF_i),\  c_j<i/n\leq c_{j+1}, 0\leq j\leq l,
		\end{align} where 
		\begin{align}\label{orignial-A3}
		\|L_j(t,\FF_0)-L_j(s,\FF_0)\|_p\leq C|t-s|
		\end{align}
		for all $t,s\in (c_j,c_{j+1}]$, $0\leq j\leq l$, some finite constant $p>1$ and some finite constant $C$.
	\end{definition}
	In the above definition, the number and locations of the break points in the errors, $\{c_j, 1\leq j\leq l\}$, are typically unknown. Stochastic Lipschitz continuity condition \eqref{orignial-A3} requires that the filters $L_j(t,\cdot)$ are smooth functions of $t$ on $(c_j,c_{j+1}]$, $j=0,\cdots, l$. Therefore at $\{c_j\}_{j=1}^l$ the process can undergo abrupt changes while between two adjacent break points the data generating mechanism evolves smoothly. As a result, the PLS processes provide a general and flexible tool to describe complex temporal dynamics that evolve both smoothly and abruptly over time. The PLS framework \eqref{eq:PLS} can be viewed as an extension of the locally stationary time series frameworks in, for example, \cite{zhou2009local} and  \cite{dahlhaus1997fitting} by allowing abrupt changes to occur in the underlying data generating mechanism. Observe that the PLS class includes natural non-stationary extensions of the classic stationary linear (such as ARMA) and
	nonlinear (such as (G)ARCH, threshold and bilinear) time series models. 
	We refer to \cite{zhou2013heteroscedasticity} and \cite{wu2018gradient} for more discussions and examples of the PLS models. 
	Throughout the paper we assume the error process $\varepsilon_{i,n}$ in model \eqref{model 2} is a PLS process with break points $0=c_0<c_1<..<c_l<c_{l+1}=1$ and filters $\{L_j(\cdot, \cdot), 0\leq j\leq l\}$ such that for some constant $p\geq 4$ the following conditions hold:
	\begin{description}  
		\item(A1) The piece-wise Lipschitz continuous condition \eqref{orignial-A3} holds. Furthermore, assume that $$\max_{0\leq j\leq l}\sup_{t\in (c_j,c_{j+1}]}\|L_j(t,\FF_0)\|_p<\infty.$$ 
		\item(A2)  For some $\chi\in (0,1),$ the dependence measure $\delta_{p}(L,i)$ satisfies 
		\begin{align}\delta_{p}(L,i):=\max_{0\leq j\leq l}\sup_{t\in (c_j,c_{j+1}]}\|L_j(t,\FF_i)-L_j(t,\FF_i^*)\|_p=O(\chi^i),\end{align}
		where $\FF_i^*=(\FF_{-1},\eta_0',\eta_{1},...,\eta_{i-1},\eta_i)$ and $(\eta_i')_{i \in \mathbb Z}$ is an $i.i.d$ copy of $(\eta_i)_{i\in \mathbb Z}$. 
		\item(A3) The long-run variance $\sigma^2(t)$ of $(\varepsilon_{i,n})_{1\leq i\leq n}$ is Lipschitz continuous on $(c_j,c_{j+1}]$ for $0\leq j\leq l$ and $\inf_{t\in[0,1]}\sigma^2(t)>0,$ where $\sigma^2(0)=\lim_{t\downarrow 0}\sigma^2(t)$ and \begin{align}\label{longrun}
		\sigma^2(t):=\sum_{k\in \mathbb Z}Cov(L_j(t,\FF_0),L_j(t,\FF_k)), c_j<t\leq c_{j+1}, 0\leq j\leq l.
		\end{align}
	\end{description}

	Condition (A1) requires the existence of $p_{th}$ moment for the errors.  The quantity $\delta_{p}(L,i)$ in condition (A2) is called ``physical dependence measures'' which quantifies the dependence of $L_j(t,\FF_i), 0\leq j\leq l$ on $\eta_0$. Condition (A2) assumes that the dependence measures decay geometrically to zero. Theoretical results of the paper can be established when $\delta_{p}(L,i)$ decays at a sufficiently fast polynomial rate.  However, substantially more involved mathematical arguments are required in this case and we shall demonstrate all our results under the geometrical decay assumption for presentational simplicity. We refer to \cite{wu2018gradient} regarding the calculations of $\delta_{p}(L,i)$ for many PLS linear and nonlinear processes. Condition (A3)  guarantees that the long-run variance is piece-wise smooth and non-degenerate over $[0,1]$.



	\subsection{Jump-Pass filters}\label{Origin}
	
	For each positive integer $k$, define the class of
	filters $\mathcal W(k)$ as the collection of functions $W$ satisfying
	\begin{align}
	&W\in \mathcal C^{1} (\mathbb R, C_{Lip}),\  \text{\it Supp}(W)\subseteq[-1,1],\ W(x)=-W(-x), \int_{0}^1 W(x)dx=1,  \notag\\&\lim_{x\downarrow -1} W'(x)=\lim_{x\uparrow 1} W'(x)=0,  \int_{-1}^1 x^uW(x)dx=0 \ \ \mbox{for}\ \  1\leq u\leq k
	\end{align}
	where $C_{Lip}>0$ is some constant. 
	A $k_{th}$, $k\geq 2$, order jump-pass filter $W(\cdot)$ is a function which satisfies 
	\begin{description}
		\item (W1) $W(\cdot)\in \mathcal W(k)$.
		\item (W2) 
		Let $F_w(x)=\int_{-1}^xW(s)ds$, and i) $|F_w(x)|$ is uniquely maximized at $0$, and $|F_w(0)|$ is at least $\bar \eta_0$ larger than all other local maximum for some  positive constant $\bar \eta_0$;  ii) there exist strictly positive constants $\bar{\eta}_1$ and $\bar{\eta}_2$ such that $F^2_w(t)-F^2_w(0)\leq -\bar{\eta}_1 t^2$ for $|t|\leq \bar{\eta}_2$, and  $W'(0)\neq 0$.
	\end{description}
	Let $\tilde G_n(t,s):=\frac{1}{ns}\sum_{i=1}^n\beta_n(i/n)W\left(\frac{i/n-t}{s}\right)$. By Proposition \ref{proposition-1-10-13} in the supplemental material, (W1) implies that, for any sufficiently small scales $s_n\rightarrow 0$ with $ns_n\rightarrow\infty$, \begin{align}\label{Jumppass1}\tilde G_n(t,s_n)=O\Big(s_n^{k+1}+\frac{1}{ns_n}\Big)\end{align} for $t \in \cup_{r=0}^{m_n}[d_r+s_n,d_{r+1}-s_n]$, where $d_1,..,d_{m_n}$ are jump points of $\beta_n(\cdot)$ and $d_0=0$, $d_{m_n+1}=1$. Hence filters with higher order $k$
	render smaller filtering bias when $t$ is sufficiently separated from the jump points,
	which is our motivation for the smoothness assumption on the filters that excludes the use of discontinuous filters such as the step functions.	Notice that (W1) implies $W(0)=0$, which leads to $\frac{\partial}{\partial x}F_w^2(x)|_{x=0}=0$. As a result, a sufficient condition for (ii) of (W2) is
	\begin{align}\label{Lip}
	-\lambda_1\leq W'(0)F_w(0)<0 
	\end{align} 
	for some sufficiently large positive constant $\lambda_1$.
	Elementary calculations by the proof of Proposition \ref{proposition-10-13-2} in the supplemental material show that, if $|t-d_r|\le s_n$, then the leading term of $\tilde G^2_n(d_r,s_n)-\tilde G^2_n(t,s_n)$ is \begin{align}(\beta_n(d_r-)-\beta_n(d_r+))^2\Big((\int_{-\infty}^0W(s)ds)^2-(\int_{-\infty}^\frac{d_r-t}{s_n}W(s)ds)^2\Big)\label{Jumppass2}.
	\end{align} 
	Hence condition (W2) guarantees that, asymptotically, $d_r$ is a local maximum point of the function $|\tilde G_n(\cdot,s_n)|$. Expressions \eqref{Jumppass1} and \eqref{Jumppass2} further demonstrate that $|\tilde G_n(\cdot,s_n)|$ is asymptotically negligible at smooth points and it is approximately proportional to the jump size at jump points. This is the reason why we call $W$ ``jump-pass filters''. Clearly jumps of a series can be detected based on the latter property.

	
	\section{The multiscale jump point detection method}\label{JumpDec}
	The MJPD statistic at each time point $t$ is defined as \begin{align}\label{new4-March-26-1}
	G(t, \tilde s_n):=\sup_{\underline {s}_n\leq s\leq \bar {s}_n}G(t,s,s^*_n):=\sup_{\underline {s}_n\leq s\leq \bar {s}_n}\frac{|H(t,s)|}{\sqrt{\sum_{i\in K(t)}H^2(i/n,s^*_n)/|K(t)|}},\  \tilde s_n=(\underline {s}_n, \bar {s}_n, s^*_n)^T,
	\end{align}
	where $\bar {s}_n=o(1), \underline {s}_n=o(\bar {s}_n)$ and $s^*_n=o(\underline {s}_n)$ are three scales whose choices will be discussed later,  
	$K(t)=\{i:s^*_n\leq |i/n-t|\leq \bar {s}_n \}$, $|\cdot|$ denotes the cardinality of a set,  and \begin{align}H(t,s)=\frac{1}{\sqrt{ns}}\sum_{j=1}^ny_jW\left(\frac{j/n-t}{s}\right)\end{align}  for some jump-pass filter $W(\cdot)\in \mathcal W(k)$. We remark that the scale $s^*_n$ is determined by $(\underline {s}_n, \bar {s}_n)$. For simplicity we write $G(t,s,s^*_n)$ as $G(t,s)$ for the rest of the paper. 
	
	Note that $G(t,s)$ is a studentized  version of the quantity $H(t,s)$ which applies the filter $W(\cdot)$ to the observed $y_i's$ in a local neighborhood of $t$. If $t$ is bounded away from any jump point $d_i$ and any break point $c_j$, dividing by $\sqrt{\sum_{i\in K(t)}H^2(i/n,s^*_n)/|K(t)|}$ in the studentization ensures that $G(t, \tilde s_n)$ is pivotal asymptotically. \textcolor{black}{ From the proof, this quantity converges to  $\sigma(t)(\int W^2(t)dt)^{1/2}$ on $T_{n,c}\cap T_{n,d}$ and it is bounded otherwise. Here $\sigma^2(t)$ is the local long-run variance of the time series which reflects the changing higher-order dynamics..} By model \eqref{model 2}, the deterministic part of $H(t,s)$ is $\sqrt{ns}\tilde G_n(t,s)$,  which is asymptotically locally maximized at the jump points. This fact indicates that locations with large values of the multiscale statistic $G(t,\tilde s_n)$ are candidate jump points. 
	Therefore the key to MJPD lies in rigorously investigating the maximum deviation of $G(t, \tilde s_n)$ over time in order to distinguish genuine jumps from fluctuations produced by the random noise. 
	
	\begin{remark}(Relation to self-normalization-based methods)\label{RevisionRemark}
	\textcolor{black}{	Since the works of \cite{shao2010self} and \cite{shao2010}, self-normalization-based methods have attracted increased research attention in the past decades due to its tuning-free characteristic compared with the traditional lag-window type long-run variance estimator. Recently, \cite{zhao2021segmenting} proposed a self-normalization-based method for segmenting time series, allowing change-point detection for a broad class of parameters. Compared with the self-normalization-based method, our statistic \eqref{new4-March-26-1} is more of a {\it studentized} statistic, of which the denominator consistently estimates the local long-run standard deviation and the limiting distribution is pivotal. 
The construction of the self-normalization-based method rests on the strictly stationary errors and the piece-wise constant trends, while in this paper we focus on piece-wise smooth trends with piece-wise locally stationary errors, i.e., we also allow structural breaks in the errors. We would like to mention that the extension of  \eqref{new4-March-26-1} to the test and estimation  of more complicated piece-wise smooth parameters,  such as the quantiles,  is nontrivial;  See Section \ref{Sec:discussion} for detailed discussion.  However,  test and estimation of piece-wise constant parameters (such as quantiles) can be tackled by \cite{zhao2021segmenting} if the error is strictly stationary.}
	\end{remark}
	
	In order to state our first result regarding the maximum deviation of $G(t,\tilde s_n)$, we introduce the following notation. For given $\tilde s_n$, let $$T_n^d=\cup_{0\leq s\leq m_n}(d_s+\bar {s}_n,d_{s+1}-\bar {s}_n), \bar T_n^d=\cup_{1\leq s\leq m_n}(d_s-\bar {s}_n,d_s+\bar {s}_n)$$ be the union of intervals containing no and one jump point, respectively. Similarly let $$T_n^c=\cup_{0\leq s\leq l}(c_s+\bar {s}_n,c_{s+1}-\bar {s}_n), \bar T_n^c=\cup_{1\leq s\leq l}(c_s-\bar {s}_n,c_s+\bar {s}_n)$$ be the union of intervals containing no and one break point in the PLS errors, respectively. Let  $\gamma_n=\min_{0\leq i\leq m_n-1}(d_{i+1}-d_i)$ if $m_n\geq 1$ and $1$ otherwise, and $\check \gamma=\min_{0\leq i\leq l-1}(c_{i+1}-c_i)$ if $l\geq 1$ and $1$ otherwise. Write \begin{align}
	\nu_{1,n}=n(s^*_n)^{2k+3}+\left((s^*_n)^{k+1}+\frac{1}{ns^*_n}\right)\left(n^{\frac{1}{4}}\log ^2n+\sqrt{ns^*_n}\left(\frac{s^*_n\log n}{\bar {s}_n}\right)^{\frac{1}{2}}\right),\\
	\nu_{2,n}=\left((s^*_n/\bar {s}_n)^{1/2}+\frac{\log n}{\sqrt{n\bar {s}_n}}\right)(s^*_n)^{-\frac{2}{p}},\\ \nu_{3,n}=\log^2 n/(ns^*_n)+\bar {s}_n-s^*_n\log s^*_n.
	\end{align}
	We assume the following condition (B):
	\begin{description}
		\item (B1) $\frac{n^{1/4}\log^2 n}{\sqrt {n\underline {s}_n}}=o(1)$, $(\nu_{1,n}+\nu_{2,n}+\nu_{3,n})\log n=o(1)$.
		\item (B2)  $\bar {s}_n\leq (\gamma_n\wedge \check{\gamma})/2:=\min(\gamma_n,\check \gamma)/2$, and  $\bar {s}_n\leq d_1\leq d_{m_n}\leq 1-\bar {s}_n$ if $m_n\geq 1$.
		\item (B3) $m_n\bar {s}_n=o(1)$ and $l$ is a fixed number where $m_n$ and $l$ is the number  of jump points in the mean and the number of break points in noise, respectively. 
		\item (B4) $\sqrt{n\bar {s}_n}\bar {s}_n^{k+1}=o(1)$ and $\frac{m_n}{n\underline {s}_n}=o(1)$. 
	\end{description}
	Condition (B1) is necessary to approximate MJPD by the maximum deviation of a certain Gaussian random field. Assumption (B2) requires that $2\bar {s}_n$ is smaller than the smallest distance between adjacent jump (break) points.
	 We should point out that the results of this paper can be extended to PLS noises with diverging number of break points, i.e., $l_n\rightarrow \infty$ with a sufficiently slow divergence rate. However,
	a substantially more complicated mathematical argument is needed. For simplicity we shall present the results of MJPD with fixed $l$ in this paper. 
	 Condition (B3) means that lengths of the intervals $\bar T_n^c$ and $\bar T_n^d$ are asymptotically negligible. As a result, the behaviour of the stochastic part of MJPD on $[0,1]$ is determined by that on $ T_n^c\cap  T_n^d$. Moreover, assumptions (B2)--(B3) admit situations in which the jump points and the break points are overlapped. \textcolor{black}{Condition (B2)  is in fact  a condition on the minimum spacing between two jumps. (B2) is new for PLS errors. If the error is locally stationary then it is in fact milder than the commonly-used  conditions under piece-wise smooth means,  see for example \cite{zhang2016testing} and \cite{chen2022inference}. (B3) puts assumptions on the  {\it upper bound} of the number of jumps in the trend and noise. } For (B4), the term $\sqrt{n\bar {s}_n}\bar {s}_n^{k+1}$ is the bias caused by the $k_{th}$ order jump-pass filter, while the term $\frac{m_n}{n\underline {s}_n}$ is due to the approximation errors of the Riemann sum of $W(\cdot)$ and its variants. Due to time series non-stationarity, the best scales to capture the jumps at different time points are usually different but will fall within $(\underline s_n, \bar s_n)$ provided that the interval is sufficiently wide. 
	To state the results of Theorem \ref{Thm1}, we define the following quantities for $W(\cdot)$:\begin{align}
	w_{11}=\int_{-1}^1 (W'(t))^2dt, ~~w_{22}=\int_{-1}^1 (W'(t)t+\frac{1}{2}W(t))^2dt,\\ u_{11}=\int_{-1}^1 (W(t))^2dt,~~
	\kappa_n=(w_{11}w_{22})^{1/2}u^{-1}_{11}(\underline {s}_n^{-1}-\bar {s}_n^{-1})(1-2\bar {s}_n),\\
	\zeta_{1,n}=\sqrt{w_{11}u^{-1}_{11}}(\bar {s}_n^{-1}+\underline {s}_n^{-1}),~~\zeta_{2,n}=2\sqrt{w_{22}u^{-1}_{11}}(\log \bar {s}_n-\log \underline {s}_n).
	\end{align} 
	Conditions B1-B4 imply that $\frac{\underline {s}_n}{  n^{-1/2}\log ^{4}n}\rightarrow \infty$.  On the other hand, if $s^*_n$ is at the order of $n^{-1/2}\log n$ which minimizes $\nu_{3,n}$, 
	then condition (B1) is reduced to 
	\begin{align*}
	    \bar {s}_n^{-1}(n^{-1/2}\log n)^{1-4/p}=o(1),~
	    n^{1-2/p} \bar {s}_n\log ^{4/p-2}n\rightarrow \infty,~
	    n^{3/2-4/p}\bar {s}_n \log n\rightarrow \infty.
	\end{align*}In particular, if $p>6$, then an upper bound $\bar {s}_n\asymp n^{-1/6}$ is allowed where
	for two real series $a_n,b_n$,  $a_n\asymp b_n$ means there exist constants $0<M_0<M_1<\infty$ such that $M_0\leq \liminf \frac{a_n}{b_n}\leq \limsup \frac{a_n}{b_n}\leq M_1$.
	\begin{theorem}\label{Thm1}
		Assume  (A1)-(A3), (B1)-(B4), (W1),  $s^*_n=o(\underline {s}_n)$, $\underline{s}_n\asymp n^{\upsilon_0}$, $\bar{s}_n\asymp n^{\upsilon_1}$ for constants $\upsilon_0,\upsilon_1$ such that $-1/2<\upsilon_0<\upsilon_1<0$. In addition assume that the filter $W(\cdot)\in \mathcal C^3[-1,1]$. 
		Then we have as $n$ and $c=c(n)$ diverge, \begin{align}\label{eq-alphac}
		\p(\sup_{t\in T_n^d} G(t,\tilde s_n)>c)=\alpha_n(c)+O\left((m_n+1)\zeta_{2,n}\exp(-c^2/2)\right)+o(1),
		\end{align}
		where\begin{align}
		\alpha_n(c)=\frac{\kappa_n c}{\sqrt 2 \pi^{3/2}}\exp(-c^2/2)+\frac{\zeta_{1,n}}{2\pi}\exp(-c^2/2)+2(1-\Phi(c)),\label{new.2018-March-26-14}
		\end{align}
		and $\Phi(\cdot)$ is the CDF of $N(0,1)$.
	\end{theorem}
	
	
 Therefore,  for a fixed $\alpha\in (0,1)$, the critical value $c_{1-\alpha}$ of MJPD can be chosen as the root of $\alpha_n(c)=\alpha$, which diverges at the rate of $\sqrt{\log n}$. Due to condition (B3), the remainder term $O((m_n+1)\zeta_{2,n}\exp(-c^2/2))$ in \eqref{eq-alphac} is negligible. The proof of Theorem \ref{Thm1} rests on a delicate Gaussian approximation technique with $m-$dependence approximation, and the volume of tubes formula in \cite{weyl1939volume} and \cite{sun1994simultaneous} for evaluating the maximum deviation of a Gaussian random field. The most important contribution of Theorem \ref{Thm1} is that it provides an asymptotic closed-form formula for the $(1-\alpha)_{th}$ quantile of MJPD, by which MJPD is applicable to large scale data sets as we discussed in the Introduction. Meanwhile, Theorem \ref{Thm1} allows a continuum of scales between $\underline {s}_n$ and $\bar {s}_n$. In comparison, the asymptotic limiting laws of most existing multiscale procedures are not directly available; see for instance \cite{horowitz2001adaptive}, 
	\cite{gao2008nonparametric} and \cite{frick2014multiscale} among others. Furthermore, the multiscale kernel-based statistics proposed by, for example, \cite{horowitz2001adaptive}, \cite{zhang2003adaptive}  and \cite{gao2008nonparametric} are sparse in the sense that at most $O(\log(n))$ scales are considered.  
	
	Observe that the  conditions of Theorem \ref{Thm1} allow a diverging number of jump points. 
	From the proof of Theorem \ref{Thm1} (where $\zeta_{1,n}=\sum_j\zeta_{j,1} $ in \eqref{New54}), in practice we shall replace $\zeta_{1,n}$ with $\zeta_{1,n}'=(1-2\bar{s}_n)\zeta_{1,n}$  to improve finite sample performance of MJPD. For a given level $\alpha$, Theorem \ref{Thm1} also provides an upper bound  for the $(1-\alpha)_{th}$ quantile of the multiscale statistic $\sup_{t\in T_n^d} G(t,\tilde s_n)$, i.e. \begin{align}\label{eq22-2018-10-15}
	M\sqrt{-2\upsilon_0\log n-2\log \alpha}\end{align} 
	where $M$ is a sufficiently large constant depending on the filter $W(\cdot)$. 
	Expression \eqref{eq-alphac} 
	motivates the following Algorithm \ref{Algorithm1} (MJPD) for jump point detection.	For convenience, we let $[a,b)=\emptyset$ and $(a,b]=\emptyset$ if $a\geq b$. Let $\eta$ be any small positive number less than $1$, say $10^{-3}$.
	
	\begin{algorithm} 
		\caption{MJPD}\label{Algorithm1}
		\begin{algorithmic}[1]
			\State Compute $G(t,\tilde s_n)$ for $0<t<1$ defined in \eqref{new4-March-26-1}.  
			\State  Choose threshold $c$ as follows. For a given $\alpha$ compute $c_{1-\alpha}$ from \eqref{new.2018-March-26-14} with $\zeta_{1,n}$ replaced by $\zeta_{1,n}'$ such that $c_{1-\alpha}$ is the root of $\alpha(c)=\alpha$. Let $c=c_{1-\alpha}$. 
			
			\State Obtain $J$,  the set of jump points by \begin{algorithmic}
				\State $I\gets [0,1]$; $J\gets \emptyset$
				\While{$max_{t\in I}G(t,\tilde s_n)\geq c_{1-\alpha}$} 
				\State 
				$\hat d\gets\argmax_{t\in  I}G(t,\tilde s_n)$; $J\gets J\cup\hat d$; $ I_{}\gets  I\cap ([\bar {s}_n, \hat d-(1+\eta)\bar {s}_n)\cup (\hat d+(1+\eta)\bar {s}_n,1-\bar {s}_n])$ 
				\EndWhile
			\end{algorithmic}
		\end{algorithmic}
	\end{algorithm}
	By construction, $|J|$ is the number of jump points. From extensive simulation studies in Section 5 and the supplemental material, it is found that Algorithm \ref{Algorithm1} performs well for time series data of length $\ge 500$. However, for small sample sizes $c_{1-\alpha}$ obtained by solving \eqref{new.2018-March-26-14} may not be accurate. In such situations we provide the following multiplier-bootstrap-assisted Algorithm \ref{Algorithm2} (SIM) to simulate $c_{1-\alpha}$ in order to enhance small sample performance of MJPD. It is found in our simulation studies that the bootstrapping time is less than 1.5 minutes when $n\le 500$. As a result the bootstrap-assisted algorithm is not too expensive to apply in small samples. However, the bootstrapping time increases drastically as sample size increases. In view of the fact that longer and longer time series are being collected in the information age, Algorithm \ref{Algorithm1} is recommended in most real applications with large sample sizes.  
	\begin{algorithm}
		\caption{SIM}\label{Algorithm2}
		\begin{algorithmic}[1]
			\State	Generate $B$ (say 5000) copies of $i.i.d.$ N(0,1) $\{V^{(r)}_i\}_{1\leq i\leq n}$, $r=1,...,B$.
			\State Let $\check H^{(r)}(t,s)=\frac{1}{\sqrt{ns}}\sum_{j=1}^nV^{(r)}_jW\left(\frac{j/n-t}{s}\right)$, and calculate 
			\begin{align}
			\check G^{(r)}(t,\tilde s_n)=\sup_{\underline {s}_n \leq s\leq \bar {s}_n}\frac{|\check H^{(r)}(t,s)|}{\sqrt{\int_{-1}^1 W^2(t)dt}}.
			\end{align}
			\State Let $\check G_1\leq \check G_2\leq...\leq \check G_B$ be the order statistics of
			$\sup_{t\in [0,1]}\check G^{(r)}(t,\tilde s_n)$. Let $\hat c_{1-\alpha}=G_{\lf B(1-\alpha)\rf}$ as the estimate of $c_{1-\alpha}$.
		\end{algorithmic}
	\end{algorithm}
	
	The next theorem discusses the asymptotic behavior of Algorithms \ref{Algorithm1} and \ref{Algorithm2}. Let $$\Delta_n:=\min_{1\le i\le m_n}|\lim_{a\downarrow d_i}\beta_n(a)-\lim_{a\uparrow d_i}\beta_n(a)|$$ denote the smallest jump size of $\beta_n(\cdot)$.
	\begin{theorem}\label{thm2}
		Under conditions of Theorem \ref{Thm1} and (W2), consider Algorithm \ref{Algorithm1}. 
		\begin{description}
			\item (a) If jump points are absent, then $\lim_{n\rightarrow \infty}\p(|J|=0)=1-\alpha$. 
			\item (b) If there are $m_n, m_n\geq 1$ jumps, and scales $\bar {s}_n$ and $\underline {s}_n$ satisfy $\frac{\sqrt{n \bar {s}_n}\Delta_n}{\sqrt{\log n-\log \alpha}}\rightarrow \infty$,$\frac{\Delta_n}{\bar {s}_n}\rightarrow \infty$, $\frac{\sqrt{n\underline {s}_n}\Delta_n}{\log n}\rightarrow \infty$, $\frac{\Delta_n}{(\bar s_n/s_n*)^{1/2}((s^*)^{k+1}+(ns^*)^{-1})}\rightarrow \infty$, $\frac{n(s^*)^2\Delta_n^2}{\bar s_n}\rightarrow \infty$ then 
			\begin{align}\label{April25-Thm4-result}
			\lim_{n\rightarrow \infty}\p(\hat m=m_n, |d_i-\hat d_i|\leq h_n, 1\leq i\leq m_n)=1-\alpha
			\end{align}
			where $\hat d_1,...,\hat d_{\hat m}$ are the estimated jump points, $\hat m$ is the estimated number of jumps, $h_n=\max\{\bar {s}_n^2\Delta_n^{-1},(\frac{\bar {s}_n}{n})^{\frac{1}{2}}\Delta_n^{-1}\log n, \bar {s}_n^{\frac{k+3}{2}}\Delta_n^{-1/2},s^*_n\}g_n$ and $g_n$ is an arbitrarily slowly diverging sequence.
			
			\item (c) The results of (a) and (b) still hold if $c_{1-\alpha}$  is replaced by $\hat c_{1-\alpha}$ of Algorithm \ref{Algorithm2}.
		\end{description}
	\end{theorem}
	Notice that $h_n=o(\bar s_n)$ hence Algorithm \ref{Algorithm1} is valid for any positive $\eta$. Part (a) of Theorem \ref{thm2} indicates that MJPD can be used as an asymptotically accurate multiscale test of the null hypothesis that the trend is smooth. The conditions on $\bar {s}_n$, $\underline {s}_n$ and $\Delta_n$ imply that MJPD is able to identify jumps with magnitude $\Delta_n$ as small as the order of 
	$n^{-1/3+\iota}$ for any $\iota>0$ when $p>12$.  
	Those conditions also guarantee that, with high probability, $G(d_r,\tilde s_n), 1\leq r\leq m_n,$ are much larger than $c_{1-\alpha}$. 
	In the expression of $h_n$, the term $\bar {s}_n^2\Delta_n^{-1}$ is due to the difference between first order derivatives of $\beta_n(\cdot)$ before and after the jumps,  and it disappears if $\beta^{(1)}_n(d_r-)=\beta^{(1)}_n(d_r+)$ for $1\leq r\leq m_n.$ The term $\bar {s}_n^{\frac{k+3}{2}}\Delta_n^{-1/2}$ dominates $\bar {s}_n^2\Delta_n^{-1}$ when $k=1$, but will be negligible if a sufficiently high order filter $W(\cdot)$ is used. Finally, the $s^*_n$ term is caused by the error in estimating the long-run variance of piece-wise locally stationary processes, which vanishes if $\{\varepsilon_{i,n}\}$ is strictly stationary. 

	Conditions (B1)-(B3) and the bandwidth conditions on $\bar {s}_n$ and $\underline {s}_n$ in (b) of Theorem \ref{thm2} put restrictions on 
	the number and magnitude of the jumps,  and on the minimum space between adjacent jumps. In the literature of multiscale inference, \cite{frick2014multiscale} makes similar assumptions on the change sizes and distances between change points. 
	Theorems \ref{Thm1} and \ref{thm2} show that by using MJPD, we are able to control the probability of incorrectly estimating the number of jumps asymptotically at level $\alpha$. Furthermore, all estimated jump points are within a distance $h_n$ of the true ones with probability approaching $1-\alpha$. 
\begin{remark}
If $\alpha=\alpha_n\rightarrow 0$ as $n\rightarrow\infty$, then Theorem \ref{Thm1} implies that $\lim_{n\rightarrow \infty}\p(\hat m=m_n)= 1$. That is, the MJPD detects the correct number of jumps with probability 1 asymptotically. On the other hand, however, the price one needs to pay with smaller $\alpha$ is that the critical value $c=c(n)$ will be larger (as seen from Theorem \ref{Thm1}) and therefore the MJPD is less sensitive to jumps with smaller sizes. This is an analogy to hypothesis testing where reducing the Type-I error rate decreases the power.  In the 
rest of the paper, we shall focus on the case when $\alpha$ is fixed unless otherwise specified.  \end{remark}

	\begin{remark}\label{Remark-SN}
		It follows from proof of Theorem \ref{Thm1} that \begin{align}
		G(d_r,\tilde s_n)
		=\frac{\sqrt {n \bar {s}_n}|\int_0^1W(t)dt|
			|\beta_n(d_r-)-\beta_n(d_r+)|}{\sigma(d_r)\sqrt{2\int_0^1 W^2(t)dt}}+O_p(1)
		\end{align} for $1\leq r\leq m_n$. 
		Therefore the quantity $$SN(W):=\frac{|\int_0^1W(t)dt|}{\sqrt{\int_0^1 W^2(t)dt}}$$ determines the signal-noise ratio of MJPD at jump points and it controls the sensitivity of MJPD to jumps. In this paper, we wish to select $W$ with the highest $SN(W)$ in a relatively large class of filters. Details are given in Section 4.1.  
		
	\end{remark}
	\subsection{Second-stage refinement}\label{Sec:Second-Stage}
	The convergence rate $h_n$ for Algorithm \ref{Algorithm1} established in Theorem \ref{thm2} is slower than the optimal rate for multiple jump point detection. Based on $\{\hat d_r, 1\leq r\leq \hat m\}$ estimated by Algorithm \ref{Algorithm1}, we propose  simple second-stage estimators $\{\tilde d_r, 1\leq r\leq \hat m\}$  which enhance the estimation accuracy of MJPD to the near optimum. For any interval $I\in \mathbb R$, introduce the notation
	\begin{align}
	S_I=\sum_{i \in \lambda(I)}y_i,  \quad \text{where}~~ \lambda(I)=\Big|\Big\{i:\frac{i}{n}\in I\Big\}\Big|.
	\end{align}  For $1\leq r\leq \hat m$,  $z_n\in [0,1]$, $\tilde \alpha>-1$ define
	\begin{align}\label{eq34-May-2019}
	l_r=\hat d_r-(2+\tilde \alpha) z_n, u_r=\hat d_r+(2+\tilde \alpha) z_n,\tilde l_r=\hat d_r-z_n,\tilde u_r=\hat d_r+z_n.
	\end{align} Next for $t\in [l_r,u_r]$, define the local cumulative sum (CUSUM) statistic $V_r(t)$ and the associated local maximizer $\tilde d_r$:  \begin{align}
	V_r(t)=S_{[l_r,t]}-\frac{\lambda([l_r,t])}{\lambda([l_r,u_r])}S_{[l_r,u_r]},\ 
	\tilde d_r=\argmax_{t\in [\tilde l_r,\tilde u_r]}|V_r(t)|.\label{eq37-March25}
	\end{align}
	Observe that the second stage estimators $\{\tilde d_r, 1\leq r\leq \hat m\}$ are obtained by applying CUSUM tests locally to shrinking neighborhoods of the estimates $\{\hat d_r, 1\leq r\leq \hat m\}$ of Algorithm \ref{Algorithm1}. We have the following theorem on the asymptotic behavior of the second-stage estimators when the jump sizes are shrinking to zero.
	\begin{theorem}\label{Thm-Second-Stage}
		Assume  conditions of Theorem \ref{Thm1}, (W2) and (b) of Theorem \ref{thm2} hold. Additionally assume
		\begin{align}\label{bandmz}
		m_n\Delta_n^{p-2}\rightarrow 0, \qquad \frac{nz_n\Delta_n^2}{\log n}\rightarrow \infty,  \qquad z_n\geq h_n,  \qquad \frac{\Delta_n}{z_n}\rightarrow \infty. 
		\end{align}
		Then for any sequence $\iota_n\rightarrow \infty$ arbitrarily slowly,  we have 
		\begin{align}
		\lim_{n\rightarrow \infty }\p\left(\hat m=m_n, \max_{1\leq i\leq m_n}|\tilde d_i-d_i|\leq \frac{\iota_n\log n}{n\Delta_n^2}\right)= 1-\alpha. 
		\end{align} 
	\end{theorem}
	It is well known that $1/(n\Delta_n^2)$ is the parametric rate of jump detection (see for instance \cite{dumbgen1991asymptotic} and \cite{muller1997two}). In this sense the rate established in Theorem \ref{Thm-Second-Stage} is optimal except a factor of logarithm. The condition  $z_n\geq h_n$ guarantees that with high probability the jumps fall into the considered vicinities  of $\{d_i, 1\leq i\leq \hat m\}$. The condition $\frac{\Delta_n}{z_n}\rightarrow \infty$ implies that the series is approximately stationary in the considered neighborhoods, while the condition $\frac{nz_n\Delta_n^2}{\log n}\rightarrow \infty$ means that the neighborhoods contain sufficient amount of data. 
	In practice, one could choose $z_n=\underline {s}_n$ and $\tilde \alpha=-0.5$ as a rule of thumb. Next, the following theorem asserts that  the second-stage estimators achieve a nearly optimal rate when the jump sizes are not shrinking to zero and under some extra mild conditions.

	\begin{theorem}\label{New.Thm4}
		Let $\iota_n$ be a series diverging arbitrarily slowly and  $\Omega_p=\sum_{i=1}^\infty \delta_p(L,i)$. 	Assume  $z_n\geq h_n,  \frac{\Delta_n}{z_n}\rightarrow \infty,$  and the conditions of Theorem \ref{Thm1}, (W2) and the conditions of Theorem \ref{thm2} (b) hold.  
		Then i) if $\Delta_n\geq \eta>0$ for some positive constant $\eta$, $n\bar {s}_n^3\rightarrow \infty$, and $nz_n(m_n^{\frac{1}{p-1}}\log n)^{-1}\rightarrow \infty$, we have
		\begin{align}
		\lim_{n\rightarrow \infty }\p\left(\hat m=m_n, \max_{1\leq i\leq m_n}|\tilde d_i-d_i|\leq \frac{\iota_nm_n^{\frac{1}{p-1}}\log n}{n}\right)= 1-\alpha;
		\end{align} 
		and (ii) if there exists some $0<\beta\leq 2$, such that	 $\frac{nz_n\Delta_n^2}{\log ^{\frac{2}{\beta}}n}\rightarrow \infty$ and
		\begin{align}\label{eq80-March25}
		\gamma:=\limsup_{p\rightarrow \infty}p^{1/2-\frac{1}{\beta}}\Omega_p<\infty,
		\end{align}
		then we have 
		\begin{align}
		\lim_{n\rightarrow \infty}\p\left(\hat m=m_n, \max_{1\leq i\leq m_n}|\tilde d_i-d_i|\leq \frac{\iota_n\log^{\frac{2}{\beta}}n}{n\Delta_n^2}\right)=  1-\alpha.
		\end{align}
	\end{theorem}
	
	i) and ii) of Theorem \ref{New.Thm4} investigate error distributions whose tails are of polynomial and geometric decays, respectively. Note that, for (ii), $\Delta_n$ need not diminish. 
	Equation \eqref{eq80-March25} is a mild condition. The following example shows how to check  \eqref{eq80-March25}  for PLS linear processes. 
	\begin{example}\label{example1}
		Suppose we have the following PLS linear errors 
		\begin{align}\label{Series-42}
		\varepsilon_{i,n}=\sum_{j=0}^\infty a_{r,j}\left(\frac{i}{n}\right)\eta_{i-j}, c_r< \frac{i}{n}\leq c_{r+1}, 0\leq r\leq l,
		\end{align}
		where as in Definition \ref{def1}, $0=c_0<c_1<...<c_l<c_{l+1}=1$ are the unknown break points in the errors. Assume that 
		$\sum_{j=0}^\infty \max_{0\leq r\leq l}\sup_{t\in (c_r,c_{r+1}]}|a_{r,j}(t)|<\infty $. 
		We show in the appendix that equation \eqref{eq80-March25} holds if
		$\|\eta_0\|_p=O( p^{\frac{1}{\beta}-\frac{1}{2}})$ for some $\beta\in(0,2)$, which is equivalent to the moment condition $\E(\exp(t|\eta_0|^{\frac{1}{\frac{1}{\beta}-\frac{1}{2}}}))<\infty$ for some positive constant $t$. 
	\end{example} 
	According to our discussions regarding the results of Theorem \ref{Thm-Second-Stage}, the rate $\frac{\log^{\frac{2}{\beta}}n}{n\Delta_n^2}$ established in Theorem \ref{New.Thm4} (ii) is optimal except a factor of logarithm. In the following we shall discuss the optimality of the results in Theorem \ref{New.Thm4} (i). First, we have the following:
	\begin{corol}\label{Corol1}
		Consider model \eqref{model 2} with i.i.d. symmetric errors $\{\varepsilon_{i,n},1\leq i\leq n\}$, and the tail probability of  $\varepsilon_{1,n}$ satisfies 
		\begin{align}\label{condition48}
		\lim_{|x|\rightarrow \infty} \frac{\p(|\varepsilon_{1,n}|\geq x)}{C^\dag x^{-p}\log^{-2} x}=1
		\end{align}
		where $C^\dag$ is a normalization constant, and $p>2$.  Assume $m_n\log ^{-2}n\rightarrow \infty$, and  $z_n\leq \bar s_n$. Then under the conditions of (i) of Theorem \ref{New.Thm4}, we have that for any $g_n=o(m_n^{\frac{1}{p-1}}\log ^{\frac{2p}{1-p}-2}n)$,
		\begin{align}
		\limsup_{n\rightarrow \infty }\p\left(\hat m=m_n, \max_{1\leq i\leq m_n}|\tilde d_i-d_i|\leq \frac{g_n}{n}\right)<1-\alpha.
		\end{align} 
	\end{corol}
	Observe that equation \eqref{condition48} implies that $\E(|\varepsilon_{1,n}|^p)<\infty$ and $\E(|\varepsilon_{1,n}|^{p+1})=\infty$. Corollary \ref{Corol1} claims that, for i.i.d. errors with tail probability \eqref{condition48}, there is a non-vanishing probability that some of the second-stage estimators $\tilde{d}_i$ will reside outside of the radius $g_n$ range of $d_i$. 
	Hence Corollary \ref{Corol1} implies that the estimation accuracy $\frac{m_n^{\frac{1}{p-1}}\log n}{n}$ in Theorem \ref{New.Thm4} (i) cannot be improved except a factor of logarithm.  
	
	To our knowledge, there have been no results on the parametric jump  point detection rate when there is a diverging number of jumps with non-shrinking jump sizes and the error distribution has polynomial tails. In the following, we explain that the rate established in Theorem \ref{New.Thm4} i) is nearly a {\it parametric} rate. 
	To this end, consider the oracle case where a): the trend is piece-wise constant; b): the number of jumps, $m_n$, is known; c): there exist $m_n$ known non-overlapping intervals and each interval contains exactly one jump point in the interior; and d): the errors are i.i.d. with parametric regular varying tails. We show that in the latter oracle case the accuracy of the local CUSUM estimators are the same as that established in i) of Theorem \ref{New.Thm4} except a factor of logarithm. The result is summarized in Corollary \ref{Corol2-May-2019}. Hence the rate $\frac{m_n^{\frac{1}{p-1}}\log n}{n}$ is nearly parametric for any CUSUM-type detection methods. 
	
	\begin{corol}\label{Corol2-May-2019}
		Assume $\beta_n(\cdot)$ is piece-wise constant with $m_n$ jump points, where $m_n$ is known. Define the associate local CUSUM estimator $\tilde d_r$ as in equation \eqref{eq37-March25} with $\hat d_r$ replaced by $d_r$ in the definition of $l_r, u_r, \tilde l_r, \tilde d_r$ in equation \eqref{eq34-May-2019}. Then under the conditions of Corollary \ref{Corol1}, we have that
		\begin{align}
		\limsup_{n\rightarrow \infty }\p\left(\max_{1\leq i\leq m_n}|\tilde d_i-d_i|\leq \frac{g_n}{n}\right)<1.
		\end{align}
		for $g_n$ defined in Corollary \ref{Corol1}.
	\end{corol}
	\begin{remark}\label{remark-new-bj}
	\textcolor{black}{As pointed out by one referee, \cite{chen2022inference} is also applicable to testing and estimating the break points of univariate time series. Since their paper mainly considers the test of break points for high dimensional time series, their assumptions for trends and component-wise series are stronger than ours. We have discussed in Section  \ref{class-psmooth} that the trends they considered are a sub-class of our piece-wise smooth functions. To accommodate the more flexible class of piece-wise smooth functions we consider a wide class of filters that can further eliminate bias. As a comparison, \cite{chen2022inference} construct the test statistics based on the difference of locally linear estimates of the left and right limits of each point, which will lead to further bias if their requirement that the left and right derivatives are equal at all order (see \eqref{f1}) is violated. Furthermore, our constructed filter allows us to further consider multi-scale tests and estimation, which is particularly important for adapting time series non-stationarity. For the errors, \cite{chen2022inference} assumes stationary vector MA($\infty$) model. In the univariate case, this will reduce to a stationary MA($\infty$) model. In this paper we allow errors to be piece-wise locally stationary. The long-run variance function is time-varying with possible jumps. Hence we specially design an innovative studentization (see the denominator of \eqref{new4-March-26-1}) which cancels the effect of the piece-wise smooth long-run variance on the test statistics) so that we could control $\alpha$, the rate of estimating the incorrect number of jumps.
	A similarity between our method and \cite{chen2022inference} is that both papers apply second-stage refinement using a localized CUSUM method. In theory, we additionally discuss the optimality of our method for errors with sub-exponential-tail error in Theorem \ref{New.Thm4} and polynomial tail in Corollary \ref{Corol1}. We also discuss computationally feasible  Algorithms for practical implementation. On the other hand,  \cite{chen2022inference} discusses the asymptotic distribution of the detected jump points which is still an open problem under time series nonstationarity.} 
\end{remark}
	\section{Implementation}\label{Sec:implement}
	\subsection{The optimal filters}\label{Filter-Construct}
	In this section we discuss the optimal filter $W(\cdot)\in \mathcal W(k)$ which satisfies conditions (W1), (W2) and optimizes the signal noise ratio $SN(W)$  defined in Remark \ref{Remark-SN}. Notice that a necessary condition for $W(\cdot)\in \mathcal W(k)$ is 
	\begin{align}
	\int_0^1 x^uW(x)dx=0, \text{\ for } u=1,3..,(2\lceil k/2\rceil -1).
	\end{align}
	For $W(\cdot)\in \mathcal W(k)$,  the following lemma gives out an upper bound of $SN(W)$.
	\begin{lemma}\label{LemmaSN}
		For any filter $W(\cdot)\in \mathcal W(k)$ satisfying (W1), we have that\begin{equation}
		SN(W)\leq \left \{\begin{array}{lr}
		1/2, \text{for $k=1,2,$}&\\3/8,\  \text{for $k=3,4$}.
		&
		\end{array}
		\right.
		\end{equation}
		Those upper bounds are almost achievable, i.e., for any $\epsilon_0>0$, there exists $W\in \mathcal W(k)$ such that $SN(W)\geq \frac{1}{2}-\epsilon_0$ if $k=2$, and  
		$SN(W)\geq \frac{3}{8}-\epsilon_0$ if $k=4$.
	\end{lemma}
	
	\begin{remark}\label{RemarkSN}
		Since $\mathcal W(k_1)\subset \mathcal W(k_2)$ for integers $k_1\geq k_2$,  $\sup_{W(\cdot)\in \mathcal W(k)}SN(W)$ is non-increasing in $k$. Therefore by Lemma \ref{LemmaSN},  $SN(W)\leq 3/8$ for all $W\in \mathcal W(k)$, $k\geq 5$.
	\end{remark}
	In this paper, we propose a class of piece-wise polynomial filters (i.e., $W(\cdot)$ is a polynomial function on subintervals of $[0,1]$ and $W(-x)=-W(x)$) derived from shifted Legendre approximations to the SN-optimised filters. Due to their low-order-polynomial form, those filters allow us to compute MJPD efficiently using fast sum updating algorithms; see Section \ref{Compu-Issue} for the details. Additionally, those filters suffer only a small loss of efficiency compared with the optimal filter in ${\cal W}(k)$. We remark here that piece-wise polynomials have been extensively investigated and broadly applied in kernel non-parametric studies. For instance, Theorem 3.1 of \cite{zhang2000minimax} proved that the minimax kernels are piece-wise polynomials by solving a variational problem.

\begin{figure}
		\centering
		\includegraphics[width=12cm,height=4cm]{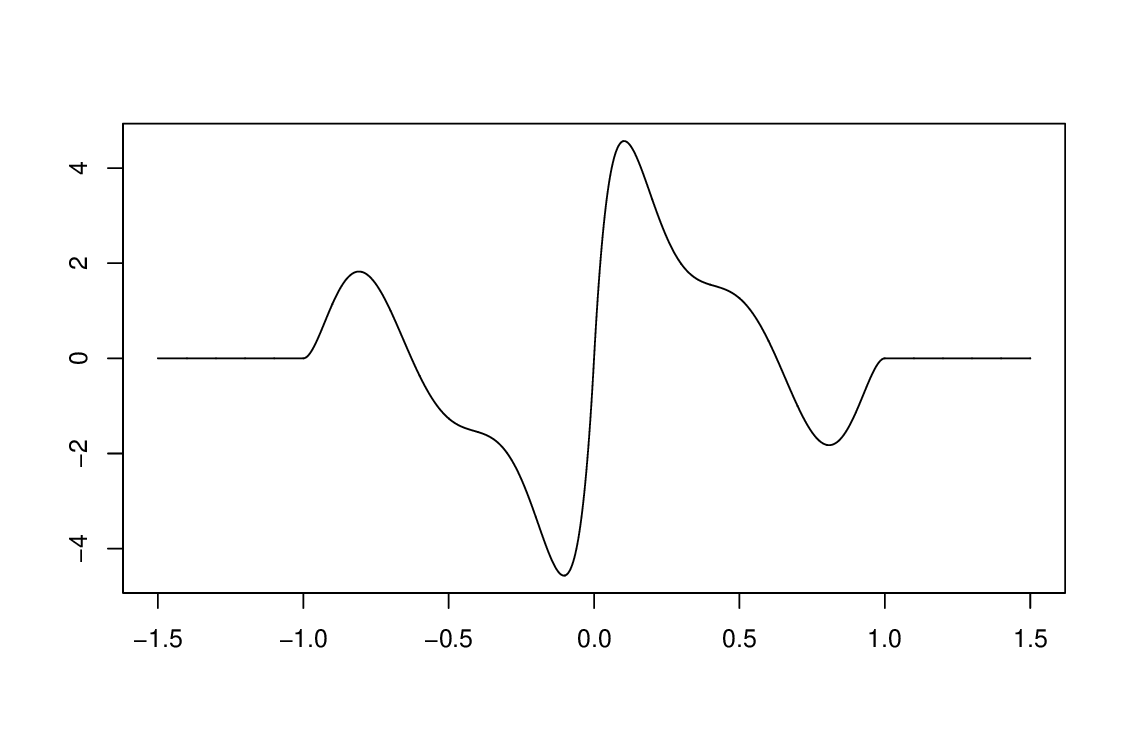}
		\vspace{-.4cm}
		\caption{\it The filter $W^*(\cdot)$ defined in equation \eqref{FilterW-order2-N6}.  }
		\label{Filter}
	\end{figure}

	Let $\mathcal {P}^n$ be the collection of all polynomials with degree $n$.  Define the class $\mathcal W_{k,N}=\{f:f\in \mathcal W(k)\mbox{ and } f\in\mathcal {P}^N \mbox{ on }[0,1]\}$. For given $k,N$, we compute the SN-optimized filter $W$ in  $\mathcal W_{k,N}$ 
	by the discrete Lagrange multiplier. The details of the filter construction are omitted and can be found in the proof of Lemma \ref{LemmaSN} in the supplemental material.  Notice that Remark \ref{RemarkSN} is in favor of small $k$ to maintain high efficiency while Theorem \ref{thm2} suggests $k\geq 2$. Moreover
	large $N$ should be avoided to maintain the smoothness of the optimal filters and the numerical stability of the fast sum updating algorithms. In order to balance all the aforementioned issues, we recommend using the optimal filter in $\mathcal W_{2,6}$.  This filter is 
	given by
	\begin{align}\label{FilterW-order2-N6}
	W^*(x)&=(-1294.2222x^6+4246.6667|x|^5\notag\\&-5320x^4+ 3188.8889 |x|^3-933.3333x^2+112|x|)\it{sgn}(x)
	\end{align}
	with $SN(W^*)\simeq 0.4606
	$. 
	Note that the highest SN for filters in ${\cal W}(2)$ is 0.5. Hence $W^*$ achieves $0.4606/0.5=92.1\%$ efficiency compared to the optimal filter in ${\cal W}(2)$. Furthermore, by Lemma \ref{LemmaSN}, $SN(W^*)$ is larger than the SN of any filter $W(\cdot)\in \mathcal W(k), k\geq 3$.  Straightforward calculations show that conditions (W1) (with $k=2$) and (W2) hold for $W^*$. This filter is displayed in Figure \ref{Filter}, 
	and is a $\mathcal C^1(\mathbb R, C_{Lip})$ function for some constant $C_{Lip}$. We remark here that although $\mathcal C^3(\mathbb R, C_{Lip})$ property is required by Theorem \ref{Thm1}, we advocate the use of a $\mathcal C^1(\mathbb R, C_{Lip})$ filter in practice. Theoretical justification lies in the smooth approximation theory (Corollary 21 of \cite{hajek2010smooth}) which guarantees that there exists a $\mathcal C^3(\mathbb R, C_{Lip})$ function $g(\cdot)$ such that $g(\cdot)$ and $g'(\cdot)$ well approximates $W^*(\cdot)$ and $[W^*]'(\cdot)$ in the $\mathcal L^2$ space. The latter approximation result indicates that the results of Theorem \ref{Thm1} are valid for $W^*$.
	
	To close this subsection, we discuss the global condition i) of (W2). Despite the fact that  this global condition holds for $W^*$,  it is not necessary satisfied by general filters $W(\cdot)\in \mathcal W(k)$.  Based on the invertibility of Hilbert matrices, in Theorem \ref{thm4-2021} of the supplemental material we show that for any $k>0$, there always exists a $q_{th}$ ($q$ is sufficiently high) order piece-wise polynomial filter that meets both conditions (W1) and (W2). 

	\subsection{Efficient computation}\label{Compu-Issue}
	At each scale $s$, a direct computation of $\{H(\frac{i}{n},s), 1\leq i\leq n\}$ requires an $O(n^2s)$ operations, which is costly for large scale inference. Thus we propose to evaluate  $\{H(\frac{i}{n},s), 1\leq i\leq n\}$ using the {\it fast sum updating  algorithm  } (c.f., e.g. \cite{langrene2019fast}) at the cost of $O(n)$ operations. In particular, to calculate MJPD with the piece-wise polynomial filter $W^*$, one has to compute terms in the form of $\sum_{i=1}^n(\frac{i}{n}-\frac{j}{n})^ay_i\mathbf{1}(|\frac{i}{n}-\frac{j}{n}|\leq s)$ for some integer $a$ and $j=1,2,..n$. To illustrate the updating algorithm, consider $a=2$ and we have 
	\begin{align}
	\sum_{i=1}^n\big(\frac{i}{n}-\frac{j}{n}\big)^2y_i\mathbf{1}\big(\big|\frac{i}{n}-\frac{j}{n}\big|\leq s\big)=\sum_{k=0}^2\sum_{i=1}^n {2\choose k}y_i\big(\frac{i}{n}\big)^k\big(\frac{-j}{n}\big)^{2-k}\mathbf 1\big(\big|\frac{i}{n}-\frac{j}{n}\big|\leq s\big).
	\end{align}
	To illustrate how the updating algorithm works, consider the term where $k=1$ for example and we have the following expression
	\begin{align}
	\sum_{i=1}^ny_i\big(\frac{i}{n}\big)\big( \frac{j}{n}\big)\mathbf 1\big(\big|\frac{i}{n}-\frac{j}{n}\big|\leq s\big)=\big(\frac{j}{n}\big)\sum_{i=1}^ny_i\big(\frac{i}{n}\big) \mathbf 1\big(\big|\frac{i}{n}-\frac{j}{n}\big|\leq s\big):=\frac{j}{n}\Theta_j,\end{align}
	where $\Theta_j=\sum_{i=1}^n\frac{iy_i}{n} \mathbf 1(|\frac{i}{n}-\frac{j}{n}|\leq s)$. For simplicity let $ns$ be an integer. Then one can compute $\Theta_j$, $1\leq j\leq n$ using $O(n)$ operations by updating based on the following identity
	\begin{align}
	\Theta_j=\Theta_{j-1}-y_{j-ns-1}\left(\frac{j-ns-1}{n}\right)+y_{j+ns}\left(\frac{j+ns}{n}\right).
	\end{align}
	We refer the readers to Section \ref{Section_Algorithm3} in the supplemental material for the detailed description of the algorithm.
	The fast sum updating algorithm has been an  attractive approach to reduce the computation complexity in nonparametric analysis, see for instance  \cite{seifert1994fast}, \cite{fan1994fast}, \cite{langrene2019fast} among others. Past studies   (e.g. \cite{fan1994fast}, \cite{seifert1994fast}) pointed out that the updating algorithm may cause numerical instability. This issue is more severe when high degree polynomials are involved (see \cite{seifert1994fast}). However, as pointed out by \cite{langrene2019fast}, the issue has been largely addressed by recent progress in computer science  (e.g.  \cite{zhu2010algorithm}).
	
	In practice, instead of directly computing $G(t,\tilde s_n )$  in equation \eqref{new4-March-26-1}  over the region $(t,s)\in [0,1]\times [\underline {s}_n, \bar {s}_n]$,  we evaluate it over a carefully designed sparse sequence of scales with which MJPD achieves the estimation accuracy in Theorem \ref{Thm1} at a computational cost of $O(n \log^{1+\epsilon}n)$ for some $\epsilon>0$.  Let $\delta_n=\lf (\log n)^{1+\epsilon}\rf$ and set the sequence $s_i=2^{g_i}$ where \begin{align}\label{scale-grid}g_i=\log_2 \underline {s}_n+(i-1) \frac{\log_2 \bar {s}_n-\log_2 \underline {s}_n}{\delta_n-1}, \ \ 1\leq i\leq \delta_n.\end{align}
	We compute 
	\begin{align}\label{142-Sep3}
	G^\dag(t, \tilde s_n):=\max_{1\leq i\leq \delta_n}\frac{|H(t,s_i)|}{\sqrt{\sum_{i\in K(t)}H^2(i/n,s^*_n)/|K(t)|}}.
	\end{align}
	\begin{theorem}\label{Save_b}
		Assume the conditions of Theorem \ref{Thm1} hold and $W''$ exists on $[-1,1]$ except on a finite number of points.  We then have that
		\begin{align}
		\sup_{t\in T_n^d} |G(t, \tilde s_n)-G^\dag(t, \tilde s_n)|=O_p(\log^{\frac{1}{2}-\epsilon} n\log \log n).
		\end{align}
	\end{theorem}
	For $t\in \bar T_n^d$, by definition we have $G^\dag(t, \tilde s_n)\leq G(t,\tilde s_n)$.  Under conditions of Theorem \ref{thm2}, similar arguments to the proof of Theorem \ref{thm2} yield that with probability tending to one,
	\begin{align}G^\dag (d_s,\tilde s_n)=G(d_s,\tilde s_n )=\frac{|H(d_s,\bar {s}_n)|}{\sqrt{\sum_{i\in K(d_r)}H^2(i/n,s^*_n)/|K(d_r)|}}, \ \forall 1\leq r\leq m_n.
	\end{align}  The above fact and Theorem \ref{Save_b} indicate that the results of Theorem \ref{thm2} remain valid if we evaluate MJPD via $G^\dag(t, \tilde s_n)$. As discussed, the computational complexity for the updating algorithm to evaluate $\{H(\frac{i}{n},s),i=1,...,n\}$ is $O(n)$  for any single scale $s>0$. Therefore, the total computational cost to calculate  $\{G^\dag (\frac{i}{n},\tilde s_n),1\leq i\leq n\}$ is $O(n\log^{1+\epsilon}n)$. Furthermore, due to the fact that $m_n\bar {s}_n=o(1)$,  the computational cost of the second stage refinement in Section \ref{Sec:Second-Stage} is $O(n)$. Hence the computational cost for MJPD is $O(n\log ^{1+\epsilon}n )$. This cost can be further reduced  through computing the $\delta_n$ number of series  $\{H(\frac{i}{n},s_u), 1\leq i\leq n\}$, $1\leq u\leq \delta_n$ independently in parallel. 
	\subsection{Choices of $\bar {s}_n$, $\underline {s}_n$ and $s^*_n$} \label{Parameter:Selection}

For an easy implementation, the following  rule-of-thumb choice: $\bar s_n=\min(1/(2L_n),n^{-1/6})$,
$\underline s_n=\min(\bar s_n/2,n^{-1/3}/2)$ $\min(1,6/\log n),$
$s_n^*=\min(n^{-1/2}\log n/6,\underline{s}_n)$ performed well in our numerical experiments, where $n$ is the sample size, and  $L_n$ is the maximum allowed number of segments determined by the users. 

In many real data applications, a data-driven choice of the tuning parameters may be desirable. In this case, we propose the following data-driven method to select $\bar{s}_n$, $\underline{s}_n$ and $s^*_n$ for researchers and practioners who wish to choose those parameters adaptively. First, if $\bar{s}_n$ and $\underline{s}_n$ are determined, $s^*_n$ can be selected via the minimum volatility (MV) method advocated by \cite{politis1999subsampling}. The MV method is useful in the literature of non-stationary time series analysis  (exemplarily \cite{zhou2013heteroscedasticity}, \cite{rho2019bootstrap}) since it is independent of any specific form of the underlying dependence structure. To implement the MV method, we consider a sequence of candidate scales $\frac{1}{6}n^{-1/2}\log ^{\frac{1}{2}}n=s^*_{1,n}<..<s^*_{i,n}..<s^*_{M,n}=\underline {s}_n$, $M>0$, and calculate the denominator of statistics $G(t,\tilde s_{n})$ defined in equation \eqref{new4-March-26-1}, that is
$$\left\{\Xi^{\frac{1}{2}}(\frac{j}{n},s^*_{i,n})\right\}_{1\leq j\leq n}:=\left\{\left(\sum_{i\in K(\frac{j}{n})}H^2(i/n,s^*_{i,n})/|K(j/n)|\right)^{\frac{1}{2}}\right\}_{1\leq j\leq n}.$$ Then for each $s_{r,n}^*$ we calculate $SE_r(\cdot)$, the standard error that measures the uniform variability of 
$\Xi^{\frac{1}{2}}(\cdot, \cdot)$ which is given by 
\begin{align*}
SE_r:=\max_{1\leq j\leq n}\big(\frac{1}{2k}\sum_{i=r-k}^{r+k}(\Xi^{\frac{1}{2}}(\frac{j}{n},s^*_{i,n})-\bar {\Xi}_r^{\frac{1}{2}}(\frac{j}{n},s^*_n) )^2\big),\bar {\Xi}_r^{\frac{1}{2}}(\frac{j}{n},s^*_n)=\frac{1}{2k+1}\sum_{i=r-k}^{r+k}\Xi^{\frac{1}{2}}(\frac{j}{n},s_{i,n}^*).
\end{align*}
Here $k$ is typically chosen as 2 or 3. We then set $s^*_n=s_{i,n}^*$ where $i=\argmin_r SE_r$. 

Meanwhile, the MV method also leads to a data-driven rule for selecting the lower and upper scales $\underline {s}_n$ and $\bar {s}_n$. The idea is that the estimated number of jump points should be stable when the pair $(\underline {s}_n, \bar {s}_n)$ is in an appropriate range. Consider two candidate sequences $\underline {s}_{1,n}^*<\underline {s}_{2,n}^*<...<\underline {s}_{k_1,n}^*$ and  $\bar {s}^*_{1,n}<\bar {s}_{2,n}^*<...<\bar {s}^*_{k_2,n}$. The range of the candidate upper and lower scales can be determined by prior knowledge of the data. Alternatively, rule-of-thumb choices of $\underline {s}_{1,n}^*$, $\underline {s}_{k_1,n}^*,$ $\bar {s}^*_{1,n}$ and $\bar {s}^*_{k_2,n}$ could be $n^{-1/3}/4$, $ n^{-1/3}/2$, $ n^{-1/6}/6$ and $ n^{-1/6}/3$, respectively.
Let $m( \underline {s}_{i,n}^*, \bar  s_{j,n}^*)$ be the number of jump points detected by MJPD using $\underline {s}_n= \underline {s}_{i,n}^*$ and $\bar {s}_n= \bar {s}_{j,n}^*$. Define \begin{align*}S(i,j,k_3)=\{(\underline {s}_{a,n}^*,\bar {s}^*_{b,n}):\underline {s}_{a,n}^*<\bar {s}^*_{b,n}, i-k_3\leq a\leq i+k_3, j-k_3\leq b\leq j+k_3, a,b\in \mathbb Z \},\\
SE(i,j)=\frac{1}{|S(i,j,k_3)|-1}\sum_{(a,b)\in S(i,j,k_3)}\left(m(\underline {s}_{a,n}^*,\bar {s}_{b,n}^*)-\frac{1}{|S(i,j,k_3)|}\sum_{(a,b)\in S(i,j,k)}m(\underline {s}_{a,n}^*,\bar {s}_{b,n}^*)\right)^2.
\end{align*} 
We then select $(\underline {s}_n,\bar {s}_n)=(\underline {s}_{i,n}^*, \bar {s}^*_{j,n})$ where $(i,j)$ is the minimizer of $SE(u,v)$ over  $k_3+1\leq u\leq k_1-k_3$ and  $k_3+1\leq v\leq k_2-k_3$. If there are multiple minimizers, then we choose, among the minimizers, the pair $(i,j)$ with the smallest $\underline {s}_{i,n}^*+\bar {s}_{j,n}^*$.  $k_3$ is typically chosen as 2 or 3. The computational complexity for searching the minimizer is $O(k^2_1k^2_2k^2_3n\log^{1+\epsilon}n)$.

	\section{Simulation studies}\label{Sec::Simu}
We study the finite sample performance of MJPD on estimating jumps over various simulated scenarios and a real data set. 
	All simulation results are averaged over 2000 iterations, and are obtained by a desktop computer with intel i7-8700 CPU. Due to page constraints, additional simulation studies for $n=5000$, 
		 a sensitivity analysis of the tuning parameters, and the accuracy and power of MJPD as a test of jumps can be found in Sections A to C of the supplementary material. 	
	
	\subsection{Identifying jump points}\label{Sec::Simuchange}

	In the simulation reports, the experiment results of Algorithm \ref{Algorithm1} and Algorithm \ref{Algorithm2} are denoted by MJPD and SIM, respectively. To implement MJPD, we use $\epsilon=0.5$ in $\delta_n$ of equation \eqref{scale-grid}, and $\alpha=0.01$. 
	The tuning parameters $\bar {s}_n$, $\underline {s}_n$ and $s^*_n$  are selected according to Section \ref{Parameter:Selection}.  
For each scenario and method, we investigate the simulated probability of detecting all jumps (denoted by ``$\hat m=m_n$''),  the average mean absolute deviation (denoted by ``MAD'') of the estimated locations when all jumps have been identified, the average number of detected jumps (denoted by ``mean $m$''), and the average computational time (denoted by ``Time'') for executing a corresponding algorithm. 
	For MJPD we also record the average time cost for a single scale (denoted by ``Single.sec''), which indicates the potential time cost of MJPD if parallel computing techniques are used. 
	For SIM, $\hat c_{1-\alpha}$ is generated separately via 5000 bootstrap samples using Algorithm \ref{Algorithm2}. 

	\subsubsection{Results of 500 sample size} \label{Sec::500_Sample_Size}
	Consider the following models for error $\varepsilon_{i,n}$, where the filtration $\FF_i=(...,\eta_{i-1},\eta_i)$ and $\{\eta_i\}_{i\in \mathbb Z}$ will be specified in each model. 
	\begin{description}
		\item(GS) $\varepsilon_{i,n}=\eta_i$, where $\eta_i's$ are i.i.d. $N(0,1)$.
		\item(PS) Let $\{\eta_i\}_{i\in \mathbb Z}$ be $i.i.d$  standardized ($\E(\eta_i)=0, \text{Var}(\eta_i)=1$ ) $\chi^2$ distribution with 3 degrees of freedom. The error $\varepsilon_{i,n}=\frac{3}{4}G_0(i/n,\FF_i)$ for $i/n\leq 0.5$ and $\varepsilon_{i,n}=\frac{5}{4}G_1(i/n,\FF_i)$ for $i/n>0.5$, where \begin{align}G_0(t,\FF_i)=0.25G_0(t,\FF_{i-1})+\eta_i,~~~
		G_1(t,\FF_i)=-0.25G_1(t,\FF_{i-1})+\eta_i.
		\end{align}
		\item(ARMA) $\varepsilon_{i,n}$ is generated from an ARMA(1,1) process with long run variance $1$, i.e.  $\varepsilon_{i,n}=X_i/2.142857$ where $X_i-0.3X_{i-1}=\eta_i+0.5\eta_{i-1}$   and $\eta_i's$ are $i.i.d.$ $N(0,1)$.
		\item(LS) Let $\{\eta_i\}_{i\in \mathbb Z}$ be $i.i.d$ Rademacher random variables and $\varepsilon_{i,n}=g(i/n)G(i/n,\FF_i)$ where \begin{align}G(t,\FF_i)=(0.5t-0.2)G(t,\FF_{i-1})+\eta_i, \quad g(t)=1+0.5t.\end{align}
		\item(PLS) Let $\{\eta_i\}_{i\in \mathbb Z}$ be $i.i.d$ $t(8)/\sqrt{4/3}$. The error $\varepsilon_{i,n}=0.9G_0(i/n,\FF_i)$ for $i/n\leq 0.4$, and $\varepsilon_{i,n}=0.9G_1(i/n,\FF_i)$ for $i/n>0.4$, where \begin{align}G_0(t,\FF_i)=0.5\sin(\pi t)G_0(t,\FF_{i-1})+\eta_i+(0.2-0.5t)\eta_{i-1},\notag \\ G_1(t,\FF_i)=(0.5-t)G_1(t,\FF_{i-1})+\eta_i+\frac{(t-0.2)^2}{2}\eta_{i-1}.\end{align}
	\end{description}
	Model (GS) and (ARMA) are stationary. (PS) is a piece-wise stationary AR(1) process driven by asymmetric innovations. Before and after the error break point $0.5$, the long run variance of model PS is unchanged.  Model (LS) is a smooth time-varying AR(1) process with discrete innovations. Model (PLS) is a piece-wise locally stationary process. Before and  after the error break point, PLS are two distinct time-varying ARMA(1,1) processes. For $\beta_n(\cdot)$ we consider a step function (I) and a piece-wise smooth function (II):
	\begin{description}
		\item (I)$\beta_n(t)=3$ for $0\leq t\leq 0.2$, $\beta_n(t)=0$ for $0.2< t\leq 0.7$ and $\beta_n(t)=-3$ for $0.7< t\leq 1$.
		\item (II) $\beta_n(t)=(5\sin(\pi t)+2.75)\mathbf 1(0\leq t\leq 0.3)+(5\sin(\pi t)-0.75)\mathbf 1(0.3<t\leq 2/3)+(5\sin(2\pi/3)+2.75)(1-10(t-2/3)^2)\mathbf 1(2/3<t\leq 1)$.
	\end{description}

	\begin{table}[htbp]
		\centering
		\tiny
		\caption{{\small Results of MJPD and SIM for $n=500$, with Mean Models I and II ($m=2$) }}
		\begin{tabular}{cccccccccccc}
			\toprule
			&       & \multicolumn{5}{c}{Mean Model I}      & \multicolumn{5}{c}{Mean Model II} \\
			\cmidrule(lr){3-7}\cmidrule(lr){8-12}
			& Measure &{ GS } &{ ARMA } &{ PS } &{ LS } &{ PLS} &{ GS } &{ ARMA } &{ PS } &{ LS } &{ PLS} \\
			\hline
			\multirow{5}[0]{*}{MJPD } 
			&	 $\hat m=m_n$ (\%) & 97.10 & 96.20 & 96.10 & 95.60 & 95.35 & 100.00 & 99.95 & 99.85 & 99.90 & 99.65 \\ 
			& MAD $(\times 10^{-3})$ & 0.347 & 0.034 & 0.379 & 1.101 & 1.517 & 0.504 & 0.348 & 0.538 & 0.451 & 0.584 \\ 
			& mean $m$ & 2.0290 & 2.0380 & 2.0390 & 2.0160 & 2.0275 & 2.0000 & 2.0005 & 2.0015 & 2.0010 & 1.9985 \\ 
			&Time ($\times 10^{-2}$) & 2.318 & 2.330 & 2.324 & 2.251 & 2.227 & 2.196 & 2.273 & 2.313 & 2.255 & 2.206 \\ 
			&Singel.sec ($\times 10^{-2}$) & 0.515 & 0.527 & 0.533 & 0.510 & 0.513 & 0.511 & 0.511 & 0.515 & 0.518 & 0.521 \\

			\hline		\multirow{4}[0]{*}{SIM }
			&	 $\hat m=m_n$ (\%)& 95.10 & 94.40 & 92.90 & 92.85 & 92.45 & 99.80 & 99.90 & 99.75 & 99.35 & 99.65 \\ 
			& MAD $(\times 10^{-3})$ & 0.363 & 0.091 & 0.347 & 1.137 & 0.689 & 0.516 & 0.343 & 0.520 & 0.457 & 0.632 \\ 
			& mean $m$ & 2.0490 & 2.0560 & 2.0710 & 2.0010 & 2.0605 & 2.0020 & 2.0010 & 2.0025 & 2.0025 & 2.0005 \\ 
			&Time ($\times 10^{-2}$)  & 2.498 & 2.503 & 2.484 & 2.474 & 2.536 & 2.502 & 2.509 & 2.494 & 2.476 & 2.489 \\

			\hline
		\end{tabular}%
		\label{Table2-May-2019}%
	\end{table}%
	The locations and magnitudes of jumps are $(0.2,0.7)$ and $3$ for mean (I),  $(0.3,2/3)$ and $3.5$ for mean (II), respectively. The sample size $n=500$. The results corresponding to I and II are displayed in  Table \ref{Table2-May-2019}. As predicted by Theorems \ref{Thm1}-\ref{New.Thm4}, the results  in Table  \ref{Table2-May-2019}  show that both MJPD and SIM are suitable for the purpose of detecting jumps in a piece-wise smooth signal under complex temporal dynamics. In this study, the time costs to generate $\hat c_{1-\alpha}$  (see SIM) for mean I, II are 92.67s and 95.84s, respectively.
	\subsection{Additional Simulation Results}\label{additional-simu}
		\textcolor{black}{Due to the page limit, we present the remaining extensive simulation results in the supplemental material. 
	 Specifically, in the supplemental material,  we compare the performance of our method with some change point methods, namely the DSMUCE (\cite{dette2020multiscale}) and PELT (\cite{killick2012optimal} ) methods which are representative  change point detection algorithms for time series. Our results show that our method performs well and is comparable with the existing methods under the classic change point setting; i.e. estimating piece-wise constant signal from strictly stationary noise, while the existing prevalent change point detection algorithms identify many spurious jumps when there is a smooth trend. Moreover, nonstationary errors lower the probability of detecting the correct number of jumps for the two conventional change point algorithms. However, we must acknowledge that the above-mentioned two change point algorithms are not designed for complex trends and noises.}

\textcolor{black}{In the supplement, we also check the performance of MJPD with $5000$ sample size, and increasing sample sizes (examining (i) the closeness between the theoretical and simulated critical values; (ii) the effect of the second-stage refinement in Section \ref{Sec:Second-Stage}) to justify the asymptotic correctness and effectiveness  of two-stage jump detection method in large samples. We then (a) conduct a sensitivity analysis of our MJPD concluding that our proposed method is relatively robust to the choices of the tuning parameters and filters and (b) examine the type 1 error and power of MJPD as a test of jumps using different filters. The results of (a) and (b) also support the superiority of our proposed filter $\mathcal W_{2,6}$. Please refer to Sections A-C of the supplementary material for the details.}

	\section{S\&P 500 analysis} \label{sp500]}
\begin{figure}[t]
		\centering
		\begin{minipage}{0.45\textwidth}
			\centering
			\includegraphics[width=6cm,height=4cm]{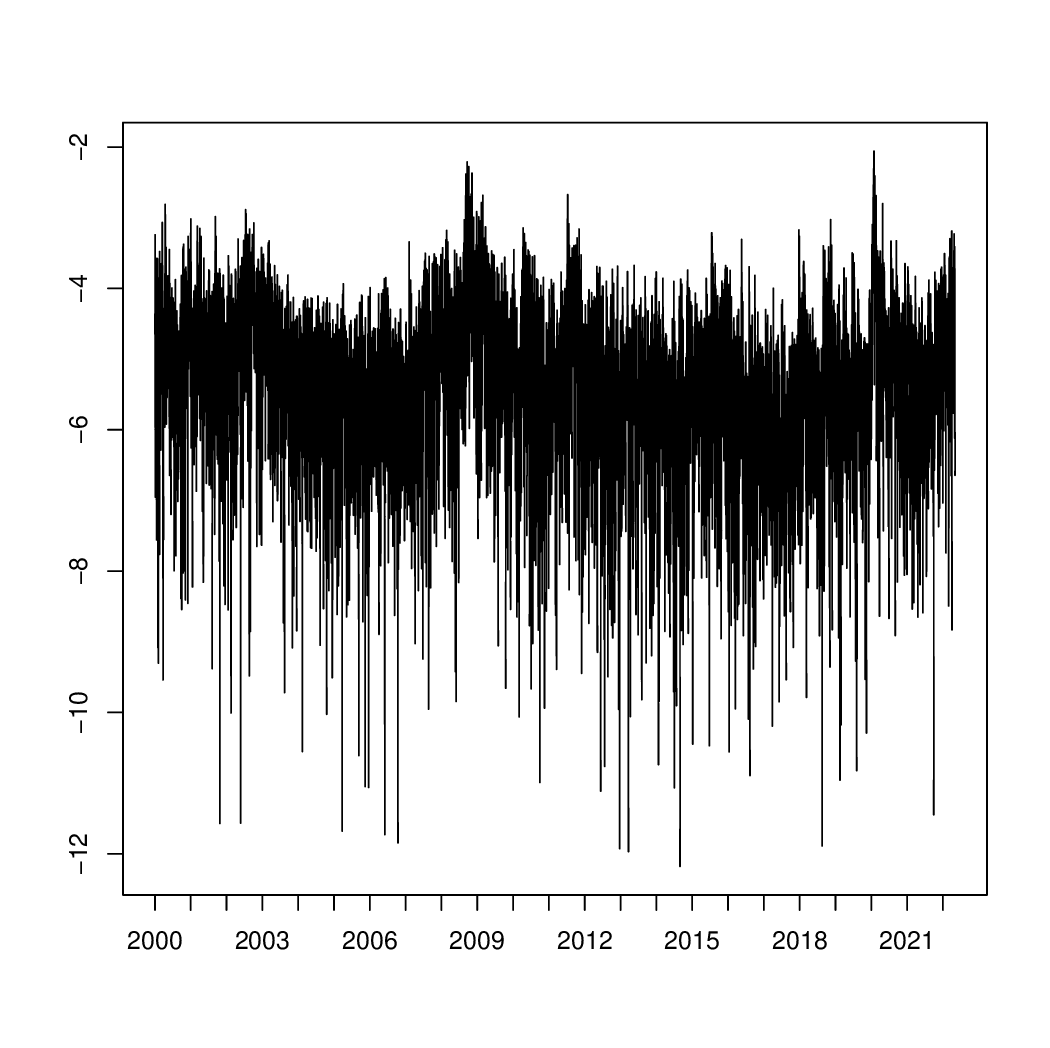}
			\vspace{-.4cm}
			\caption{{\small \it log absolute return of SP500 }}
				\label{SP500-0}
		\end{minipage}
		\begin{minipage}{0.45\textwidth}
			\centering
			\includegraphics[width=6cm,height=4cm]{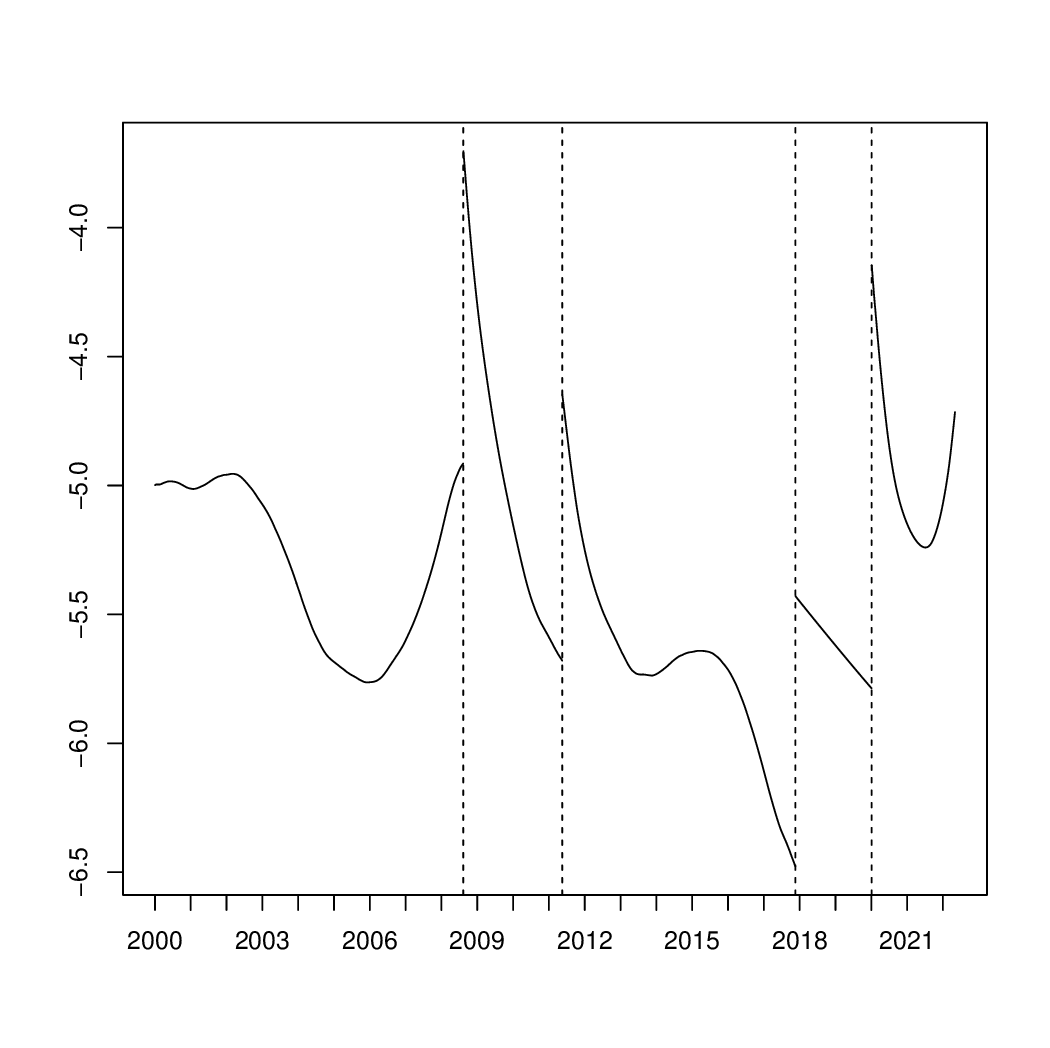}
			\vspace{-.4cm}
			\caption{{\small \it Jump points and the fitted trends}} 
				\label{SP500}
		\end{minipage}
	
	\end{figure}

	We now study the daily closing value of S\&P 500 Index between 31 Dec. 1999 and 22 June. 2022, which is denoted by $p_t$. Consider the daily log return 
	\begin{align}
	r_t=\log p_t-\log p_{t-1}.
	\end{align}  
	\cite{stuaricua2005nonstationarities} studied the non-stationarities of S\&P 500 returns and concluded that the dynamics of this series are mostly concentrated in the shifts of the unconditional variance. As a result, they consider the following model:
	\begin{align}
	y_t=\log |r_t|=\mu(t)+\sigma(t)\varepsilon_t, \label{model-SP500}
	\end{align}
	with time varying functions $\mu(\cdot)$ and $\sigma(\cdot)$ and error $\varepsilon_t$.  Observe that the dynamics of the spread of $r_t$ are reflected in the time-varying pattern of the function $\mu(\cdot)$. Assuming that $\varepsilon_t$ are $i.i.d.$ with mean $0$ and variance $1$, \cite{stuaricua2005nonstationarities} studied the continuously and significantly changing dynamics of the series by approximating model  \eqref{model-SP500} locally via stationary models. Those authors proposed a test-based method to  construct the homogeneity intervals of $\mu(\cdot)$ for model \eqref{model-SP500}. The data is displayed in Figure \ref{SP500-0}.
	
	To reflect better the non-stationarity of the data, we assume that $\varepsilon_t$ is a piece-wise locally stationary process 
	and $\mu(t)$ is a piece-wise smooth function. We aim to identify those intervals where $\mu(t)$ is smoothly changing.  
	For this purpose we apply MJPD to model \eqref{model-SP500}.
	In the analysis we exclude $r_{759}$, $r_{2012}$ and $r_{4283}$ since they amount to $0$. 
	We choose $\underline {s}_n= 0.0174
	$, $\bar {s}_n=  0.05 $, and $s^*_n=0.00766
	$  via the MV method stated in Section A.2 of the supplementary material. We consider $\alpha=0.01$, and  obtain $ c_{1-\alpha}=  4.658
	$ by Theorem \ref{Thm1}. We also apply Algorithm \ref{Algorithm2} in the main article to obtain the simulated $\hat c_{1-\alpha}=  4.677$ with $2571.36$ seconds computation time. Both choices of $\hat c_{1-\alpha}$  detect  $2178_{th}$, $2877_{th}$, $4524_{th}$, $5062_{th}$ 
	observations as  jumps, which correspond to Aug. 29, 2008, Jun. 9, 2011, Dec. 26, 2017 and Feb. 18, 2020. The first date is near the critical date Sep. 7, 2008 when Fannie and Freddie, two large home mortgage companies created by the U.S. Congress, were nationalized by the US Government. The second is close to  ``August 2011 stock markets fall" due to European sovereign debt crisis. The third is close to Jan. 22, 2018 when US announced tariffs on solar panels and washing machines which marked the beginning of China–United States trade war. The final date is related to the onset of Covid-19. On Feb. 25, 2020,  CDC warned public that COVID-19 was ``Heading Toward Pandemic Status." 
	We present the fitted piece-wise smooth $\mu(t)$ in Figure \ref{SP500} , where $\hat{\mu}(t)$ is obtained by performing
	local linear kernel regression on the  subseries between the estimated jump. It can be observed that there is a surge in volatility around the estimated dates which can be characterized as jumps. Other variations of $\mu(t)$ are better characterized by smooth changes according to the MJPD algorithm. 
	
	We further 
	apply the test in \cite{dette2015change} to the five periods to check the constancy of the auto-covariance structure of $y_t$. We identify no evidence of structural breaks in the lag-3 autocorrelations at $10\%$ significance level. For the variance, our test implies a structural change in the second period with $p$-value $4.04\%$. For the lag-1 autocorrelation, test in \cite{dette2015change}  yields a $p$-values of $3.45\%$, $8.41\%$ and  $2.26\%$ for the second, third and fifth period, respectively. For the lag-2 autocorrelation, test in \cite{dette2015change} obtains a $p$-value $5.51\%$ for the first period. The test results indicate that the auto-covariance structure of $y_t$ is non-stationary in this case. As a result, jump detection algorithms based on stationary errors may not work accurately for this data set.
	
\begin{figure}[t]
		\centering
		\begin{minipage}{0.45\textwidth}
			\centering
			\includegraphics[width=6cm,height=4cm]{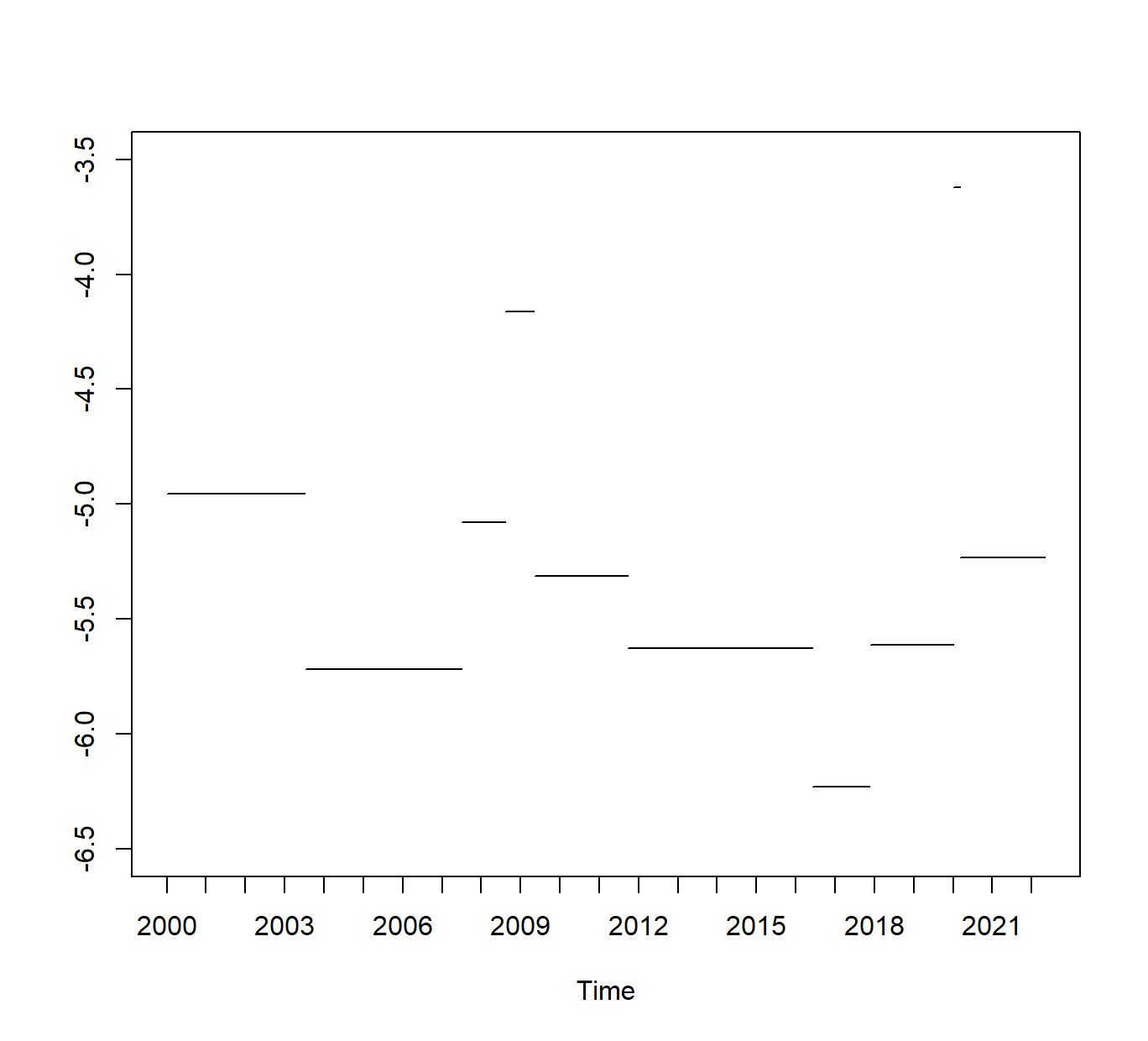}
			\vspace{-.4cm}
			\caption{{\small \it SP500, Fitted mean by DSMUCE }}
				\label{SP500-constant-0}
		\end{minipage}
		\begin{minipage}{0.45\textwidth}
			\centering
			\includegraphics[width=6cm,height=4cm]{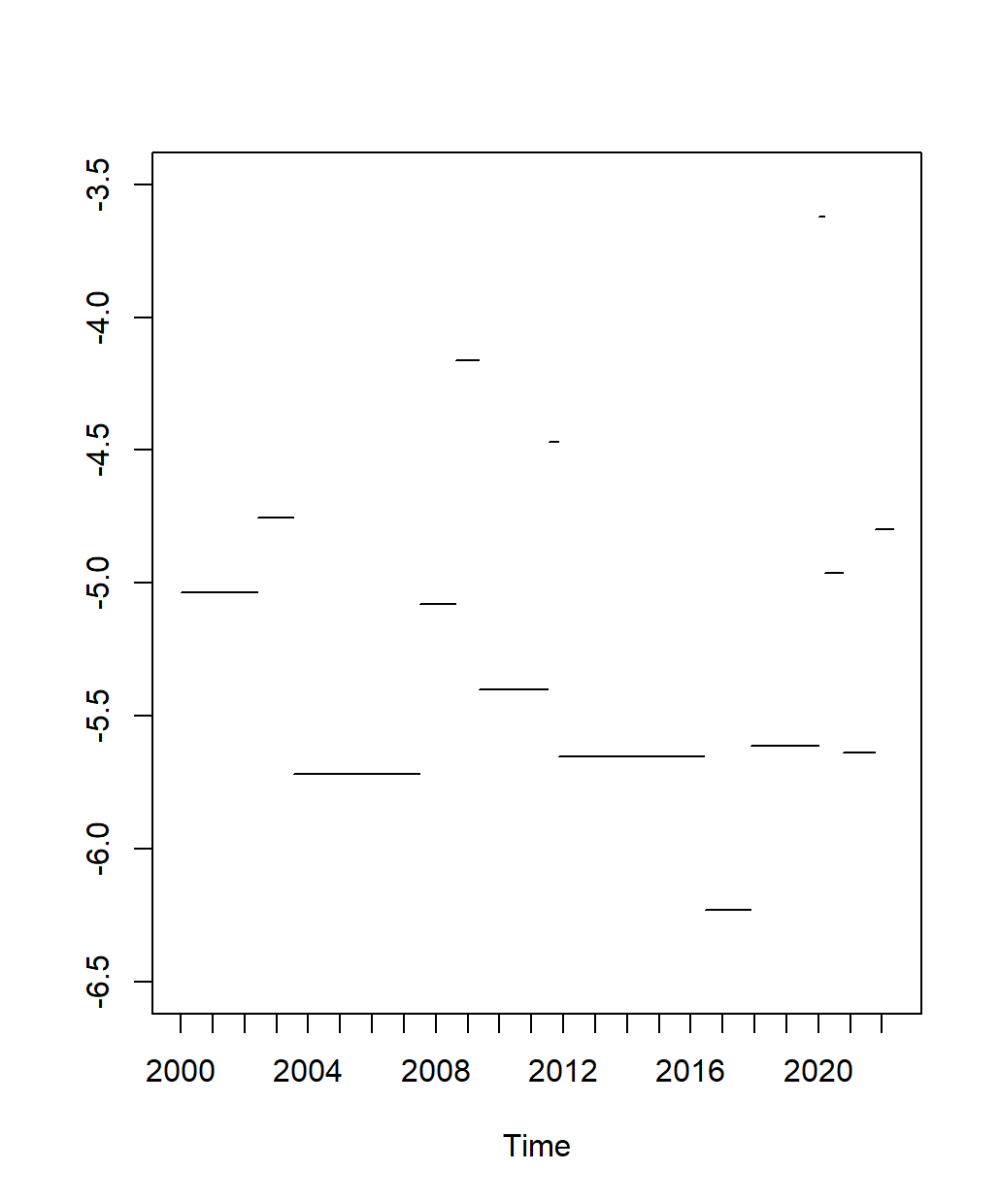}
			\vspace{-.4cm}
			\caption{{\small \it  SP500, Fitted mean by PELT}} 
				\label{SP500-constant}
	\end{minipage}

\end{figure}

   \textcolor{black}{ Finally, we apply the popular change points algorithms DSMUCE and PELT to the data which identify many more jumps than our results. We display the results in Figures  \ref{SP500-constant-0} and  \ref{SP500-constant}. Following the simulation results in Section \ref{compare} of the supplemental material, we believe that many jumps in Figures \ref{SP500-constant-0} and  \ref{SP500-constant} are likely to be produced by smooth trends or non-stationary noises and hence are spurious. Furthermore, Figure \ref{SP500} shows interesting trends of the log absolute returns between the phenomenal and well-interpreted jumps which are worth investigating by financial investors. }                                                                                                                                                          
	
	\section{Summary and discussions}\label{Sec:discussion}
	In this paper, we proposed a multiscale method for jump testing and estimation under complex temporal dynamics where the covariance and high-order structures of the time series can experience both smooth changes and jumps over time, 
	which is composed of two steps. Multisacle and self-normalized applications of an optimal jump-pass filter to the observed time series is utilized in the first step to test the existence of jumps, detect the number of jumps and preliminarily determine the jump locations. The second step consists of a local CUSUM procedure that refines the jump locations estimated from the first step. The MJPD method is asymptotically correct,  detecting the correct number of jump points with a pre-specified probability asymptotically and  locating the jump points, if they exist, within a nearly parametric range for a wide class of trend functions under piece-wise locally stationary errors. Computationally, thanks to the closed form formula of the limiting distribution, the fast sum updating algorithm and the efficient sparsification of the scales, the MJPD method requires a nearly linear $O(n\log ^{1+\epsilon}n)$ run time to execute. 
	
	It has been increasingly common 
	to encounter time series data with complexly evolving data generating mechanisms in various applications. Though jump detection for dependent data has attracted some attention, few results are available for multiple jump detection for time series models with non-stationary covariance and higher order structures in the errors which motivated us to investigate the heteroscedasticity-robust MJPD methodology and its asymptotic theory.  On the other hand, compared with the popular piece-wise constant assumption on the parameters of interest in the change point detection literature, the piece-wise smooth modelling of such parameters formulated in, for example, \cite{muller1992change} and \cite{qiu2003jump} seems to be more appropriate in many situations under complex temporal dynamics and hence we adapt it in this article. 

	The MJPD method can be easily extended to testing and estimating jumps in $\mu_{i,n}:=\E[h(\vec{y}_{i,n})]$, where $\vec{y}_{i,n}$ is an observed $d$-dimensional piece-wise locally stationary time series and $h$ : $\mathbb{R}^d\rightarrow\mathbb{R}$ is a known function. Examples of this kind include moments, auto-covariances and level-crossing probabilities of a univariate time series and cross covariances of a multivariate time series. On the other hand, it is a non-trivial task to extend MJPD to cases where the parameter of interest cannot be written directly in the form of $\E[h(\vec{y}_{i,n})]$ such as in generalized estimating equations or maximum likelihood estimations.  We shall investigate this extension in a future research endeavour. Furthermore, high dimensional jump detection has attracted some attention recently (\cite{chen2022inference}). In this article we focus on the MJPD methodology and its theory for a one-dimensional parameter function. It is of great interest to investigate non-stationarity-robust jump detection methods and their theoretical properties for high-dimensional time series. We hope that the MJPD method will shed some light on the latter high dimensional problem and we shall leave it to a future research. 
\vspace{-0.3cm}	
\begin{acks}[Acknowledgments]
The authors would like to thank the anonymous referees, an Associate
Editor and the Editor for their constructive comments that improved the
quality of this paper.  Weichi Wu was supported by NSFC 12271287 and  11901337. Zhou's research was supported by NSERC of Canada. 
\end{acks}

\cleartheorem{lemma}
	
\cleartheorem{corol}
	\newtheorem{corol}{Corollary}
	\newtheorem{lemma}{Lemma}
	\renewcommand{\thelemma}{\Alph{section}.\arabic{lemma}}
	\renewcommand{\theproposition}{\Alph{section}.\arabic{proposition}}
	\renewcommand{\thealgorithm}{\Alph{section}.\arabic{algorithm}}
	\renewcommand{\theequation}{\Alph{section}.\arabic{equation}}
	\renewcommand{\thetheorem}{\Alph{section}.\arabic{theorem}}
	\renewcommand{\thetable}{\Alph{section}.\arabic{table}}
	\renewcommand{\thecorol}{\Alph{section}.\arabic{corol}}
	\renewcommand{\thesection}{\Alph{section}}  
	\renewcommand{\theremark}{\Alph{section}.\arabic{remark}}  
	
	\renewcommand{\thefigure}{\Alph{section}.\arabic{figure}}  
	\setcounter{section}{0}
		\setcounter{equation}{0}
			\setcounter{theorem}{0}
				\setcounter{lemma}{0}
				\setcounter{proposition}{0}
						\setcounter{corol}{0}
							\setcounter{remark}{0}
								\setcounter{figure}{0}
									\setcounter{figure}{0}	\setcounter{algorithm}{0}
\begin{frontmatter}
	\title{Supplement to "Multiscale jump testing and estimation under complex temporal dynamics"}
	\runtitle{Multiscale jump detection}
	
	\begin{aug}
		\author[A]{\inits{}\fnms{Weichi}~\snm{Wu}\ead[label=e1]{wuweichi@mail.tsinghua.edu}}
		\author[B]{\inits{}\fnms{Zhou}~\snm{Zhou}\ead[label=e2]{zhou@utstat.toronto.edu}}
		\address[A]{Center for Statistical Science, Department of Industrial Engineering, Tsinghua University, China\printead[presep={,\ }]{e1}}
		
		\address[B]{Department of Statistical Science, University of Toronto, Canada\printead[presep={,\ }]{e2}}
	\end{aug}
	
	\begin{abstract}
		This supplemental material contains 
		additional simulation results for 5000 sample size in Section \ref{5000Sample-Result}, a sensitivity analysis in Section \ref{Sensativity-Check}, 
		the performance of MJPD as a test of jumps in Section \ref{Multiscale-Test}, a detailed description of the fast sum updating algorithm in Section  \ref{Section_Algorithm3}, and detailed proofs of the theoretical results of the main article in Section \ref{Technical}. Section \ref{Technical} also contains an additional theoretical result on the existence of a general order $k$ filter (Theorem \ref{thm4-2021}). 
	\end{abstract}

\end{frontmatter}

\section{Simulation results for estimating break points}
\subsection{Performance of some existing change point methods under complex trends and non-stationary noises}\label{compare}
\textcolor{black}{
	In this section we apply DSMUCE and PELT to mean model I and II with errors defined in Section \ref{Sec::500_Sample_Size} of the main article. DSMUCE \cite{dette2018multiscale} is an  extension of SMUCE \cite{frick2014multiscale} that is designed for
	piece-wise constant signals, where SMUCE is an algorithm which minimizes the number
	of change points while penalizing a multiscale goodness-of-fit statistic applicable to independent noise. DSMUCE is its dependent extension, suitable for stationary dependent error.  PELT combines dynamic programming together with pruning steps to accurately detect the change points with an expected linear computational
	cost. We acknowledge that there are many other interesting jump/change point detection methods in the literature, but due to the page limit we focus on the two methods.}

{\color{black}Tables \ref{TableA1} and \ref{TableA2} display the simulation results. Our results show that both DSMUCE and PELT identify many spurious jumps when there is a smooth trend. Moreover, nonstationary errors (especially PS errors) lower the probability of detecting the correct number of jumps for the two conventional change point algorithms. However, we must acknowledge that the above-mentioned two change point algorithms are not designed for complex trends and noises.}

\begin{table}[ht]
	\scriptsize
	\centering
	\begin{tabular}{rrrrrrrrrrr}
		\toprule
		& \multicolumn{5}{c}{Mean Model I}      & \multicolumn{5}{c}{Mean Model II} \\
		\cmidrule(lr){2-6} \cmidrule(lr){2-6} \cmidrule(lr){7-11}
		& GS & ARMA & PS & LS & PLS2 & GS & ARMA & PS & LS & PLS2 \\ 
		\midrule
		$\hat m=m_n$ (\%) & 99.95 & 100.00 & 56.45 & 75.90 & 93.70 & 0.00 & 0.00 & 0.00 & 0.00 & 0.00 \\ 
		MAD $(\times 10^{-3})$ & 0.331 & 0.047 & 0.348 & 0.889 & 0.384 & NA  & NA  & NA  &  NA &  NA \\ 
		mean $m$  & 2.0005 & 2.0000 & 3.0805 & 2.3715 & 2.1130 & 7.4175 & 7.3255 & 8.5690 & 7.6855 & 7.6815 \\ 
		Time ($\times 10^{-2}$)& 0.049 & 0.035 & 0.054 & 0.063 & 0.043 & 0.057 & 0.065 & 0.053 & 0.057 & 0.062 \\ 
		\bottomrule
	\end{tabular}
	\caption{Results of PELT for $n=500$, with Mean Models I and II ($m=2$) } \label{TableA1}
\end{table}

\begin{table}[ht]
	\scriptsize
	\centering
	\begin{tabular}{rrrrrrrrrrr}
		\toprule
		& \multicolumn{5}{c}{Mean Model I}      & \multicolumn{5}{c}{Mean Model II} \\
		\cmidrule(lr){2-6} \cmidrule(lr){7-11}
		& GS & ARMA & PS & LS & PLS2 & GS & ARMA & PS & LS & PLS2 \\ 
		\midrule
		$\hat m=m_n$ (\%) & 99.85 & 99.85 & 53.30 & 94.30 & 93.95 & 0.00 & 0.00 & 0.00 & 0.00 & 0.00 \\ 
		MAD($\times 10^{-3}$) & 0.349 & 0.045 & 0.314 & 0.855 & 0.382 &  NA & NA  & NA  & NA  & NA  \\ 
		mean $m$ & 2.0025 & 2.0015 & 2.9740 & 2.0575 & 2.0735 & 5.2875 & 5.3545 & 5.9510 & 5.1275 & 5.3675 \\ 
		Time ($\times 10^{-2}$) & 0.342 & 0.348 & 0.337 & 0.355 & 0.353 & 0.325 & 0.327 & 0.351 & 0.352 & 0.345 \\ 
		\bottomrule
	\end{tabular}
	\caption{Results of DSMUCE for $n=500$, with Mean Models I and II ($m=2$) }\label{TableA2}
\end{table}

\subsection{Simulation results of 5000 sample size} \label{5000Sample-Result}
To investigate the performance of  MJPD at 5000 sample size, we consider
\begin{description}
	\item ($I'_n$) $\beta_n(t)=1+1.99\sum_{0\leq u\leq 8}\frac{(-1)^u+1}{2}\mathbf 1 \left(b_u< t\leq b_{u+1} \right),$ where $b_u=u/9$ for $0\leq u\leq 7$ and $u=9$, and $b_8=25/27$.
	\item ($II_n'$) $\beta_n(t)=5\sin(2\pi t)+1.99\sum_{0\leq u\leq 8}\frac{(-1)^u+1}{2}\mathbf 1 \left(\frac{u}{9}< t\leq \frac{u+1}{9} \right)$,
	
\end{description}and error $\varepsilon_{i,n}= 1.1G(\frac{i}{n},\FF_i)$, where $G(\cdot,\FF_i)$ follows models $PS_n'$, $LS_n'$, $PLS_n'$, $GS$, and $ARMA$. Among them,  $GS$ and $ARMA$ are  defined in Section \ref{Sec::500_Sample_Size}  of the main article, $PLS_n'$ can be found in detail in Section \ref{Sec::Increasing} below, 
and  $PS_n'$, $LS_n'$ are defined as follows.
\begin{description}
	\item($PS_n'$) $G(t,\FF_i)=a(t)G(t,\FF_{i-1})+\eta_i$ for $a(t)=1.25b(t)$ where $b(t)$ equals
	$$-0.3\mathbf 1(0<t\leq 1/4)+0.1\mathbf 1(1/4<t\leq 2/3)+0.2\mathbf 1( 2/3<t\leq 3/4)-0.1\mathbf 1(3/4<t\leq 1)$$ and $\{\eta_i\}$ are $i.i.d.$ $(\chi^2(3)-3)/\sqrt 6$.
	\item ($LS_n'$) $G(t,\FF_i)=(3(t-0.5)^2-0.3)G(t,\FF_{i-1})+\eta_i+(0.2-0.4t)\eta_{i-1}$, where $\{\eta_i,i\in \mathbb Z\}$ are $i.i.d.$ $t(8)/\sqrt{4/3}$.
\end{description} 
Observe that the number of jumps is $8$ with magnitude $1.99$, and the distances between jump points in mean $I_n'$ are non-equal. The errors we consider include stationary, locally stationary and piece-wise (locally) stationary processes driven by  Gaussian as well as heavy-tailed and asymmetric innovations. We summarize the corresponding experiment results in Table \ref{5000mutemp}, which demonstrate the ability of  MJPD to accurately estimate jumps of piece-wise smooth signals under complex temporal dynamics. 

From Table \ref{5000mutemp} we find that  the computational cost of MJPD is less than 0.06 seconds for each single scale and is less than 0.4 seconds overall for $n=5000$.  We remark here that at the time of writing, our algorithm is written in R using basic package and code. 
The computational cost of MJPD should be improved if advanced packages, parallel computing and faster languages such as Python are used.

\begin{table}[htbp]
	\centering
	\scriptsize
	\caption{Performance of MJPD for Mean Model $I_n'$ and $II_n'$ where $m=8$, $n=5000$.}
	\begin{tabular}{ccccccccccc}
		\toprule
		& \multicolumn{5}{c}{Mean $I'_n$}          & \multicolumn{5}{c}{Mean $II'_n$} \\
		\cmidrule(lr){2-6} \cmidrule(lr){7-11}
		& { GS    } & { ARMA  } & { $PLS_n'$ } & { $LS_n'$ } & { $PS_n'$} & { GS    } & { ARMA  } & { $PLS_n'$ } & { $LS_n'$ } & { $PS_n'$} \\
		\midrule
		$\hat m=m_n$ $(\%) $  & 99.90 & 99.95 & 96.30 & 98.85 & 100.00 & 99.05 & 98.75 & 95.10 & 98.00 & 99.25 \\ 
		MAD $( 10^{-4})$  & 2.200 & 1.632 & 2.530 & 2.177 & 1.463 & 2.218 & 1.675 & 2.994 & 2.705 & 1.518 \\ 
		mean $m$ & 7.9990 & 7.9995 & 7.9670 & 7.9915 & 8.0000 & 8.0095 & 8.0125 & 7.9830 & 8.0020 & 8.0075 \\ 
		Time ($10^{-2}$) & 38.973 & 39.841 & 39.960 & 38.901 & 39.938 & 39.776 & 39.788 & 39.689 & 39.931 & 39.735 \\ 
		Single.sec ($10^{-2}$) & 6.408 & 6.240 & 6.252 & 5.957 & 6.272 & 6.235 & 6.250 & 6.294 & 6.246 & 6.243 \\

		\bottomrule
	\end{tabular}%
	\label{5000mutemp}%
\end{table}%

		\subsection{Results of increasing sample sizes}\label{Sec::Increasing}
		
		In this section we study the behaviour of MJPD when the sample size is gradually increased from $500$ to $5000$. For this purpose define the series $k_n=\lf(\frac{4}{3}\log n/6)^5\rf$, $K_n=k_n/2$ and $\Delta_n=4/((\log n/6)^2)$.  
		We consider the following piece-wise linear $\beta_n(\cdot)$:
		\begin{align}
			\beta_n(t)=10t+\Delta_n\sum_{0\leq u\leq  k_n , u \  \text{even}}\mathbf 1 \left(\frac{u}{ k_n +1}< t\leq \frac{u+1}{ k_n +1} \right),
		\end{align}
		which has an increasing number $k_n$ of jumps with diminishing jump size $\Delta_n$.  We consider error $\varepsilon_i=1.1G(\frac{i}{n},\FF_i)$ where $G(\cdot,\FF_i)$ (denoted by ``$PLS_n'$") is a PLS process given by
		\begin{align}
			G(t,\FF_i)=0.6(1+0.7g(t))\sum_{j=0}^\infty\left(0.2\cos(2\pi t)+0.2g(t)\right)^j\eta_{i-j},
		\end{align}
		where $\{\eta_i\}_{i\in \mathbb Z}$ are $i.i.d.$ standard normal, and
		\begin{align}\label{mean_III}
			g(t)=\sum_{0\leq u\leq (1\vee \lf K_n\rf ), u \  \text{even}}\mathbf 1 \left(\frac{u}{ K_n +1}< t\leq \frac{u+1}{ K_n +1}\right) 
		\end{align}
		is a step function. Process $PLS_n'$ has an increasing number ($\lf K_n\rf$) of breaks. The Monte-Carlo experiment results are displayed in Table \ref{tabelincreasefixed}.
		Rows 2-4 in Table \ref{tabelincreasefixed}  record the number of jumps in mean, the number of breaks in the errors, and the magnitude of the jumps, respectively. Row 5 contains the average computational costs for MJPD at different sample sizes, which fit well with the theoretical $O(n\log^{\frac{3}{2}} n)$ computational complexity.  
		Row 6 shows the time  to generate  $\hat c_{1-\alpha}$ for  SIM  using bootstrap at different sample sizes. The results show that to obtain threshold $\hat c_{1-\alpha}$, the computational cost of Algorithm \ref{Algorithm2} is much more expensive than that of Algorithm \ref{Algorithm1} for large samples. Hence in practice we recommend using Algorithm \ref{Algorithm1} to perform jump estimation when $n>500$. Rows 7 and 8 display the simulated probabilities of correctly identifying all jumps for both MJPD and SIM. Those results demonstrate the correctness of MJPD for jump detection under complex temporal dynamics. 
		
		We then study the improvement in accuracy when the second stage refinement in Section \ref{Sec:Second-Stage} is applied. In  Row 9, ``MAD1'' represents the simulated average estimation MAD of one-stage MJPD obtained by directly applying Algorithm \ref{Algorithm1} while ``MAD2'' in Row 10 stands for the simulated average estimation MAD when the second-stage refinement proposed in \eqref{eq37-March25} is applied to MJPD.  The results conclude that MAD2 is much less than MAD1 at all sample sizes which demonstrates the benefits of applying the second stage refinement. 
		
		
		Table \ref{Table-critcompare} compares the theoretical $c_{1-\alpha}$ derived from  Theorem \ref{Thm1} to the simulated $\hat c_{1-\alpha}$ via the bootstrap procedure in Algorithm \ref{Algorithm2} for  different $\alpha's$ and sample sizes. The scenarios considered are the same as those of Table \ref{tabelincreasefixed}. It can be seen that the critical values for MJPD and SIM are very close for $n\ge 500$.
		\begin{table}[htbp]
			\centering
			\scriptsize
			\caption{Simulation results for increasing sample size with increasing number of jumps. }
	\begin{tabular}{lrrrrrrrrrr}
		\toprule
		$n$ ($\times 100$) & 5   & 10  & 15  & 20  & 25  & 30  & 35  & 40  & 45  & 50 \\
		\hline
		\# jumps in mean  &2& 3 &4& 4& 5& 6& 6& 7 &7& 8
		\\
		\# breaks in errors  &1& 1& 2& 2& 2& 3& 3& 3& 3& 4
		
		\\
		Jump	sizes & 3.73  & 3.02  & 2.69  & 2.49  & 2.35  & 2.25  & 2.16  & 2.09  & 2.04  & 1.99 \\
		
		Time (s) & 0.024 & 0.056  &0.099 & 0.133&  0.179 & 0.218 & 0.281  &0.325&  0.372  &0.417

		\\
		Boots ($\times 100$s) & 1.07     & 2.51    &  4.47      &6.10   &   8.59   &  10.08  &   13.00 & 15.04    & 17.26     &19.33
		\\
		MJPD (\%)
		
		& 100.00 & 99.80 & 99.00 & 99.55& 98.45 & 98.70  &98.85 & 98.35 & 98.00 & 95.05

		\\
		SIM (\%) 
		
		&100.00  &99.80 & 99.00&  99.50  &98.55 & 98.70 & 98.85 & 98.50 & 97.75 & 95.65

		\\
		MAD1 ($\times 10^{-3}$) 
		
		
		& 1.774 	& 1.498	&  1.413  	&1.270 	& 1.099  	&0.892 		&0.750 	& 0.757 	& 0.703  	&0.690

		\\
		MAD2 ($\times 10^{-3}$) 
		
		
		&  1.074&  0.151&  0.272&  0.288 & 0.369 & 0.316 & 0.190 & 0.152 & 0.191  &0.323 
		\\

		\bottomrule
	\end{tabular}%
	\label{tabelincreasefixed}%
\end{table}%
\begin{table}[htbp]
	\centering
	\scriptsize
	\caption{Theoretical and Simulated $\hat c_{1-\alpha}$ for scenarios considered in Table \ref{tabelincreasefixed}.}
	\begin{tabular}{ccrrrrrrrrrr}
		
		\toprule
		&$n$ & 500     & 1000     & 1500     & 2000    & 2500     & 3000     & 3500     & 4000     & 4500     & 5000 \\
		\hline
		\multirow{2}[0]{*}{Scales} & $\bar {s}_n$  & 
		
		0.167& 0.125& 0.100 &0.100& 0.083& 0.071& 0.071 &0.062& 0.062 &0.056

		\\
		&$\underline {s}_n$  
		&	0.061& 0.043& 0.036& 0.031& 0.028& 0.026& 0.024& 0.023& 0.022& 0.020

		\\
		\hline
		\multirow{2}[0]{*}{$\alpha=0.1$} & MJPD  & 
		
		3.672 & 3.809 & 3.875  &3.931 & 3.960  &3.980 & 4.008  &4.023  &4.041 & 4.068

		\\
		& SIM  
		& 3.623 &3.792 &3.881 &3.914 &3.955 &3.955& 4.000 &4.035 &4.042 &4.068

		\\
		\hline
		\multirow{2}[0]{*}{$\alpha=0.05$} & MJPD 
		& 3.870 &3.999& 4.062 &4.115& 4.142 &4.160& 4.188 &4.201 &4.219 &4.246

		\\
		
		& SIM 
		
		& 3.838 &3.984 &4.057 &4.113& 4.140& 4.136 &4.188 &4.208& 4.220 &4.238	
		\\
		\hline	
		
		\multirow{2}[0]{*}{$\alpha=0.01$} & MJPD 
		
		& 4.289 &4.404&4.462 &4.509& 4.534 &4.550 &4.576 &4.588& 4.604& 4.628

		\\
		& SIM
		
		&4.286& 4.388& 4.462 &4.546& 4.574 &4.521& 4.623& 4.561 &4.620 &4.589
		
		\\
		\bottomrule
	\end{tabular}%
	\label{Table-critcompare}
\end{table}%

\section{Sensitivity Analysis}\label{Sensativity-Check}
\subsection{Sensitivity to tuning parameters}\label{Tuning-robust}
In this section, we investigate the sensitivity our MJPD to the choice of tuning parameters.  We consider the scenario of  Model  II with errors $ARMA$, and $PLS$ defined in Section  \ref{Sec::500_Sample_Size} of the main article. Consider the rule of thumb choices of $\bar s_n$, $\underline s_n$ and $s_n^*$ recommended in Section \ref{Parameter:Selection} and we investigate the performance of MJPD when these parameters are enlarged or shrinked. Denote by $\bar s^0_n$, $\underline s_n^0$ and $s_n^{*0}$ the rule of thumb scales. Let $\bar s+=1.25\bar s_n^0$, $\bar s-=\bar s_n^0/1.25$, $\underline s+=1.25\underline s_n^0$ and $\underline s-=\underline s_n^0/1.25$. We examine the results of MJPD 	with lower and upper scales equaling to $(\underline s_n^0, \bar s+ )$ , $(\underline s_n^0,\bar s-) $ , $(\underline s+,\bar s_n^0) $, and $(\underline s-,\bar s_n^0)$ respectively.  The results are presented in Table \ref{robust-Check}. From the table, we observe that the results are relatively stable; in particular, the effect of $\underline s$ on MJPD is much smaller than that of $\bar s$. We then use $MV$ method to choose $\bar s_n$ over $[\bar s-, \bar s+]$, $\underline s_n=\underline s_n^0(\bar s_n)$ and $s_n^*$ over  $[s_n^{*0}(\underline s_n)/2, 2s_n^{*0}(\underline s_n)]$ and display the results under the column of $MV$ in table \ref{robust-Check}. Here $\underline s_n^0(\bar s_n)$ is the rule of thumb choices of $\underline s_n$ given $\bar s_n$ and $s_n^{*0}(\underline s_n)$ is the rule of thumb choices of $s_n^*$ given $\underline s_n$. Comparing with results in Table \ref{Table2-May-2019} of the main article we shall see that MJPD is reasonably stable with respect to different choices of maximum and minimal scales as long as those scales deviate moderately from the ones selected by the rule of thumb or the MV method.  
\begin{table}[ht]
	\centering
	\tiny
	\caption{Sensitivity analysis of MJPD using $(\underline s_n^0, \bar s\pm)$, $(\underline s\pm,\bar s_n^0)$ and the $MV$ method. }
	\begin{tabular}{rrrrrrrrrrr}
		\toprule
		&\multicolumn{5}{c}{ARMA}&\multicolumn{5}{c}{PLS}\\
		\cmidrule(lr){2-6} \cmidrule(lr){7-11}
		& $(\underline s_n^0, \bar s+ )$&  $(\underline s_n^0,\bar s-) $ & $(\underline s+,\bar s_n^0) $ & $(\underline s-,\bar s_n^0)$ & MV & $(\underline s_n^0, \bar s+ )$ &$(\underline s_n^0,\bar s-) $ & $(\underline s+,\bar s_n^0) $ & $(\underline s-,\bar s_n^0)$ & MV \\ 
		\midrule
		$\hat m=m_n $ $(\%)$ & 100.00 & 94.50 & 100.00 & 100.00 &99.9 & 100.00 & 94.60 & 99.85 & 99.25 & 99.4 \\ 
		MAD $(\times 10^{-4})$ & 3.408 & 3.392 & 3.422 & 3.392 & 3.478 & 5.305 & 5.907 & 5.537 & 5.706 & 6.169 \\ 
		mean $m$& 2.0000 & 2.0585 & 2.0000 & 2.0000 & 2.0001 & 2.0000 & 2.0190 & 1.9995 & 1.9955 & 1.996 \\ 
		\bottomrule	
	\end{tabular}\label{robust-Check}
\end{table}


\subsection{Sensitivity to filters}\label{Tuning-filters}
It is important to check the influence of filters on  the proposed algorithm. For this purpose, we further examine another filter 
$\tilde W_1\in \mathcal W_{2,4}$, which is
\begin{align}\label{FilterW-order2-N5}
	\tilde W_1(x)=(-120x^4+ 300|x|^3-240x^2+60|x|)\it{sgn}(x),
\end{align} 
with  $SN(\tilde W_1)\simeq  0.418$.  In fact, $\tilde W_1$ is optimal in terms of signal-noise ratio over the class $\mathcal W_{2,5}$, not only  $\mathcal W_{2,4}$. Using this filter, we apply MJPD to Models I and II with errors $GS$, $ARMA$, $PS$, $LS$ and $PLS$ defined in Section  \ref{Sec::500_Sample_Size} of the main article. The sample size is 500 and  the results are summarized in Table \ref{different filters}. Comparing with Table \ref{Table2-May-2019} in the main article, we conclude that our test results are stable when optimal filters in different polynomial classes are applied, and the filter suggested in \eqref{FilterW-order2-N6} slightly outperforms that in \eqref{FilterW-order2-N5} due to its higher signal-noise ratio.
\begin{table}[htbp]
	\centering
	\scriptsize
	\caption{Results of MJPD using the filter $\tilde W_1$  with Mean Model I ($m=2$)  and  Mean Model II ($m=2$)}
	\begin{tabular}{ccccccccccc}
		\toprule
		& \multicolumn{5}{c}{Mean Model I}      & \multicolumn{5}{c}{Mean Model II} \\
		\cmidrule(lr){2-6} \cmidrule(lr){7-11}
		&{ GS } &{ ARMA } &{ PS } &{ LS } &{ PLS} &{ GS } &{ ARMA } &{ PS } &{ LS } &{ PLS} \\
		

		\midrule	
		$\hat m=m_n$ (\%) & 96.05 & 94.40 & 94.80 & 92.00 & 92.55 & 99.85 & 99.90 & 99.65 & 99.70 & 98.80 \\ 
		MAD $(\times 10^{-3})$ & 0.360 & 0.053 & 0.368 & 1.532 & 2.947 & 0.500 & 0.345 & 0.516 & 0.410 & 0.664 \\ 
		mean $m$ & 2.0365 & 2.0560 & 2.0500 & 1.9860 & 2.0145 & 2.0015 & 2.0010 & 2.0025 & 2.0030 & 1.9930 \\ 
		Time ($\times 10^{-2}$) & 2.183 & 2.358 & 2.350 & 2.389 & 2.380 & 2.426 & 2.378 & 2.432 & 2.450 & 2.436 \\

		\bottomrule
	\end{tabular}%
	\label{different filters}%
\end{table}%

\section{Type I error and power when MJPD is used as a test}\label{Multiscale-Test}
When the existence of jumps is concerned, MJPD can be used as a multiscale test. Specifically, if MJPD detects no jumps, we accept the null hypothesis of a smooth trend. To study the MJPD based test, consider the model with mean
\begin{align}\label{Smoothmodel}
	\beta_n(t)=\cos(\pi t)+d\mathbf 1(0<t\leq 0.5)
\end{align}
and  error $\varepsilon_{i,n}/2$ where $\varepsilon_{i,n}$ is the PLS process such that $\varepsilon_{i,n}=G_0(i/n,\FF_i)$ for $i/n\leq 0.6$, and $\varepsilon_{i,n}=G_1(i/n,\FF_i)$ for $i/n>0.6$  where 
\begin{align}\ \ \ \ G_0(t,\FF_i)=(0.5t-0.2)G_0(t,\FF_{i-1})+\eta_i,\  G_1(t,\FF_i)=0.6\cos(2\pi t)G_1(t,\FF_{i-1})+\eta_i,\label{PLS1}\end{align}
$\FF_i=(\eta_{-\infty},...\eta_i)$ and  the series $\{\eta_i\}_{i\in \mathbb Z}$ is $i.i.d.$ $t(8)/\sqrt{4/3}$.  We investigate the performance of MJPD  via optimal filters in the class $\mathcal W_{2,6}$ defined in \eqref{FilterW-order2-N6} and the optimal filter in  $\mathcal W_{2,5}$ defined in \eqref{FilterW-order2-N5}. Hence the results of this subsection also serves as a small sensitivity analysis of MJPD with respect to the filter. 
We first study the case that $d=0$ by applying MJPD to Model \eqref{Smoothmodel}, fixing $\alpha=0.05$ and $\alpha=0.1$. In this case, $\alpha$ is the nominal level of the test. We increase the sample size $n$ from $500$ to $3000$. 
We observe from Table \ref{Table-Type-1} that the simulated type I errors are close to the nominal levels for MJPD, which is consistent with our theoretical findings. Furthermore, it can be seen that the simulation results are relatively stable across the two filters.

\begin{table}[htbp]
	\centering
	\scriptsize
	\caption{Simulated rejection probabilities for Model \eqref{Smoothmodel} with nominal level 5\% and 10\%, via filters in $\mathcal W_{2,6}$ and $\mathcal W_{2,5}$.}
	\begin{tabular}{ccrrrrrrrr}
		\toprule
		\multicolumn{2}{c}{}&\multicolumn{4}{c}{Filter \eqref{FilterW-order2-N6} in $\mathcal W_{2,6}$}&\multicolumn{4}{c}{Filter \eqref{FilterW-order2-N5}    in $\mathcal W_{2,5}$}\\
		\cmidrule(lr){3-6} \cmidrule(lr){7-10}
		$n$	&$\alpha$ & 500     & 1000     & 1500     & 3000    & 500     & 1000     & 1500     & 3000     \\
		\midrule
		\multirow{2}[0]{*}{MJPD} & 5\% &0.065& 0.055 &  0.063& 0.0525 & 0.061 &  0.064 & 0.0655& 0.067
		
		\\
		& 10\% & 0.105 &0.100 & 0.1065 &0.104 & 0.109  &0.1115 &0.1145& 0.1085

		\\
		
		%
		

		\bottomrule
	\end{tabular}%
	\label{Table-Type-1}
\end{table}%

\begin{figure}[t]
	\centering
	\includegraphics[width=14cm,height=4cm]{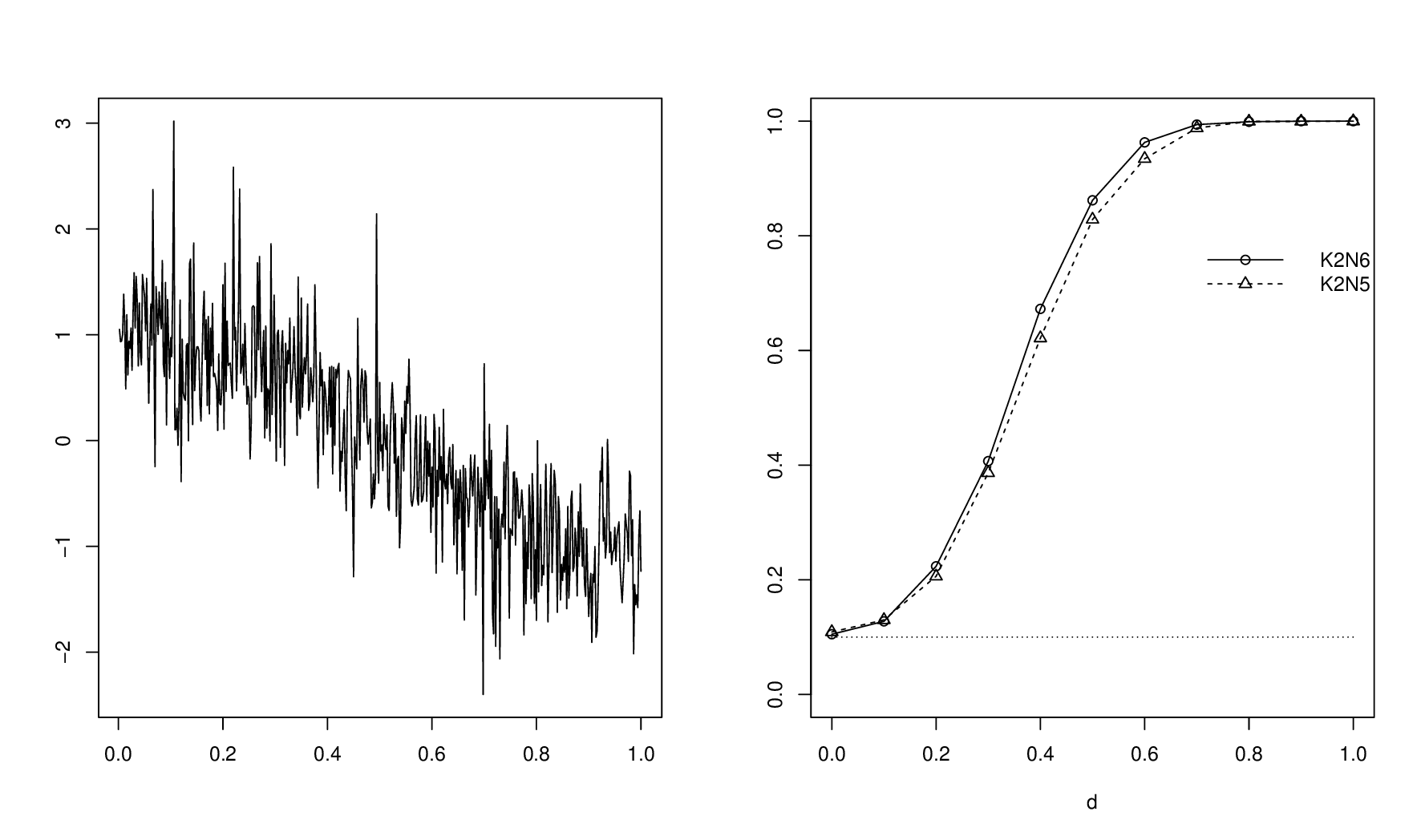}
	\vspace{-.4cm}
	\caption{\it Left panel: typical sample path of Model \eqref{Smoothmodel} with $d=0$, error \eqref{PLS1} and sample size $500$. Right panel: simulated rejection probabilities versus different values of $d$, for sample size $500$ and optimal filters in $\mathcal W_{2,5}$ and $\mathcal W_{2,6}$. }
	\label{SmoothFigure}
\end{figure}

We then examine the power of the MJPD test by varying  $d$ over $(0,1)$ at sample size 500, with optimal filters in $\mathcal W_{2,5}$ and $\mathcal W_{2,6}$. We present the corresponding results in  Figure \ref{SmoothFigure}. 
The right panel of  Figure \ref{SmoothFigure} displays the simulated rejection probabilities when $d$ increases, with nominal level $\alpha=0.1$. 
The lines ``$K2N6$'' and ``$K2N5$'' represent results of MJPD using optimal filters in the class $\mathcal W_{2,6}$ and $\mathcal W_{2,5}$, respectively. Figure  \ref{SmoothFigure} demonstrates a decent power performance of MJPD as well as the power enhancement when signal-noise ratio increases (Note that  the optimal filter in  $\mathcal W_{2,6}$ has a higher SN ratio). 

\section{Fast Sum Updating Algorithm }\label{Section_Algorithm3}
We provide the detailed fast sum updating  algorithm in Algorithm \ref{Algorithm3} to calculate  \begin{align} H\big(\frac{j}{n},s\big)=\frac{1}{\sqrt{ns}}\sum_{i=1}^nW\left(\frac{i/n-j/n}{s}\right )y_i,~~
	1+\lf ns\rf\leq j\leq n-\lf ns \rf
\end{align} for any given scale $s\in(0,1)$ in $O(n)$ time.

\begin{algorithm}
	\caption{Fast Sum Updating Algorithm }\label{Algorithm3}
	\begin{algorithmic}[1]
		\State Input $\mathbf y=(y[1],...,y[n])$, scale $s$.     
		\State  Set $Coeff[1]= 112.0000$, $Coeff[2]=  -933.3333$, $Coeff[3]=3188.8889$, $Coeff[4]=-5320.0000  $, $Coeff[5]=4246.6667$, $Coeff[6]=-1294.2222$.\\
		Set $Length=n-\lf ns\rf$.
		Set $a_l[1]=0$ for $l=0,1,..,6$, $i=1,..., Length$. Set $a_l[1]=\sum_{i=1}^{1+\lf ns\rf}y[i](\frac{i-1}{ns})^l$. Set $lend=1$, $rend=2+\lf ns\rf$.
		
		\For{$2\leq i\leq Length$}
		\\ Set $temp_0=0$, $temp_l=\sum_{k=0}^{l-1}{l\choose{k}}a_k[i-1](\frac{-1}{ns})^{l-k}$ for $l=1,2,..,6$.\\
		$a_l[i]=a_l[i-1]+temp_l-(\frac{-1}{ns})^ly[lend]-(\frac{\lf ns \rf}{ns})^ly[rend],$ for $l=0,1,...,6$.
		\\  $lend=lend+1$, $rend=rend+1$, $i=i+1$.
		\EndFor
		\State
		Denote by $\mathbf a_l$ the vector of $(a_l[1],...a_l[n])^T$ for $l=0,...6$. 
		Construct vectors $\mathbf A_0=\mathbf a_0$, and $\mathbf A_l=(A_l[1],...,A_l[n])^T$ by $A_l[i]=\sum_{k=0}^l{l\choose k}(-)^ka_{l-k}[i]$ for $l=1,...,6$.
		\State 
		Calculate $\mathbf W_1$ and $\mathbf W_2$ with $i_{th}$ entry $W_1[i]$ and $W_2[i]$, $1\leq i\leq length-\lf ns \rf$ by
		\begin{align*}i'=i+\lf ns \rf,
			W_1[i]=\sum_{k=1}^6 Coeff[k]a_k[i'], 
			W_2[i]=\sum_{k=1}^6 (-1)^{k+1}Coeff[k]A_k[i].\end{align*}
		\State Output: $(\mathbf W_1+\mathbf W_2)/\sqrt{ns}$.
	\end{algorithmic}
\end{algorithm}

In Algorithm \ref{Algorithm3}, the coefficients $Coeff[k]$, $1\leq k\leq 6$  are determined by the optimal fiter $W^*$ over $\mathcal W_{2,6}$ in \eqref{FilterW-order2-N6} of the main article. $\mathbf W_1$ is the result of input $\mathbf y$ filtered by positive part of the filter, and $\mathbf W_2$ is the convolution related to the negative part of the filter. The algorithm has utilized the fact that the filter $W^*$ is an odd function. 



\section{Technical Appendix}\label{Technical}

In this section, we provide detailed proofs for  theorems, lemmas, propositions and corollaries. In addition, in Theorem \ref{thm4-2021} we provide a theory for the existence of a general order $k$ filter. For the sake of brevity, throughout this section, we omit the subscript $n$ of $y_{i,n}$, $\varepsilon_{i,n}$, $\bar s_n$, $\underline s_n$, $s^*_n$ and $m_n$ if it causes no confusion. For any two $p-$dimensional vectors $\mathbf a=(a_1,...,a_p)^T$ and $\mathbf b=(b_1,...,b_p)^T$, write $\langle \mathbf a,\mathbf b\rangle=\sum_{i=1}^pa_ib_i$. Define the projection operator $\pp_i(\cdot)=\E(\cdot|\FF_i)-\E(\cdot|\FF_{i-1})$, where $\FF_i$ is the filtration in Definition \ref{def1} in the main article. Write $\max(a,b)=a\vee b$ and $\min(a,b)=a \wedge b$.  Recall the definition of $T_n^d$, $\bar T_n^d$, $T_n^c$ and $\bar T_n^c$ defined in Section \ref{JumpDec} of the main article. In the following proofs, let $M$ be a generic sufficiently large constant that varies from line to line if no confusion arises. Meanwhile, the decomposition $H(t,s)=\sqrt{ns}\tilde G_n(t,s)+\tilde H(t,s)$ is often used, where $\tilde G_n(t,s)$ is in Section \ref{Origin} of the main article, and 
\begin{align}
	\tilde H(t,s)=\frac{1}{\sqrt{ns}}\sum_{j=1}^n\varepsilon_{j} W\left(\frac{j/n-t}{s}\right).
\end{align}
Moreover, the following quantities are carefully examined, which are 
\begin{align}\label{April-10-119}\Xi(t)=\sum_{i\in K(t)}H^2(i/n,s^*)/|K(t)|,~~\tilde \Xi(t)=\sum_{i\in K(t)}\tilde H^2(i/n,s^*)/|K(t)|.\end{align}

Before proving the theorems, we first give out two propositions which have been used frequently in the proofs.

\begin{proposition}\label{proposition-1-10-13}
	Assume (W1) and that  $\beta(\cdot)\in \mathcal M(m, \Delta_n, \gamma_n, k)$.  Let $s_n$ be a positive sequence of real numbers such that $s_n\rightarrow 0$, $(ns^2_n)^{-1}=O(1)$ and $s_n\leq \gamma_n/2$, then we have that uniformly for $t \in \cup_{r=0}^m[d_r+s_n,d_{r+1}-s_n]$,
	\begin{align}\label{disc-constant}
		\tilde G_n(t,s_n)&=O\Big(s_n^{k+1}+\frac{1}{ns_n}\Big).
	\end{align}
	Moreover uniformly for  $t\in[d_r-s_n,d_r+s_n]$, $1\leq r\leq m$, it follows that
	\begin{align}
		\label{April20_8}
		\tilde G_n(t,s_n)&=(\beta(d_r-)-\beta(d_r+))\int_{-\infty}^{\frac{d_r-t}{s_n}}W(u)du\notag
		\\&-\big(\sum_{v=1}^{k}((\beta^{(v)}(d_r-)-\beta^{(v)}(d_r+))/v!\big)\int_{-\infty}^{\frac{d_r-t}{s_n}}(t+s_nu-d_r)^vW(u)du)+O(s_n^{k+1}+\frac{1}{ns_n}).
	\end{align} 
	In particular, uniformly for $1\leq r\leq m$,
	\begin{align}\tilde G_n(d_r,s_n)&=(\beta(d_r-)-\beta(d_r+))\int_{-1}^0W(u)du+O\Big(s^{}_n+\frac{1}{ns_n}\Big).\label{dr-value}
	\end{align}
\end{proposition}
The proof of this proposition rest upon the property that $\int_{-1}^1 u^sw(u)du=0$, $0\leq s\leq k$. Then a direct application of  Taylor expansion and Riemann sum approximation will show that  $\tilde G_n(t,s_n)$ is negligible.\\

{\it Proof.}
Taylor expansion of $\beta$ yields that for all $t\in \cup_{r=0}^m[d_r+s_n,d_{r+1}-s_n] $,
\begin{align}\label{disc-constant-01}
	&\tilde G_n(t,s_n):=\frac{1}{ns_n}\sum_{i=1}^n\beta(i/n)W\left(\frac{i/n-t}{s_n}\right)\notag\\
	=&\frac{1}{ns_n}\sum_{i=1}^n\left(\beta(t)+\sum_{v=1}^{k}\frac{\beta^{(v)}(t)(i/n-t)^v}{v!}+\frac{\beta^{(k)}(\zeta_{i,t})(i/n-t)^{k}-\beta^{(k)}(t)(i/n-t)^{k}}{k!}\right)W\left(\frac{i/n-t}{s_n}\right)\notag\\
	=&O((s_n^{k+1}+\frac{1}{ns_n})),
\end{align}
where  $\zeta_{i,t}$ is a number between $i/n$ and $t$, and the last equality is due to condition (W1). This shows \eqref{disc-constant}.  Straightforward calculations show that for $t\in[d_r-s_n,d_r+s_n]$, $1\leq r\leq m$,
\begin{align}
	\tilde G_n(t,s_n)&=\int _{-\infty}^\infty \beta(t+s_nu)W(u)du+O(\frac{1}{ns_n})\notag\\
	&=\int_{-\infty}^{\frac{d_r-t}{s_n}}\beta(t+s_nu)W(u)du+\int_{\frac{d_r-t}{s_n}}^\infty\beta(t+s_nu)W(u)du+O(\frac{1}{ns_n}).\label{Sep4}
\end{align}
By using similar arguments to \eqref{disc-constant-01}, we have that 
\begin{align}
	&\int_{-\infty}^{\frac{d_r-t}{s_n}}\beta(t+s_nu)W(u)du\notag\\
	&=\beta(d_r-)\int_{-\infty}^{\frac{d_r-t}{s_n}}W(u)ds+(\sum_{v=1}^{k}(\beta^{(v)}(d_r-)/v!)\int_{-\infty}^{\frac{d_r-t}{s_n}}(t+s_nu-d_r)^vW(u)du)+O(s_n^{k+1}+\frac{1}{ns_n}),\label{April17-4}\\
	&\int_{\frac{d_r-t}{s_n}}^\infty\beta(t+s_nu)W(u)du\notag\\&=-\beta(d_r+)\int_{-\infty}^{\frac{d_r-t}{s_n}}W(u)du-(\sum_{v=1}^{k}(\beta^{(v)}(d_r+)/v!)\int_{-\infty}^{\frac{d_r-t}{s_n}}(t+s_nu-d_r)^vW(u)du)+O(s_n^{k+1}+\frac{1}{ns_n}),\label{April17-5}
\end{align}
uniformly for $t\in[d_r-s_n,d_r+s_n]$, $1\leq r\leq m$, where for the last equality we have used the fact that $\int_{\mathbb R} u^vW(u)du=0$ with $v=1,...,k$. Combining \eqref{April17-4} and \eqref{April17-5} we 
show \eqref{April20_8}.
	~Finally, by letting $t=d_r, 1\leq r\leq m$ we shall see that \eqref{dr-value} follows from expression \eqref{April20_8}, which finishes the proof. \hfill $\Box$
	

	\begin{proposition}\label{proposition-10-13-2} For $\beta\in \mathcal M(m, \Delta_n, \gamma_n,k)$, define $\Delta_{r,n}=\beta(d_r-)-\beta(d_r+)$. Assume (W2), $\frac{|\Delta_{r,n}|}{s_n+\frac{1}{ns_n}}\rightarrow \infty$ and that the conditions of Proposition \ref{proposition-1-10-13} hold. Then uniformly for $|t-d_r|\leq s_n$, $1\leq r\leq m$ we have that
		\begin{align}\label{April20-9}
			\tilde G^2_n(t,s_n)-\tilde G^2_n(d_r,s_n)=\Delta^2_{r,n}\Big((\int_{-\infty}^\frac{d_r-t}{s_n}W(s)ds)^2-(\int_{-\infty}^0W(s)ds)^2\Big)+O(\Omega_n(t)),
		\end{align}
		where\begin{align}
			\Omega_n(t)=|\Delta_{r,n}|s_n^{-1}(d_r-t)^2+|\Delta_{r,n}||d_r-t|+|\Delta_{r,n}|s_n^{k+1}+\frac{|\Delta_{r,n}|}{ns_n},
		\end{align}and
		$(\int_{-\infty}^\frac{d_r-t}{s_n}W(s)ds)^2-(\int_{-\infty}^0W(s)ds)^2\leq 0$.
	\end{proposition}


	{\it Proof}.
	By (W1) and a direct calculation using \eqref{April20_8} of Proposition \ref{proposition-1-10-13} we have that  uniformly for $|t-d_r|\leq s_n$, $1\leq r\leq m$,
	\begin{align}
		\tilde G_n(t,s_n)-\tilde G_n(d_r,s_n)=\Delta_{r,n}(\int_{-\infty}^{\frac{d_r-t}{s_n}}W(s)ds-\int_{-\infty}^0W(s)ds)\notag\\+O(|\beta^{(1)}(d_r-)-\beta^{(1)}(d_r+)||d_r-t|+s_n^{k+1}+\frac{1}{ns_n}),\label{March5-65}\\
		\label{March5-66}\tilde G_n(t,s_n)+\tilde G_n(d_r,s_n)=\Delta_{r,n}(\int_{-\infty}^{\frac{d_r-t}{s_n}}W(s)ds+\int_{-\infty}^0W(s)ds)\notag\\+O(s_n+\frac{1}{ns_n}).
	\end{align}
	It follows from $\tilde G^2_n(t,s_n)-\tilde G^2_n(d_r,s_n)=(\tilde G_n(t,s_n)-\tilde G_n(d_r,s_n))(\tilde G_n(t,s_n)+\tilde G_n(d_r,s_n))$ that
	\begin{align}\label{Squareddiff}
		\tilde G^2_n(t,s_n)-\tilde G^2_n(d_r,s_n)=\Delta^2_{r,n}\left((\int_{-\infty}^\frac{d_r-t}{s_n}W(s)ds)^2-(\int_{-\infty}^0W(s)ds)^2\right)+O(A_{r,n}+B_{r,n}+C_{r,n}),
	\end{align}
	where by (W2) and Taylor expansion, \begin{align}
		A_{r,n}=|\Delta_{r,n}|\left|\frac{d_r-t}{s_n}\right|^2\left(s_n+\frac{1}{ns_n}\right),B_{r,n}=(|d_r-t|+s_n^{k+1}+\frac{1}{ns_n})(s_n+\frac{1}{ns_n}),\label{A16Sep4}\\ C_{r,n}=|\Delta_{r,n}||\beta^{(1)}(d_r-)-\beta^{(1)}(d_r+)||d_r-t|+|\Delta_{r,n}|s_n^{k+1}+\frac{|\Delta_{r,n}|}{ns_n}.\label{A17Sep4}
	\end{align} Then \eqref{April20-9} follows from \eqref{Squareddiff},\eqref{A16Sep4}, \eqref{A17Sep4} and the definition of class $\mathcal{M}(m,\Delta_n,\gamma_n,k)$.  Finally, the conclusion that $(\int_{-\infty}^\frac{d_r-t}{s_n}W(s)ds)^2-(\int_{-\infty}^0W(s)ds)^2\leq 0$ follows from (W2).
	\hfill $\Box$

	
	\ \\\noindent{\bf Proof of Theorem \ref{Thm1}}. 
	By Lemma \ref{Lemma1}, Lemma \ref{Lemma3} and Lemma \ref{Lemma4}, we have that
	\begin{align}\label{April13.120}
		\sup_{t\in T_n^d}| \Xi(t)-\E \tilde \Xi(t)|=O_p(\nu_{1,n}+\nu_{2,n}),
	\end{align}
	where 	\begin{align}\E\tilde \Xi(t)=\sigma^2(t)\int W^2(t)dt+O(\nu_{3,n})\label{Xi(t)approx}
	\end{align} uniformly for $t\in T_n^c$, and $\E\tilde \Xi(t)=g(t)+O_p(\nu_{3,n})$ for $t\in \bar T_n^c$ for some bounded real function $g(t)$ such that $M_1\leq g(t)\leq M_2$ with $M_1>0,M_2>0$.
	By equation \eqref{disc-constant}, Proposition \ref{GaussianApprox} and summation by parts formula, we have that there exists a sequence of $i.i.d$ standard normals $V_i$ such that
	\begin{align}\label{approxV1_1}
		\sup_{t\in  T_n^d, s\in [\underline s, \bar s]}|H(t,s)-V(t,s)|=O_p\left(\frac{n^{1/4}\log^2 n}{\sqrt {n\underline s}}\right),
	\end{align}
	where \begin{align}\label{approxV1_2}
		V(t,s)=\frac{1}{\sqrt{ns}}\sum_{j=1}^n\sigma(j/n)V_jW\left(\frac{j/n-t}{s}\right),
	\end{align}
	and $\sigma(t)$ is the long run variance defined in \eqref{longrun} of the main article. 
	
	Define $ G^\diamond (t,\tilde s)=\sup_{\underline s\leq s\leq \bar s}\frac{|V(t,s)|}{\sqrt{\E \tilde \Xi(t)}}$. By similar arguments to Proposition B.2. of \cite{dette2015change} and Proposition \ref{techpropo-2019}, we obtain that $\sup_{t\in T_d,s\in [\underline s,\bar s]}|V(t,s)|=O_p(\log n)$. Thus by the definition of $G(t,\tilde s)$ we get
	\begin{align}\label{new.17}
		\sup_{t\in T_n^d}|G(t,\tilde s)- G^\diamond(t,\tilde s)|=O_p\left((\nu_{1,n}+\nu_{2,n})\log n+\frac{n^{1/4}\log^2 n}{\sqrt {n\underline s}}\right)
	\end{align}
	As a result, it suffices to study the limiting behavior of $\sup_{t\in T_n^d}| G^\diamond(t,\tilde s)|$. We first consider the quantity $\sup_{t\in T_n^d\cap T_n^c}| G^\diamond(t,\tilde s)|$.
	

	
	For this purpose define an $n-$dimensional vector 
	$T(t,s)=(\sigma(j/n)W((j/n-t)/s)(ns)^{-1/2})_{1\leq j\leq n}$ with its Euclidean norm $|T(t,s)|=\sqrt{\sum_{j=1}^n(\sigma(j/n)W((j/n-t)/s)(ns)^{-1/2})^2}$.~By Taylor expansion and equation \eqref{Xi(t)approx} we have that 
	\begin{align}
		\sup_{t\in T_n^d \cap T_n^c,\underline s\leq s\leq \bar s}||T(t,s)|^2-\E\tilde \Xi(t)|=O(\bar s+\frac{1}{n\underline s}+\nu_{3,n}).
	\end{align}
	Define $ G^\circ(t,\tilde s)=\sup_{\underline s\leq s\leq \bar s}\frac{|V(t,s)|}{|T(t,s)|}$ then 
	\begin{align}\sup_{t\in  T_n^d\cap T_n^c}|G^\circ(t,\tilde s)-G^\diamond(t,\tilde s)|=O_p((\bar s+\frac{1}{n\underline s}+\nu_{3,n})\log n)=o_p(1).\label{circ_sep4}\end{align} Therefore it suffices to evaluate the distribution of $G^\circ(t,\tilde s)$.
	Let $\mathbf V=(V_1,...,V_n)^T$ where $V_i,1\leq i\leq n$ are i.i.d. standard normals defined in equation \eqref{approxV1_2}.
	Define an $n\times 1$ vector $\check T(t,s)=T(t,s)/|T(t,s)|$ such that its $j_{th}$ element is \begin{align}\check T_j(t,s):=\sigma(j/n)W((j/n-t)/s)(ns)^{-1/2}/|T(t,s)|.\end{align}  Then we have the following representation of $G^\circ(t,s)$, which is
	\begin{align}G^\circ(t,s)=\langle \check T(t,s),\mathbf V\rangle.\end{align}
	Notice that for given $\bar s$, there exists $q$ $(q\leq m+l)$ disjoint intervals $I_j, 1\leq j\leq q$ such that 
	\begin{align}
		T_n^d\cap T_n^c=\cup_{1\leq j\leq q} I_j,~~ I_a\cap I_b=\emptyset \text{\ \ for $1\leq a\neq b\leq q$}.
	\end{align}
	Let $b_{j,1}$ and $b_{j,2}$ be the left and right end point of the corresponding interval $I_j$, respectively. 
	By Proposition 2 of \cite{sun1994simultaneous} which utilized Weyl's formula for the volume of tubes, we have that for $c\rightarrow \infty$ and for $1\leq j\leq q$
	\begin{align}
		\p(\sup_{t\in I_j} G^\circ(t,s)>c)=\alpha_j(c),
	\end{align}
	and $\alpha_j(c)$ has the following form
	\begin{align}\label{F37}
		\alpha_j(c)=\frac{\kappa_jc}{\sqrt 2 \pi^{3/2}}\exp(-c^2/2)+\frac{\zeta_j}{2\pi}\exp(-c^2/2)+2(1-\Phi(c))+o(\exp(-c^2/2)),
	\end{align}
	where $\Phi(c)$ represents the CDF of a standard normal. The constants $\kappa_j$ and $\zeta_j$ are  
	\begin{align}
		\kappa_j=(w_{11}w_{22})^{1/2}u^{-1}_{11}(b_{j,2}-b_{j,1})(\underline s^{-1}-\bar s^{-1})(1+O(\bar s+\frac{1}{n\underline s})),
		\zeta_j=\zeta_{j,1}+\zeta_{j,2},\\
		\zeta_{j,1}=\sqrt{w_{11}u^{-1}_{11}}(\bar s^{-1}+\underline s^{-1})(b_{j,2}-b_{j,1})(1+O(\bar s+\frac{1}{n\underline s})),\label{New54}\\
		\zeta_{j,2}=2\sqrt{w_{22}u^{-1}_{11}}(\log \bar s-\log \underline s)(1+O(\bar s+\frac{1}{n\underline s})).\end{align}
	The detailed calculations of $\kappa_j$ and $\zeta_j$ are discussed after the proof of this theorem.  Here $\kappa_j$, $\zeta_j$, $\zeta_{j,1}$ and $\zeta_{j,2}$ should depend on the sample size $n$, which is omitted from the subscript for the sake of brevity.
	
	It follows from the fact $(m+l)\bar s\rightarrow 0$ that $\sum_{j=1}^q |I_j|\rightarrow 1$. This result and the fact that $\sup_{t\in I_j} G^\diamond(t,s)$ are independent of $\sup_{t\in I_u} G^\diamond(t,s)$ for $u\neq j$ lead to that
	as $c\rightarrow \infty$,
	\begin{align}\label{new.30}
		\p(\sup_{t\in  T_n^d \cap T_n^c} G^\circ(t,s)>c)=&\p(\max_{1\leq j\leq q }\sup_{t\in I_j} G^\circ(t,s)>c)
		\notag\\&=1-\Pi_{s=1}^q(1-(\alpha_j(c)))=\sum_{j=1}^q\alpha_j(c)(1+o(1))\notag&\\
		& =\alpha_n(c)(1+o(1)).
	\end{align}
	Observing that for any random variables $X$ and $X'$,
	\begin{align}\label{F42}
		|\p(X>c)-\p(X'>c)|\leq \p(|X'-X|>\delta)+\p(|X-c|\leq \delta)
	\end{align}
	Pluging $\sup_{t\in  T_n^d \cap T_n^c} G^\diamond(t,s)$ to  $X$ and $\sup_{t\in  T_n^d \cap T_n^c} G^\circ(t,s)>c)$ to  $X'$, and taking $\delta=\iota_n$ with $\iota_n=o(1)$ but $\iota_n/ ((\bar s+\frac{1}{n\underline s}+\nu_{3,n})\log n)\rightarrow \infty$.
	This and \eqref{circ_sep4} yield that as $n$ and $c$ diverging, 
	\begin{align}\label{New.A34}
		|\p(\sup_{t\in  T_n^d \cap T_n^c} G^\diamond(t,s)>c)-\alpha_n(c)(1+o(1))|\notag\\=
		o(1)+(\alpha_n((c-\iota_n))-\alpha_n((c+\iota_n)))(1+o(1))=o(1).
	\end{align}
	where the first $o(1)$ term in the second line is due to \eqref{circ_sep4}, and the  $o(1)$ after the second equation can be verified by applying the formula of $\alpha_n(c)$, i.e., \eqref{F37} and \eqref{new.30} to $\alpha_n(c+\iota_n)-\alpha_n(c-\iota_n)$.
	For $t\in T_n^c$, by Lemma \ref{Lemma4}, equation \eqref{April13.120} and similar arguments to the evaluation of  $\sup_{t\in I_j} G^\diamond(t,s)$ we get 
	\begin{align}\label{new.31}
		&\notag\p(\sup_{t\in  \bar T_n^c\cap T_n^d} G^\diamond(t,s)>c)\leq \p(\sup_{t\in \bar T_n^c\cap T_n^d} \sup_{s\in [\underline s,\bar s]}|V(t,s)|>M_1c)\\&=\bigg(\frac{\kappa'M_1c}{\sqrt 2 \pi^{3/2}}\exp(-(M_1c)^2/2)\notag \\&+\frac{\zeta'}{2\pi}\exp(-(M_1c)^2/2)+2(1-\Phi(M_1c))+o(\exp(-c^2/2))\bigg)(1+o(1))=o(\alpha_n(c)),
	\end{align}
	for $\kappa'$ satisfies that there exist positive real numbers $a_1,a_2$, $a_1<\kappa'/(m\bar s/\underline s)<a_2$, and $\zeta'=\zeta'_1+\zeta'_2$ where $\zeta'_1$ is on the order of $m\bar s/\underline s$ and $\zeta'_2$ is on the order of  $\log \frac{\bar s}{\underline s}$. Notice that $\sup_{t\in T_n^d} G^\diamond(t,s)=\max\{\sup_{t\in T_n^d\cap T_n^c}G^\diamond(t,s),\sup_{t\in T_n^d\cap \bar T_n^c} G^\diamond(t,s) \}$, so that
	\begin{align}\label{April.new.132}
		\p(\sup_{t\in T_n^d\cap T_n^c} G^\diamond(t,s)>c)&\leq \p(\sup_{t\in T_n^d} G^\diamond(t,s)>c)\notag\\ &\leq \p(\sup_{t\in T_n^d\cap T_n^c} G^\diamond(t,s)>c)+\p(\sup_{t\in T_n^d\cap \bar T_n^c} G^\diamond(t,s)>c).
	\end{align} Therefore the theorem follows from expressions \eqref{new.17}, \eqref{New.A34} and the argument yielding \eqref{New.A34} using \eqref{F42}, \eqref{new.31} and \eqref{April.new.132}. \hfill $\Box$
	
	\ \\\noindent{\bf Calculations of $\kappa_j$ and $\zeta_{j,1}, \zeta_{j,2}$}.\ 
	It follows from Proposition 2  and Section 3 of \cite{sun1994simultaneous} that
	\begin{align}
		\kappa_j=\int_{t\in I_j, s\in [\underline s, \bar s]} {\det}^{1/2}(A^TA) ds dt
	\end{align}
	where the $n\times 2$ matrix $A=(\frac{\partial }{\partial t}\check T(t,s), \frac{\partial }{\partial s}\check T(t,s)),$ and the notation $\det$ denotes the 
	determinant.
	Tedious but straightforward calculations show that for $t\in \bar T_n^c$,
	
	\begin{align}\label{Kappa-Zeta_0}
		\left|T(t,s)\right|^2=\sigma^2(t)\int W^2(u)du+O(s+\frac{1}{ns}),\\\label{Kappa-Zeta_1}
		\frac{\partial}{\partial t}\left|T(t,s)\right |^2=o(1),
		~~\frac{\partial}{\partial s}\left|T(t,s)\right |^2=o(1).
	\end{align}
	Together with the fact that \begin{align}
		\frac{\partial \left|T(t,s)\right|^2}{\partial t}=2\left|T(t,s)\right|\frac{\partial \left|T(t,s)\right|}{\partial t},\frac{\partial \left|T(t,s)\right|^2}{\partial s}=2\left|T(t,s)\right|\frac{\partial \left|T(t,s)\right|}{\partial s}, \\\label{Kappa-Zeta_2}
		\frac{\partial }{\partial t}\check T_j(t,s)=\frac{-\sigma(\frac{j}{n})W'(\frac{j/n-t}{s})}{\sqrt{ns}s\left|T(t,s)\right|}-\frac{\sigma(\frac{j}{n})W(\frac{j/n-t}{s})\frac{\partial}{\partial t}\left|T(t,s)\right|}{\sqrt{ns}\left|T(t,s)\right|^2},\\
		\frac{\partial }{\partial s}\check T_j(t,s)=\frac{-\sigma(\frac{j}{n})W'(\frac{j/n-t}{s})(j/n-t)}{\sqrt{ns}s^2\left|T(t,s)\right|}-\frac{\sigma(\frac{j}{n})W(\frac{j/n-t}{s})}{2\sqrt{ns}s\left|T(t,s)\right|}-\notag\\\frac{\sigma(\frac{j}{n})W(\frac{j/n-t}{s})\frac{\partial}{\partial s}\left|T(t,s)\right|}{\sqrt{ns}\left|T(t,s)\right|^2},\label{Kappa-Zeta_3}
	\end{align}
	and tedious but straightforward calculations show that
	\begin{align}
		{\det}(A^TA)=((w_{11}w_{22}-w^2_{12})/(s^4u^2_{11}))(1+O(\frac{1}{ns}+s)).
	\end{align}
	where  $w_{12}=\int (W'(t))^2tdt+\frac{1}{2}\int W'(t)W(t)dt=0$ due to the fact that $W(t)$ is an odd function. The calculations also rely on the positiveness and boundedness of $w_{11}$, $w_{22}$ and $u_{11}$. Finally, following  Proposition 2  and Section 3 of \cite{sun1994simultaneous}, the quantities $\zeta_{j,1}$ and $\zeta_{j,2}$ are calculated by  \eqref{Kappa-Zeta_0}, \eqref{Kappa-Zeta_1}, \eqref{Kappa-Zeta_2} and \eqref{Kappa-Zeta_3}, of which the details are omitted for the sake of brevity. \hfill $\Box$
	\begin{lemma}\label{Lemma1}
		Under conditions of Theorem \ref{Thm1}, we have that 
		$$\sup_{t\in T_n^d\cup_{1\leq i\le m}\{d_i\}}|\Xi(t)-\tilde {\Xi}(t)|=O_p\left(n(s^*)^{2k+3}+\left((s^*)^{k+1}+\frac{1}{ns^*}\right)\left(n^{\frac{1}{4}}\log ^2n+\sqrt{ns^*}\left(\frac{s^*\log n}{\bar s}\right)^{\frac{1}{2}}\right)\right).$$
	\end{lemma}
	
	{\it Proof.} Straightforward calculations show that $\sup_{t\in T_n}|\Xi(t)-\tilde {\Xi}(t)|\leq I+II,$ where
	\begin{align*}
		&I=\sup_{t\in T_n^d\cup_{1\leq i\le m}\{d_i\}}\left|\sum_{i\in K(t)}\left(\frac{1}{\sqrt{ns^*}}\sum_{j=1}^n\beta(j/n)W\left(\frac{j/n-i/n}{s^*}\right)\right)^2/|K(t)|\right|,\\
		&II=2\sup_{t\in T_n^d\cup_{1\leq i\le m}\{d_i\}}\left|\sum_{i\in K(t)}\left(\frac{1}{\sqrt{ns^*}}\sum_{j=1}^n\beta(j/n)W\left(\frac{j/n-i/n}{s^*}\right)\right)\left(\frac{1}{\sqrt{ns^*}}\sum_{j=1}^n\varepsilon_jW\left(\frac{j/n-i/n}{s^*}\right)\right)/|K(t)|\right|.
	\end{align*}
	By \eqref{disc-constant}, we have that \begin{align}
		I&=O(n(s^*)^{2k+3}+\frac{1}{ns^*}),\label{I-(9)}\\
		II&\leq C(\sqrt{ns^*}(s^*)^{k+1}+\frac{1}{\sqrt{ns^*}})\sup_{t\in T_n^d\cup_{1\leq i\le m}\{d_i\}}\left|\frac{1}{\sqrt{ns^*}}\sum_{j=1}^n\varepsilon_j\sum_{i\in K(t)}W\left(\frac{i/n-j/n}{s^*}\right)/|K(t)|\right|\notag\\
		&:=C(\sqrt{ns^*}(s^*)^{k+1}+\frac{1}{\sqrt{ns^*}})\sup_{t\in T_n^d\cup_{1\leq i\le m}\{d_i\}}\left|\frac{1}{\sqrt{ns^*}}\sum_{j=1}^n\varepsilon_jG(j,t)\right|,\label{II-bound}
	\end{align}
	where $G(j,t)=\sum_{i\in K(t)}W\left(\frac{i/n-j/n}{s^*}\right)/|K(t)|$.
	By Proposition \ref{GaussianApprox}, there exist a series of $i.i.d.$ standard normals $\{V_i\}_{i\in \mathbb Z}$ such that $$\max_{1\leq i\leq n}\Big|\sum_{s=1}^i\varepsilon_s-\sum_{s=1}^i\sigma(s/n)V_s\Big|=o_p(n^{1/4}\log^2n).$$ Then by summation by parts formula, we have that
	\begin{align}\label{gussian-sumbyparts}
		&\sup_{t\in T_n^d\cup_{1\leq i\le m}\{d_i\}}\left|\sum_{j=1}^n\varepsilon_jG(j,t)-\sum_{j=1}^n\sigma(j/n)V_jG(j,t)\right|\notag\\
		&\leq \max_{i\leq n}\left|\sum_{s=1}^i\varepsilon_s-\sum_{s=1}^i\sigma(s/n)V_s\right|\sup_{t\in [\bar s, 1-\bar s]}\left(\left|G(1,t)\right|+\sum_{j=2}^n\left|G(j,t)-G(j-1,t)\right|\right)\notag\\
		&=o_p(n^{1/4}\log^2n),
	\end{align}
	where the bound of $(n^{1/4}\log^2n)$ is due to the following facts:
	\begin{description}
		\item(i) $G(j,t)=0$ if $|j/n-t|>s^*+\bar s$ which is  due to the definition of $K(t)$
		\item(ii) $\sup_{t\in[\bar s,1-\bar s]}|G(j,t)-G(j-1,t)|\leq M(n\bar s)^{-1}$ for a sufficiently large constant $M$. This is due to the mean value theorem which yields that  
		\begin{align}
			G(j,t)-G(j-1,t)=\frac{-1}{|K(t)| ns^*}\sum_{i\in K(t)}W'\left(\frac{\frac{i}{n}-\frac{j-\theta}{n}}{s^*}\right)
		\end{align}
		for some $\theta\in[0, 1]$.
	\end{description}  
	
	\noindent Combining \eqref{II-bound} and \eqref{gussian-sumbyparts} we obtain that
	\begin{align}\label{II-(12)}
		II\leq C(\sqrt{ns^*}(s^*)^{k+1}+\frac{1}{ns^*})\sup_{t\in T_n^d\cup_{1\leq i\le m}\{d_i\}}\left|\frac{1}{\sqrt{ns^*}}\sum_{j=1}^n\sigma(j/n)V_jG(j,t)\right|\notag
		\\+o_p\left(\left(\left(s^*\right)^{k+1}+\frac{1}{ns^*}\right)n^{1/4}\log^2n\right).
	\end{align}
	Observe that  $K(t),0<t<1$ is a set of indices which has $d$ different values such that $d\leq (4^2-1)n=15n$. Let $0<t_1<...<t_d<1$ be $d$ different points such that $K(t_s)$  be the $d$ different values.
	Then $\frac{1}{\sqrt{ns^*}}\sum_{j=1}^n\sigma(j/n)V_jG(j,t_s)$, $1\leq s\leq d$ are $d$ centered Gaussian random variables, such that
	\begin{align}\label{Discrete}
		\max_{1\leq s\leq d}\left|\frac{1}{\sqrt{ns^*}}\sum_{j=1}^n\sigma(j/n)V_jG(j,t_s)\right|=\sup_{0<t< 1}\left|\frac{1}{\sqrt{ns^*}}\sum_{j=1}^n\sigma(j/n)V_jG(j,t)\right|.
	\end{align}
	By using fact (i) and the fact that $\max_{1\leq j\leq n}\sup_{0\leq t\leq 1}|G(j,t)|\leq \frac{M s^*}{\bar s}$, we see that the variances of those centered Gaussian random variables are bounded, i.e. \begin{align}
		\frac{1}{ns^*}\sum_{j=1}^n\sigma^2(j/n)G^2(j,t_s)\leq M\frac{n\bar s}{ns^*}\left(\frac{s^*}{\bar s}\right)^2=M\frac{s^*}{\bar s} \end{align} for some sufficiently large constant $M$. Therefore Proposition \ref{max-gaussian} gives that 
	\begin{align}\label{II-(13)}
		\max_{1\leq s\leq d}\left|\frac{1}{\sqrt{ns^*}}\sum_{j=1}^n\sigma(j/n)V_jG(j,t_s)\right|=O_p(( s^*/ \bar s)^{1/2}\log^{1/2}n).
	\end{align}
	Then the lemma holds in view of \eqref{I-(9)},  \eqref{II-bound}, \eqref{II-(12)} and  \eqref{II-(13)}. \hfill $\Box$
	
	Recall $\FF_i=(\eta_{-\infty},...,\eta_i)$. Write $\FF_{j-m,j}=(\eta_{j-m},\eta_{j-m+1},...,\eta_j)$. In the remaining of the supplemental material, for the sake of brevity let $p'= p/2$ where $p$ is defined in condition (A1) and (A2) of the main article.
	\begin{lemma}\label{tildeXi}
		Under conditions of Theorem \ref{Thm1}, we have
		\begin{align}\label{tildeXibound}
			\sup_{t\in [\bar s, 1-\bar s]}\| \tilde \Xi(t)-\mathbb{E} \tilde \Xi(t)\|_{p'}\leq M(p')^{1/2}\left((s^*/\bar s)^{1/2}+\frac{\log n}{\sqrt{n\bar s}}\right)
		\end{align}
	\end{lemma}
	{\it Proof.} 
	Write  $\tilde \varepsilon_j=\E(\varepsilon_j|\FF_{j-m,j})$ and \begin{align}\tilde \Xi_m(t)=\sum_{i\in K(t)}\tilde H_m^2(i/n,s^*)/|K(t)|, \\ \text{where }\tilde H_m(t,s^*)=\frac{1}{\sqrt{ns^*}}\sum_{j=1}^n\tilde \varepsilon_jW\left(\frac{j/n-t}{s^*}\right). \notag\end{align}
	It is not hard to see that $\{\tilde \varepsilon_i, i\in \mathbb Z\}$ are generated from a PLS process with nonlinear filters satisfying conditions (A1)-(A3). Let $\tilde \varepsilon_{i,(i-l)}$ be the random variable by changing the innovation $\eta_{i-l}$ of $\tilde \varepsilon_i$ to its $i.i.d$ copy $\eta_{i-l}'$. Observe that we have the following observation by properties of conditional expectation which is for any $u>0$, $u\in \mathbb Z$,
	\begin{description}
		\item (a) $\|\tilde \varepsilon^2_u\|_{p'}<\infty$.
		\item (b) $\|\pp_{u-l} \tilde \varepsilon^2_u\|_{p'}\leq \|\tilde \varepsilon^2_u-\tilde \varepsilon^2_{u,u-l}\|_{p'}\leq \|\tilde \varepsilon_u-\tilde \varepsilon_{u,u-l}\|_p\|\tilde \varepsilon_u+\tilde \varepsilon_{u,u-l}\|_p=O(\chi^l)$,
		\item (c) $\pp_{u-l}\tilde \varepsilon_u \tilde\varepsilon_v=0$ if $l>2m$, $\forall v\in \mathbb Z$.
		\item (d)	$\|\tilde \varepsilon_{j}-\varepsilon_j\|_{p}\leq M\chi^m$, for $j\in \mathbb Z$.
	\end{description} 
	Here (d) can be proved using the same proof of Theorem 1 in \cite{zhou2014inference}.
	On the other hand, straightforward calculations show that 
	\begin{align}
		\tilde \Xi_m(t)&=\frac{1}{ns^*}\sum_{u=1}^n\sum_{v=1}^n\tilde \varepsilon_u\tilde \varepsilon_v\sum_{i\in K(t)}W\left(\frac{u/n-i/n}{s^*}\right)W\left(\frac{v/n-i/n}{s^*}\right)/|K(t)|\notag\\
		&:=\frac{1}{ns^*}\sum_{u=1}^n\sum_{v=1}^n\tilde \varepsilon_u\tilde \varepsilon_vG(u,v,t)\label{new-tildexi}\\
		&:=\frac{1}{ns^*}\sum_{u=1}^n\tilde \varepsilon_u^2 G(u,u,t)+\frac{2}{ns^*}\sum_{u=1}^n\sum_{v=1}^{u-1}\tilde \varepsilon_u \tilde \varepsilon_vG(u,v,t):=A(t)+B(t),\label{tildexi}
	\end{align}
	where $A(t)$ and $B(t)$ are defined in an obvious manner, and \begin{align}\label{April24-92}
		G(u,v,t):=\sum_{i\in K(t)}W\left(\frac{u/n-i/n}{s^*}\right)W\left(\frac{v/n-i/n}{s^*}\right)/|K(t)|.
	\end{align}
	Observe that for all $t\in[0,1]$, we have that \begin{description}
		\item (i) $G(u,v,t)=0$ if $|u/n-t|>(s^*+\bar s)$ or $|v/n-t|>(s^*+\bar s)$ or $|u/n-v/n|>2s^*$.
		\item (ii) $|G(u,v,t)|\leq Cs^*(\bar s)^{-1}$.
	\end{description}
	Notice that (i) implies for each $t$, there are at most $O(n^2\bar s s^*)$ of non-zero $G(u,v,t)$.
	By decomposing $\Xi(t)$ in the same way as the decomposition \eqref{new-tildexi} of $\tilde \Xi(t)$,  property (d) and the Cauchy inequality imply that
	\begin{align}
		\|\tilde \Xi(t)-\tilde \Xi_m(t)\|_{p'}=\frac{1}{ns^*}\Big\|\sum_{u=1}^n\sum_{v=1}^n G(u,v,t)(\tilde \varepsilon_u\tilde \varepsilon_v-\varepsilon_u\varepsilon_v)\Big\|_{p'}\leq M n\bar s\chi^m.
	\end{align}
	Thus by the property of $\mathcal L^{p'}$ space and the triangle inequality, we have that
	\begin{align}
		\|\tilde \Xi(t)-\E \tilde \Xi(t)-(\tilde \Xi_m(t)-\E\tilde \Xi_m(t))\|_{p'}\leq C n\bar s\chi^m.\label{Xi(t)-2}
	\end{align}
	To establish the bound of \eqref{tildeXibound}, the remaining task is to evaluate the bound of $\|\tilde \Xi_m(t)-\E\tilde \Xi_m(t)\|_{p'}$, which is, by \eqref{tildexi}, further bounded by $\|A(t)-\E A(t)\|_{p'}$ and $\|B(t)-\E B(t)\|_{p'}$.
	For $A(t)-\E A(t)$, we have that 
	\begin{align}\label{A(t)-1}
		\|A(t)-\E(A(t))\|_{p'}\leq \frac{1}{ns^*}\sum_{l=0}^\infty\left\|\sum_{u=1}^n\pp_{u-l}\tilde \varepsilon_u^2G(u,u,t)\right\|_{p'}.
	\end{align}
	Furthermore, proposition \ref{Prop-Burk}, property (b) of $\tilde \varepsilon_i$ and properties (i) (ii) of $G(u,v,t)$ show that
	\begin{align}
		\left \|\sum_{u=1}^n\pp_{u-l}\tilde \varepsilon_{u}^2G(u,u,t) \right\|_{p'}^2\leq Cp'\sum_{u=1}^n\|\pp_{u-l}\tilde \varepsilon_u^2G(u,u,t)\|^2_{p'}\leq Cp' n\bar s \chi^{2l} (s^*/\bar s)^2,
	\end{align} where $C$ is the constant defined in Proposition \ref{Prop-Burk}.
	This, together with \eqref{A(t)-1} and property (a) of $\tilde \varepsilon_i$ shows that
	\begin{align}\label{A(t)-2}
		\|A(t)-\E(A(t))\|_{p'}\leq M(p')^{1/2}(n\bar s)^{-1/2}.
	\end{align}
	For $B(t)-\E B(t)$, property (c) of $\tilde \varepsilon_i$ shows that
	\begin{align}\label{B(t)-1}
		\|B(t)-\E B(t)\|_{p'}\leq  \frac{2}{ns^*}\sum_{l=0}^{2m}\left\|\sum_{u=1}^{n}\pp_{u-l}\tilde \varepsilon_u\sum_{v=1}^{u-1}\tilde \varepsilon_vG(u,v,t)\right\|_{p'}:=\frac{2}{ns^*}\sum_{l=0}^{2m} \tilde s_l(t),
	\end{align}
	where $\tilde s_l(t)$ is defined in an obvious manner. 
	By proposition \ref{Prop-Burk} and Property (i) of $G(u,v,t)$, it follows that 
	\begin{align}
		\tilde s_l^2(t)\leq Cp'\sum_{u\in K'(t)}\left\|\pp_{u-l}\tilde \varepsilon_u\sum_{v=1}^{u-1}\tilde \varepsilon_vG(u,v,t)\right\|_{p'}^2,
	\end{align}
	where $K'(t)=[\lf nt-ns^*-n\bar s\rf\vee 1,\lceil nt+ns^*+n\bar s\rceil\wedge n]$.
	Straightforward calculations using Proposition \ref{Prop-Burk}, the triangle inequality, the Cauchy inequality, properties (i) (ii) of $G(u,v,t)$ and nonstationary extension of Theorem 1 of \cite{wu2005nonlinear} show that 
	\begin{align}
		\left\|\pp_{u-l}\tilde \varepsilon_u\sum_{v=1}^{u-1}\tilde \varepsilon_vG(u,v,t)\right\|_{p'}\leq \left\|\tilde \varepsilon_u\sum_{v=1}^{u-1}\tilde \varepsilon_{v}G(u,v,t)-\tilde \varepsilon_{u,(u-l)}\sum_{v=1}^{u-1}\tilde \varepsilon_{v,(u-l)}G(u,v,t)\right\|_{p'}\leq I+II, 
	\end{align}
	where
	\begin{align}
		I=&\left\|(\tilde \varepsilon_{u,(u-l)}-\tilde \varepsilon_u)(\sum_{v=1}^{u-l-1}+\sum_{v=u-l}^{u-1})\tilde \varepsilon_{v,(u-l)}G(u,v,t)\right\|_{p'}\leq M\chi^ls^*(\bar s)^{-1}(\min(u,ns^*))^{1/2} ,\\
		II=&\left\|\tilde \varepsilon_u(\sum_{v=1}^{u-l-1}+\sum_{v=u-l}^{u-1})(\tilde \varepsilon_{v,(u-l)}-\tilde \varepsilon_v)G(u,v,t)\right\|_{p'}\notag\\
		=&\left\|\tilde \varepsilon_u(\sum_{v=u-l}^{u-1})(\tilde \varepsilon_{v,(u-l)}-\tilde \varepsilon_v)G(u,v,t)\right\|_{p'}\leq M(s^*/\bar s)\min(l,ns^*,\sum_{s=1}^{l-1}\chi^s).
	\end{align}
	As a result, we have that 
	\begin{align}
		\tilde s_l^2(t)\leq Mp'\chi^{2l}(s^*/\bar s)^2n\bar s (ns^*
		)+Mp'n\bar s(s^*/\bar s)^2. 
	\end{align}
	By plugging the above equation into \eqref{B(t)-1}, we have that
	\begin{align}\label{B(t)-2}
		\|B(t)-\E B(t)\|_{p'}\leq M(p')^{1/2}\left( \left(\frac{s^*}{\bar s}\right)^{1/2}+2m/\sqrt {n\bar s}\right).
	\end{align}
	Therefore \eqref{A(t)-2} and \eqref{B(t)-2} lead to that
	\begin{align}\label{Xi(t)-3}
		\|\tilde \Xi_m(t)-\E \tilde \Xi_m(t)\|_{p'}\leq M(p')^{1/2}\left( \left(\frac{s^*}{\bar s}\right)^{1/2}
		+2m/\sqrt {n\bar s}\right).
	\end{align}
	Finally  by	taking $m=a\log n$ such that $a\log \chi<-3$, the lemma follows from \eqref{Xi(t)-2} and \eqref{Xi(t)-3}. \hfill $\Box$ 
	
	\begin{lemma}\label{Lemma3}
		Assume conditions of Lemma \ref{tildeXi} hold. Then we have that
		\begin{align}
			\sup_{t\in[\bar s, 1-\bar s] } |\tilde \Xi(t)-\mathbb{E} \tilde \Xi(t)|=O_p \left((p')^{1/2}\left((s^*/\bar s)^{1/2}+\frac{\log n}{\sqrt{n\bar s}}\right)(s^*)^{-\frac{1}{p'}}\right)
		\end{align}
	\end{lemma}
	{\it Proof.} Write $\Psi(t)=\tilde \Xi(t)-\E\tilde \Xi(t)$. Recall the definition of $G(u,v,t)$ in the proof of Lemma \ref{tildeXi}.
	Observe that there exists a sufficiently large positive constant $C$ such that
	\begin{description}
		\item (a) $|G(u,v,t_1)-G(u,v,t_2)|\leq C(t_1-t_2)(\bar s)^{-1}$ for $n^{-1}\leq|t_1-t_2|\leq \bar s.$
		\item (b) $|G(u,v,t_1)-G(u,v,t_2)|\leq C s^*(\bar s)^{-1}$. 
		\item (c) $|G(u,v,t_1)-G(u,v,t_2)|\leq C(n\bar s)^{-1}$ for $|t_1-t_2|\leq n^{-1}.$
	\end{description}
	Since $s^*\leq \bar s$, by considering the case that $|t_1-t_2|\geq \bar s\geq s^*$, (a) and (b) lead to that
	\begin{align}\label{c-star}
		|G(u,v,t_1)-G(u,v,t_2)|\leq C(t_1-t_2)(\bar s)^{-1}~~ \text{for}~~ t_1,t_2\in [\bar s, 1-\bar s].
	\end{align}
	It follows from  \eqref{c-star}, (i) (ii) in the proof of Lemma \ref{tildeXi} and a similar argument of proof of  Lemma \ref{tildeXi} that,
	for $t_1,t_2\in [\bar s, 1-\bar s]$ and a sufficient large constant $M$,
	\begin{align}
		\sup_{t_1,t_2\in [\bar s, 1-\bar s],|t_1-t_2|\geq n^{-1}}\left\|\frac{\Psi(t_1)-\Psi(t_2)}{|t_1-t_2|}\right\|_{p'}\leq M(p')^{1/2}\left((s^*/\bar s)^{1/2}+\frac{\log n}{\sqrt{n\bar s}}\right)(s^*)^{-1},\label{new.eq34}\\
		\sup_{t_1,t_2\in [\bar s, 1-\bar s],|t_1-t_2|\leq n^{-1}}\left\|\Psi(t_1)-\Psi(t_2)\right\|_{p'}\leq M(p')^{1/2}\left((s^*/\bar s)^{1/2}+\frac{\log n}{\sqrt{n\bar s}}\right)(ns^*)^{-1}.\label{new.eq35}
	\end{align}
	For any series of integers $c_n\rightarrow \infty$, $c_n=o(n)$ define $t_s=\bar s+s\eta_n, \eta_n=\frac{1-2\bar s}{c_n}, 1\leq s\leq (c_n-1)$. By the triangle inequality we have the following decomposition,
	\begin{align}
		\sup_{t\in [\bar s, 1-\bar s]}|\Psi(t)|\leq \max_{1\leq s\leq c_n-1}|\Psi(t_s)|+\max_{1\leq s\leq c_n-1}\sup_{|t-t_s|\leq \eta_n}|\Psi(t_s)-\Psi(t)|.
	\end{align}
	First, by Lemma \ref{tildeXi} and the triangle inequality we have
	\begin{align}\label{New.A.74-Sep}
		\big\| \max_{1\leq s\leq c_n-1}|\Psi(t_s)|\big\|_{p'}\leq 	\big(\sum_{1\leq s\leq c_n-1} \big\| \Psi(t_s)\big\|^{p'}_{p'}\big)^{\frac{1}{p'}}
		\leq Mc^{\frac{1}{p'}}_n(p')^{1/2}\left((s^*/\bar s)^{1/2}+\frac{\log n}{\sqrt{n\bar s}}\right).
	\end{align}
	A further application of the triangle inequality yields that
	\begin{align}
		&\max_{1\leq s\leq c_n-1}|\Psi(t_s)-\Psi(t)|\leq \max_{1\leq i\leq c_{n-1}}\tilde A_i+\max_{1\leq i\leq c_n-1} \tilde B_i,\label{April-182}\\
		&\tilde A_i=\bigg(\sum_{s=\lceil n(t_i-\eta_n)\rceil+1}^{\lfloor n(t_i+\eta_n)\rfloor }\left|\Psi(\frac{s}{n})-\Psi(\frac{s-1}{n})\right|\notag
		\\&+\left|\Psi(\frac{\lceil n(t_i-\eta_n)\rceil }{n})-\Psi(t_i-\eta_n)\right|+\left|\Psi(t_i+\eta_n)-\Psi(\frac{\lfloor n(t_i+\eta_n)\rfloor }{n})\right|\bigg),
		\\&\tilde B_i=\sup_{|t-t_i|\leq \eta_n}\left(\left|\Psi(t)-\Psi(\frac{\lf n t\rf}{n })\right|+\left|\Psi(\frac{\lf n t\rf+1}{n})-\Psi(t)\right|\right)
	\end{align}
	By equation \eqref{new.eq34} and \eqref{new.eq35}, we have that
	\begin{align}
		\|\tilde A_i\|_{p'}\leq M(p')^{1/2}\left((s^*/\bar s)^{1/2}+\frac{\log n}{\sqrt{n\bar s}}\right)(s^*)^{-1}\eta_n.
	\end{align} 
	Using similar argument to \eqref{New.A.74-Sep} we obtain
	\begin{align}\label{April-185}
		\big\| \max_{1\leq s\leq c_n-1}|\tilde A_i|\big\|_{p'}
		\leq c_n^{\frac{1}{p'}} M(p')^{1/2}\left((s^*/\bar s)^{1/2}+\frac{\log n}{\sqrt{n\bar s}}\right)(s^*)^{-1}\eta_n.
	\end{align}
	
	On the other hand, for $B_i$ we have that
	\begin{align}
		&\max_{1\leq i\leq c_{n-1}} B_i\leq \sup_{t\in (0,1)} (Z_1(t)+Z_2(t)),\\
		&\text{where~~} Z_1(t)=\left|\Psi(t)-\Psi(\frac{\lf n t\rf}{n})\right|,~Z_2(t)=\left|\Psi(\frac{\lf n t\rf+1}{n})-\Psi(t)\right|.\label{A82Sep6}
	\end{align}
	In the following, we shall show that for $s=1,2$,
	\begin{align}\label{S83Sep27}
		\|\sup_{0<t< 1}|Z_s(t)|\|_{p'}\leq M(p')^{1/2}\left((s^*/\bar s)^{1/2}+\frac{\log n}{\sqrt{n\bar s}}\right)(ns^*)^{-1}n^{\frac{1}{p'}},\end{align}
	such that
	\begin{align}\label{April-187}
		\max_{1\leq i\leq c_{n-1}} B_i=O_p\big((p')^{1/2}\big((s^*/\bar s)^{1/2}+\frac{\log n}{\sqrt{n\bar s}}\big)(ns^*)^{-1}n^{\frac{1}{p'}}\big).
	\end{align}
	Then the lemma follows from the estimates \eqref{New.A.74-Sep}, \eqref{April-185} and \eqref{April-187} by letting $c_n=(s^*)^{-1}$.
	Now we show \eqref{A82Sep6} holds.
	Notice that $\tilde \Xi(t)$ has a  expansion similar to that of $\tilde \Xi_m(t)$ in \eqref{new-tildexi}
	\begin{align}
		\tilde \Xi(t)=\frac{1}{ns^*}\sum_{u=1}^n\sum_{v=1}^n \varepsilon_u \varepsilon_vG(u,v,t)
	\end{align}
	with $G(u,v,t)$ defined in \eqref{April24-92}. As a consequence, we have that
	\begin{align}
		Z_1(t)=\frac{1}{ns^*}\sum_{u=1}^n\sum_{v=1}^n \big(\varepsilon_u \varepsilon_v-\E(\varepsilon_u \varepsilon_v)\big)G^\circ(u,v,t),\\\text{where}~
		G^\circ(u,v,t)=G(u,v,t)-G(u,v,t_n'), ~t'_n=\frac{\lf n t\rf}{n}
	\end{align}
	Notice that by property (a) and property (i) of Lemma \ref{tildeXi}, using similar argument to the proof of Lemma \ref{tildeXi}, we have that
	\begin{align}
		\sup_{t\in [\bar s, 1-\bar s]}\| Z_1(t)\|_{p'}\leq M(p')^{1/2}\left((s^*/\bar s)^{1/2}+\frac{\log n}{\sqrt{n\bar s}}\right)(ns^*)^{-1}
	\end{align}
	It follows from a similar argument to \eqref{Discrete} that there exist $0<v_1<...<v_d<1$ for $d\leq 15n$ such that \begin{align}\label{Discrete1}
		\max_{1\leq s\leq d}\left|Z_1(v_s)\right|=\sup_{0<t< 1}\left|Z_1(t)\right|. \end{align}
	As a consequence, we have that
	\begin{align}
		\|\sup_{0<t< 1}|Z_1(t)|\|_{p'}=&\| \max_{1\leq s\leq d}|Z_1(v_s)|\|_{p'}\leq \Big(\sum_{s=1}^d\|Z_1(v_s)|\|^{p'}_{p'}\Big)^{\frac{1}{p'}}
		\notag\\&\leq M(p')^{1/2}\left((s^*/\bar s)^{1/2}+\frac{\log n}{\sqrt{n\bar s}}\right)(ns^*)^{-1}n^{\frac{1}{p'}}.
	\end{align}
	Using the same arguments we shall see that $$\|\sup_{0<t< 1}|Z_2(t)|\|_{p'}\leq M(p')^{1/2}\left((s^*/\bar s)^{1/2}+\frac{\log n}{\sqrt{n\bar s}}\right)(ns^*)^{-1}n^{\frac{1}{p'}},$$ therefore \eqref{S83Sep27} holds and the proof is completed. \hfill $\Box$

	\begin{lemma}\label{Lemma4}
		Under the conditions of Theorem \ref{Thm1}, we have that: 
		i) Uniformly for $t\in T_n^c$
		\begin{align}
			\E\tilde \Xi(t)=\sigma^2(t)\int W^2(t)dt+O_p(\log^2 n/(ns^*)+\bar s-s^*\log s^*).
		\end{align}
		If we further assume that the long-run variance in condition (A3) has a Lipschitz continuous first order derivative, then $\bar s$ in the above rate can be improved to $\bar s^2$.
		
		ii)Uniformly for $t\in \bar{T}_n^c$, $\E\tilde \Xi(t)=g(t)+O_p(\nu_{3,n})$ for $t\in \bar T_n^c$ for some bounded real function $g(t)$ such that $M_1\leq g(t)\leq M_2$ with constants $M_1>0,M_2>0$. 
	\end{lemma}
	{\it Proof.} Proof of (i).
	It follows from the proof of Lemma 5 of \cite{zhou2010simultaneous} that
	\begin{align}\label{2019-April-173}
		\E(\varepsilon_i\varepsilon_j)=O(\chi^{|i-j|}).
	\end{align}
	Let $r=r(n)\rightarrow \infty$ be a diverging series such that $r(n)=o(ns^*)$. Since $W(\frac{i/n-t}{s^*})W(\frac{j/n-t}{s^*})=0$ if $|i/n-t|\geq s^*$ or $|j/n-t|\geq s^*$ or $|i/n-j/n|\geq s^*$, we have uniformly for $t\in T_n^c$, \begin{align}\label{2019-April-174}
		\E(\tilde H^2(t,s^*))=\frac{1}{ns^*}\E\left(\sum_{i=1}^n\sum_{j=1}^n
		\varepsilon_i\varepsilon_jW\left(\frac{i/n-t}{s^*}\right)W\left(\frac{j/n-t}{s^*}\right)\mathbf 1(|i-j|\leq r)\right)+R,
	\end{align}
	where \begin{align}\label{2019-April-175}
		|R|\leq \frac{M\sum_{s=r+1}^{\lf ns^*\rf}(\lf ns^*\rf+1-s) \chi^s}{ns^*}\leq M \chi^{r+1}.
	\end{align}
	On the other hand, the Cauchy inequality and conditions (A1) yield that uniformly for $t\in (c_s+\bar s, c_{s+1}-\bar s)$ and $|i/n-t|\leq s^*, |j/n-t|\leq s^*$,
	\begin{align}
		\E(L_s(t,\FF_i)L_s(t,\FF_j)-L_s(i/n,\FF_i)L_s(j/n,\FF_j))=O(\min\{s^*, \chi^{|i-j|}\}),
	\end{align}
	which further leads to that
	\begin{align}\label{Aug-11-115}
		&\frac{1}{ns^*}\E\left(\sum_{i=1}^n\sum_{j=1}^n
		\varepsilon_i\varepsilon_jW\left(\frac{i/n-t}{s^*}\right)W\left(\frac{j/n-t}{s^*}\right)\mathbf 1(|i-j|\leq r)\right)\notag\\
		&=\frac{1}{ns^*}\E\left(\sum_{i=1}^n\sum_{j=1}^n
		L_s(t,\FF_i)L_s(t,\FF_j)W\left(\frac{i/n-t}{s^*}\right)W\left(\frac{j/n-t}{s^*}\right)\mathbf 1(|i-j|\leq r)\right)+O(-s^*\log s^*)\notag
		\\&:=I+O(-s^*\log s^*).
	\end{align}
	Furthermore, 
	let $r=a\log n$ for some sufficiently large constant $a$, then mean value theorem and straightforward calculations show that
	\begin{align}
		I=&\frac{1}{ns^*}\sum_{u=1}^n\E(L_s^2(t,\FF_u))W^2\left(\frac{u/n-t}{s^*}\right)+\frac{2}{ns^*}\sum_{u=1}^r\sum_{i=u+1}^{n-u}\E(L_s(t,\FF_i)L_s(t,\FF_{i+u}))W^2\left(\frac{i/n-t}{s^*}\right)\notag\\
		&+\frac{1}{ns^*}\sum_{u=1}^r(\sum_{i=1}^u+\sum_{i=n-2u+1}^{n-u})\E(L_s(t,\FF_i)L_s(t,\FF_{i+u}))W^2\left(\frac{i/n-t}{s^*}\right)+O(\frac{r^2}{ns^*})\notag\\
		&=\int W^2(v)dv\left(\E(L_s^2(t,\FF_0)+2\sum_{u=1}^r\E(L_s(t,\FF_0)L_s(t,\FF_u))\right)+O(\frac{1}{ns^*}+\frac{r^2}{ns^*})\notag\\&=\sigma^2(t)\int W^2(u)du+O(\frac{1}{ns^*}+\chi^r+\frac{r^2}{ns^*}),\label{Aug-11-116}
	\end{align}
	where we have used the definition of long-run variance in the last equality. 
	Then by equation \eqref{2019-April-174}, \eqref{2019-April-175}, \eqref{Aug-11-115} and \eqref{Aug-11-116} we obtain
	\begin{align}\label{2019-april-179}
		\E(\tilde H^2(t,s^*))=\sigma^2(t)\int W^2(u)du+O\left(-s^*\log s^*+\frac{\log^2n}{ns^*}\right).
	\end{align}
	By definition of $\tilde \Xi(t)$, and the Lipschitz continuity of $\sigma^2(t)$, it follows that
	\begin{align}
		\E(\tilde \Xi(t))&=\E\left(\frac{\sum_{i\in K(t)}\sigma^2(\frac{i}{n})\int W^2(u)du}{|K(t)|}\right)+O(-s^*\log s^*+\frac{\log ^2n}{ns^*})
		\\&=\sigma^2(t)\int W^2(u)du+O(-s^*\log s^*+\frac{\log ^2n}{ns^*}+\bar s).
	\end{align}
	Since $K(t)$ is a set of $i$ that is symmetric around $t$, the above bias $\bar s$ is reduced to $\bar s^2$ if $\sigma^2(t)$ has a Lipschitz continuous first order derivative. 
	
	Proof of ii). For $t\in [c_s-\bar s, c_s- s^*)\cap  (c_s+ s^*, c_s- \bar s]$, $1\leq s\leq l$, equation \eqref{2019-april-179} holds and the corresponding results follow. For $t\in [c_s-s^*,c_s+s^*]$, $1\leq s\leq l$, let $\theta=\frac{t-(c_s-s^*)}{2s^*}\in [0,1]$ such that $t=c_s-s^*+2s^*\theta$. 
	Consider the case that $r>a\log n$ for some sufficiently positive $a$. Elementary calculations show that for $t\in [c_s-s^*,c_s+s^*]$,
	\begin{align}\label{2019-April-182}&\E(\tilde H^2(t,s^*))-R=\frac{1}{ns^*}\E\bigg(\sum_{i=1}^n\sum_{j=1}^n
		\varepsilon_i\varepsilon_jW\left(\frac{i/n-t}{s^*}\right)W\left (\frac{j/n-t}{s^*}\right) \mathbf 1(|i-j|\leq n\theta)\times\notag\\&\bigg (
		\mathbf 1(1\leq i,j\leq \lf nc_s\rf)+\mathbf 1(\lf nc_s\rf\leq i,j\leq n)+2\mathbf 1(1\leq i\leq \lf n c_s\rf)\mathbf 1(\lf nc_s\rf\leq j\leq n)\bigg)\bigg)\notag\\&=I_n+II_n+III_n,
	\end{align}  where $R$ is the remaining term defined by equation \eqref{2019-April-175},
	and \begin{align*}
		I_n=\frac{1}{ns^*}\E\left(\sum_{|i-j|\leq r, i,j\in \mathcal I_1}
		\varepsilon_i\varepsilon_jW\left(\frac{i/n-c_s}{s^*}+1-2 \theta\right)W\left(\frac{j/n-c_s}{s^*}+1-2 \theta\right)\right),\\ 
		II_n=\frac{1}{ns^*}\E\left(\sum_{|i-j|\leq r, i,j\in\mathcal I_2}
		\varepsilon_i\varepsilon_jW\left(\frac{i/n-c_s}{s^*}+1-2 \theta\right)W\left(\frac{j/n-c_s}{s^*}+1-2 \theta\right)\right),\\ 
		III_n=\frac{2}{ns^*}\E\left(\sum_{|i-j|\leq r, i\in \mathcal I_1, j\in \mathcal I_2}
		\varepsilon_i\varepsilon_jW\left(\frac{i/n-c_s}{s^*}+1-2 \theta\right)W\left(\frac{j/n-c_s}{s^*}+1-2 \theta\right)\right),\\
		\mathcal I_1= [\lceil nc_s-2ns^*+2\theta ns^*\rceil, \lf nc_s\rf ],\mathcal I_2= [ \lf nc_s\rf+1,\lceil nc_s+2\theta ns^*\rceil ].
	\end{align*}
	For $I_n$, observe that for $i,j\in \mathcal I_1$,
	\begin{align}
		\E(L_{s-1}(c_s-,\FF_i)L_{s-1}(c_s-,\FF_j)-L_{s-1}(i/n,\FF_i)L_{s-1}(j/n,\FF_j))=O(\min\{s^*, \chi^{|i-j|}\}),
	\end{align}
	Using similar arguments to the proof of part (i) with the above fact, we get
	\begin{align}
		\E(I_n)=\sigma^2(c_s-)\int_{-1}^{1-2\theta} W^2(t)dt+O\left(-s^*\log s^*+\frac{\log^2n}{ns^*}\wedge (2-2\theta) \right).
	\end{align}
	Similarly we have
	\begin{align}
		\E(II_n)= \sigma^2(c_s+)\int_{1-2\theta}^1 W^2(t)dt+O\left(-s^*\log s^*+\frac{\log^2n}{ns^*}\wedge (2\theta) \right).
	\end{align}
	Using equation \eqref{2019-April-173} we have that 
	\begin{align}
		\E(III_n)=O(\frac{\log n}{ns^*})=o(1).
	\end{align}
	By the definition of the long-run variance $\sigma^2(t)$ and the limiting results of expectations of $I_n, II_n$ and $III_n$, we have that 
	\begin{align}
		\E(\tilde H^2(t,s^*))=\sigma^2(c_s-)\int_{-1}^{1-2\theta} W^2(t)dt+\sigma^2(c_s+)\int_{1-2\theta}^1 W^2(t)dt+o(1).
	\end{align}
	uniformly for $t\in [c_s-s^*,c_s+s^*]$, $1\leq s\leq l$.
	This fact together with the definition of $\tilde \Xi(t)$ show that ii) of the lemma holds. Hence the lemma follows. \hfill $\Box$
	
	
	\begin{lemma}\label{G_dr}
		Under the condition (b) of Theorem \ref{thm2}, we have that with $n\rightarrow \infty$, 
		\begin{align}
			\p(\min_{1\leq r\leq m}|G(d_r,\tilde s)|\leq c_{1-\alpha})\rightarrow 0
		\end{align}
	\end{lemma}
	{\it Proof.} By Slutsky's theorem and Lemmas \ref{Lemma1}-\ref{Lemma4}, it suffices to show that
	\begin{align}
		\p(\min_{1\leq r\leq m}\sup_{\underline s\leq s\leq \bar s}\frac{|H(d_r, s)|}{\sigma(d_r)(\int_{-1}^1 W^2(u)du)^{1/2}}\leq c_{1-\alpha})\rightarrow 0
	\end{align}
	By basic properties of probability, summation by parts formula in the proof of equation (41) of \cite{zhou2010nonparametric} and Proposition \ref{GaussianApprox}, it further suffices to show 
	\begin{align}\label{eq-G_dr}
		\p\left(\min_{1\leq r\leq m}\frac{|\sqrt{ns}\tilde G_n(d_r,\bar s)+\check H(d_r,\bar s)|}{\sigma(d_r)\sqrt{\int_{-1}^1W^2(u)du}}\leq c_{1-\alpha}\right)\rightarrow 0,
	\end{align}
	where $$\check H(d_r,\bar s)=\frac{1}{\sqrt{n \bar s}}\sum_{j=1}^n\sigma(j/n)V_jW\left(\frac{j/n-d_r}{\bar s}\right) $$
	and $(V_i)_{1\leq i\leq n}$ are a series of $i.i.d.$ $N(0,1)$ random variables. As a consequence,  $\{\check H(d_r,\bar s)\}_{1\leq r\leq m}$ are normal random variables such that $\max_{1\leq r\leq m}Var(\check H(d_r,\bar s))\leq M$. Then the LHS of equation \eqref{eq-G_dr} is bounded by 
	\begin{align}
		\sum_{1\leq r\leq m}\Big(&\p(\check H(d_r,\bar s)\leq \sigma(d_r)c_{1-\alpha}\sqrt{\int_{-1}^1W^2(u)du}-\sqrt{n\bar s}\tilde G(d_r,\bar s))\notag\\&-\p(\check H(d_r,\bar s)\leq -\sigma(d_r)c_{1-\alpha}\sqrt{\int_{-1}^1W^2(u)du}-\sqrt{n\bar s}\tilde G(d_r,\bar s))\Big)\rightarrow 0,
	\end{align}
	where the convergence to $0$ is due to Proposition \ref{proposition-1-10-13}, the fact that $c_{1-\alpha}\leq M\sqrt{\log n-\log \alpha}$ and the properties of tail probability of normal random variables. \hfill $\Box$
	\ \\\ \ \\\ \\
	\noindent{\bf Proof of Theorem \ref{thm2}.} 
	We emphasis that in the following proof, we omit subscript $n$ of $m_n$ to ease the notation. Please bear in mind that $m$ can depend on $n$.
	
			First, (a) follows from Theorem \ref{Thm1}.

			Second, in order to show (b) we define the events \begin{align}
				A_n=\{\hat m=m, |d_i-\hat d_i|\leq \bar s, 1\leq i\leq m\},\\
				S_n=\{\sup_{h_n< |t-d_r|\leq \gamma}\sup_{\underline s\leq s\leq \bar s}|G(t,s)|< \sup_{\underline s\leq s\leq \bar s}|G(d_r,s)|,1\leq r\leq m\},\end{align}
			where the parameter $\gamma$ is short for $\gamma_n\wedge \check \gamma$ which has been used in condition (B2) of the main article.
			Then to show (b), it is equivalent to show 
			\begin{align}
				\lim_{n\rightarrow \infty}\p(A_n\cap  S_n)= 1-\alpha.
			\end{align} 
			
			Notice that  
			Theorem \ref{Thm1} and Lemma \eqref{G_dr} imply  that $\lim_{n\rightarrow \infty}\p(A_n)=1-\alpha$. As a result it suffices to show $\lim_{n\rightarrow \infty}\p( S_n)=1$, or equivalently
			\begin{align}
				\lim_{n\rightarrow \infty}\p(\sup_{h_n<  |t-d_r|\leq \gamma}\sup_{\underline s\leq s\leq \bar s}|G(t,s)|\geq \sup_{\underline s\leq s\leq \bar s}|G(d_r,s)|,1\leq r\leq m)=0
			\end{align}
			Due to the fact that
			\begin{align}
				\sup_{|t-d_r|\in (h_n, \gamma ]}\sup_{ s\in [\underline s, \bar s]}(G^2(t,s)-G^2(d_r,s))\geq 	\sup_{|t-d_r|\in (h_n,\gamma]}
				\sup_{s\in [\underline s, \bar s]}G^2(t,s)-\sup_{ s\in [\underline s, \bar s]}G^2(d_r,s),
			\end{align}
			to prove (b) it suffices to show
			\begin{align}
				\lim_{n\rightarrow \infty}\p\left(\max_{1\leq r\leq m}\sup_{|t-d_r|\in (h_n, \gamma ]}\sup_{ s\in [\underline s, \bar s]}(G^2(t,s)-G^2(d_r,s))>0\right)=0
			\end{align}
			By the definition of $G(t,s)$, the above is equivalent  to 
			\begin{align}\label{April21-46}
				\lim_{n\rightarrow \infty}\p\left(\max_{1\leq r\leq m}\sup_{|t-d_r|\in (h_n, \gamma]}\sup_{s\in [\underline s,\bar s]}\left(\frac{H^2(t,s)\Xi(d_r)-\Xi(t)H^2(d_r,s)}{\Xi(d_r)\Xi(t)}\right)\geq 0\right)=0.
			\end{align}
			Since  $\Xi(d_r)\Xi(t)\geq 0$, to show \eqref{April21-46}, it is equivalent to show
			\begin{align}\label{2019-April-193}
				\lim_{n\rightarrow \infty}\p\left(\max_{1\leq r\leq m}\sup_{|t-d_r|\in (h_n, \gamma]}\sup_{s\in [\underline s,\bar s]}\left(H^2(t,s)\Xi(d_r)-\Xi(t)H^2(d_r,s)\right)\geq 0\right)=0.
			\end{align}
			Since $H^2(t,s)>0$ and $\Xi(t)>0$, by using the decomposition
			\begin{align}
				H^2(t,s)\Xi(d_r)-\Xi(t)H^2(d_r,s)=H^2(t,s)(\Xi(d_r)-\Xi(t))+\Xi(t)(H^2(t,s)-H^2(d_r,s)) ,
			\end{align} we shall see that to prove \eqref{2019-April-193}  and hence Theorem \ref{thm2}, it suffices to show that 
			\begin{align}\label{2019-01-15}
				\lim_{n\rightarrow \infty}\p\left(\max_{1\leq r\leq m}\sup_{|t-d_r|\in (h_n, \gamma]}\sup_{s\in [\underline s,\bar s]}(H^2(t,s)-H^2(d_r,s))\geq 0\right)=0,\\
				\lim_{n\rightarrow \infty}\p\left(\max_{1\leq r\leq m}\sup_{|t-d_r|\in (h_n, \gamma]}(\Xi(d_r)-\Xi(t))\geq 0\right)=0. \label{2019-04-07}
			\end{align}
			In the following of the proof, we shall prove expressions \eqref{2019-01-15}, \eqref{2019-04-07} in two steps. 
			
			
			Step 1, proof of equation \eqref{2019-01-15}. 
			
			Recall $\tilde G_n(t,s)$ defined in Section \ref{Origin} of the main article. Note that \begin{align}
				H(t,s)=\sqrt{ns}\tilde G_n (t,s)+\frac{1}{\sqrt{ns}}\sum_{i=1}^n\varepsilon_iW(\frac{i/n-t}{s}),
			\end{align}
			which together with \eqref{April20-9} we have that for $1\leq r\leq m$,
			\begin{align}
				\max_{1\leq r\leq m}\sup_{h_n< |t-d_r|\leq  \gamma}\sup_{s\in [\underline s,\bar s]}&(H^2(t,s)-H^2(d_r,s))= \notag\\&\max_{1\leq r\leq m}	\sup_{h_n< |t-d_r|\leq \gamma}\sup_{s\in [\underline s,\bar s]}(I_r(t,s)+II_r(t,s)+III_r(t,s)),\end{align}
			where 
			\begin{align}
				&I_r(t,s)=ns[\tilde G_n^2(t,s)-\tilde G_n^2(d_r,s)],\\&II_r(t,s)=2\tilde G_n(t,s)\sum_{i=1}^n\varepsilon_iW(\frac{i/n-t}{s})-2\tilde G_n(d_r,s)\sum_{i=1}^n\varepsilon_iW(\frac{i/n-d_r}{s}),\\
				&III_r(t,s)=\frac{1}{ns}(\sum_{i=1}^n\varepsilon_iW(\frac{i/n-t}{s}))^2-\frac{1}{ns}(\sum_{i=1}^n\varepsilon_iW(\frac{i/n-d_r}{s}))^2.
			\end{align}
			By condition (W2) and Proposition \ref{proposition-10-13-2}, 
			there exists a strictly positive constant $\eta$ 	such that uniformly for $h_n<|t-d_r|\leq \gamma$, $s\in[\underline s,\bar s]$, $1\leq r\leq m$, we have
			\begin{align}
				I_r(t,s)\leq -\eta\left(\frac{|t-d_r|}{s}\wedge 1\right)^2ns\Delta^2_{r,n},
			\end{align}
			where  $\Delta_{r,n}=\beta(d_r-)-\beta(d_r+)$.
			Notice that $II_r(t,s)$ can be decomposed as \begin{align}
				II_r(t,s)=II_{1,r}(t,s)+II_{2,r}(t,s), \\II_{1,r}(t,s)=2\tilde G_n(t,s)(\sum_{i=1}^n\varepsilon_iW(\frac{i/n-t}{s})-\sum_{i=1}^n\varepsilon_iW(\frac{i/n-d_r}{s})),\\
				II_{2,r}(t,s)=2\sum_{i=1}^n\varepsilon_iW(\frac{i/n-d_r}{s})(\tilde G_n(t,s)-\tilde G_n(d_r,s)).
			\end{align}
			Notice that \begin{align}\label{April21-54}
				II_{1,r}(t,s)=\frac{-2\tilde G_n(t,s)}{s}\int_{d_r}^t(\sum_{i=1}^n\varepsilon_iW'(\frac{i/n-u}{s})du),
			\end{align}
			then by using the convention that $0/0=0$, we have that 
			\begin{align}\label{2019-April=209}
				\max_{1\leq r\leq m}\sup_{|t-d_r|\in (h_n,s]}\frac{|II_{1,r}(t,s)|}{|2\sqrt{ns}\tilde G_n(t,s)||\frac{t-d_r}{s}|}\leq \sup_{u\in [0,1]}\left|\frac{\sum_{i=1}^n\varepsilon_iW'(\frac{i/n-u}{s})}{\sqrt{ns}}\right|, \\
				\max_{1\leq r\leq m}\sup_{|t-d_r|\in (s,\bar s]}\frac{|II_{1,r}(t,s)|}{|2\sqrt{ns}\tilde G_n(t,s)||}\leq \max_{1\leq r\leq m}\sup_{t\in [0,1]}\frac{1}{\sqrt{ns}}\left|\sum_{i=1}^n\varepsilon_i\left(W(\frac{i/n-t}{s})-W(\frac{i/n-d_r}{s})\right)\right|, 
			\end{align}
			which together yield that
			\begin{align}
				\max_{1\leq r\leq m}\sup_{|t-d_r|\in (h_n,\gamma]}\frac{|II_{1,r}(t,s)|}{|2\sqrt{ns}\tilde G_n(t,s)|(|\frac{t-d_r}{s}|\wedge 1)}\leq U(s)+V(s),
			\end{align}
			where 
			\begin{align}
				U(s)=\sup_{u\in [0,1]}\left|\frac{\sum_{i=1}^n\varepsilon_iW'(\frac{i/n-u}{s})}{\sqrt{ns}}\right|,
				V(s)=\max_{1\leq r\leq m}\sup_{t\in [0,1]}\frac{1}{\sqrt{ns}}\left|\sum_{i=1}^n\varepsilon_i\left(W(\frac{i/n-t}{s})-W(\frac{i/n-d_r}{s})\right)\right|.
			\end{align}
			
			Using similar arguments to \eqref{approxV1_1}, Proposition \ref{GaussianApprox}, 
		Proposition \ref{techpropo-2019}, and Proposition B.2 of \cite{dette2015change} we have that $\sup_{s\in [\underline s, \bar s]}U(s)=O_p(\log n)$, $\sup_{s\in [\underline s, \bar s]}V(s)=O_p(\log n)$,
		and therefore
		\begin{align}\label{new.214}
			\max_{1\leq r\leq m}\sup_{|t-d_r|\in (h_n,\gamma]}\sup_{s\in [\underline s, \bar s]}\frac{|II_{1,r}(t,s)|}{|2\sqrt{ns}\tilde G_n(t,s)|(|\frac{t-d_r}{s}|\wedge 1)}=O_p(\log n).
		\end{align}
		Similar arguments applying to $II_{2,r}(t,s)$ we have
		\begin{align}\label{new.217}
			\max_{1\leq r\leq m}\sup_{|t-d_r|\in (h_n,\gamma]}\sup_{s\in [\underline s, \bar s]}\left|\frac{|II_{2,r}(t,s)|}{2\sqrt{ns}|\tilde G_n(t,s)-\tilde G_n(d_r,s)|}\right|=O_p(\log n).
		\end{align}
				By using the fact that $A^2-B^2=(A+B)(A-B)$, similar argument to \eqref{April21-54} and triangle inequalities we have the following two expressions:
				\begin{align*}
					\max_{1\leq r\leq m}\sup_{|t-d_r|\in (h_n,\gamma]}\sup_{s\in [\underline s, \bar s]}\left|\frac{III_{r}(t,s)}{\frac{1}{\sqrt{ns}}\sum_{i=1}^n\varepsilon_i(W(\frac{i/n-t}{s})-W(\frac{i/n-d_r}{s}))}\right|=O_p(\log n),\\
					\max_{1\leq r\leq m}\sup_{|t-d_r|\in (h_n,\gamma]}\sup_{s\in [\underline s, \bar s]}\left|\frac{\frac{1}{\sqrt{ns}}\sum_{i=1}^n\varepsilon_i(W(\frac{i/n-t}{s})-W(\frac{i/n-d_r}{s}))}{|\frac{d_r-t}{s}|\wedge 1}\right|=O_p(\log n),
				\end{align*}
				which lead to
				\begin{align}\label{new.218}
					\max_{1\leq r\leq m}\sup_{|t-d_r|\in (h_n,\gamma]}\left|\frac{III_{r}(t,s)}{|\frac{d_r-t}{s}|\wedge 1}\right|=O_p(\log^2 n).
				\end{align}
				By straightforward calculations using \eqref{new.214},\eqref{new.217},\eqref{new.218}, 
				Propositions \ref{proposition-1-10-13} and \ref{proposition-10-13-2}, we have that expression \eqref{2019-01-15} follows.
				
				
				
				Step 2. Proof of equation \eqref{2019-04-07}.
				
				By definitions of $\Xi(t)$, $\tilde \Xi(t)$, $\tilde H(t,s)$ we shall see that uniformly for  $1\leq r\leq m$ and $h_n\leq |t-d_r|\leq \bar s$,
				\begin{align}
					\notag\Xi(t)-\Xi(d_r)&=\frac{(\sum_{i\in K(t)}-\sum_{i\in K(d_r)})H^2(i/n,s^*)}{2(n\bar s-ns^*)}\left (1+O(\frac{1}{n\bar s})\right)\\
					&=\frac{\sum_{i\in K(t)\setminus K(d_r)}H^2(i/n,s^*)I(i,t,d_r)}{2(n\bar s-ns^*)}\left (1+O(\frac{1}{n\bar s})\right)
				\end{align}
				where $K(t)\setminus K(d_r)=\{i: i\in K(d_r), i\not \in K(t)\}\cup\{i: i\in K(t), i\not \in K(d_r)\}$
				and $I(i,t,d_r)=\mathbf 1(i\in K(t))-\mathbf 1(i\in K(d_r))$. By definition, it follows that 
				\begin{align}
					&\sum_{i\in K(t)\setminus K(d_r)}H^2(\frac{i}{n},s^*)I(i,t,d_r)=\sum_{i\in K(t)\setminus K(d_r)}\tilde H^2(\frac{i}{n},s^*)I(i,t,d_r)
					\notag\\&+2\sum_{i\in K(t)\setminus K(d_r)}\sqrt{ns^*}\tilde G_n(\frac{i}{n},s^*)\tilde H(\frac{i}{n},s^*)I(i,t,d_r)\notag\\&
					+\sum_{i\in K(t)\setminus K(d_r)}ns^*\tilde G_n^2(\frac{i}{n},s^*)I(i,t,d_r).\label{April-10-226}
				\end{align}
				
				Observe that, for $|t-d_r|\in [s^*,\bar s]$, $K(t)$ contains at least one of the intervals $[d_r-s^*,d_r]$ and $[d_r,d_r+s^*]$.
				~Hence by Proposition \ref{proposition-1-10-13}, we have that there exists a positive constant $\eta>0$ such that
				\begin{align}\label{April10-2019-01}
					\min_{1\leq r\leq m}\inf_{h_n< |t-d_r|\leq \bar s}\sum_{i\in K(t)} \frac{\tilde G_n^2(\frac{i}{n},s^*)}{\Delta^2_{r,n}}\geq \eta (ns^*).
				\end{align}
				On the other hand, Proposition \ref{proposition-1-10-13} implies that
				\begin{align}\label{April10-2019-02}
					\max_{1\leq r\leq m}\sum_{i\in K(d_r)}\tilde G_n^2(\frac{i}{n},s^*)\leq M(n\bar s)((s^*)^{k+1}+(ns^*)^{-1})^2.
				\end{align}
				Equations \eqref{April10-2019-01} and \eqref{April10-2019-02} together lead to 
				\begin{align}\label{April-10-231}
					\min_{1\leq r\leq m}\min_{|t-d_r|\in (h_n,\bar s]}\sum_{i\in K(t)\setminus K(d_r)}\frac{ns^*\tilde G_n^2(\frac{i}{n},s^*)I(i,t,d_r)}{\Delta_{r,n}^2}\geq \eta' (ns^*)^2
				\end{align}
				for some constant $\eta'>0$.
				
				Moreover, by applying Lemmas \ref{Lemma1}-\ref{Lemma4} and the definition of $ \tilde H$, we shall see that with probability tending to 1
				\begin{align}\label{April-10-232}
					\sup_{t\in [\bar s,1-\bar s]}\left|\frac{\sum_{i\in K(t)\setminus K(d_r)}\tilde H^2(\frac{i}{n},s^*)I(i,t,d_r)}{2(n\bar s-ns^*)}\right|\leq M
				\end{align}
				for some large positive constant $M$.
				Moreover, notice that
				\begin{align}\label{April-10-233}
					\sup_{1\leq r\leq m }\sup_{|t-d_r|\in (h_n,\bar s]}\left|\sum_{i\in K(t)\setminus K(d_r)}\sqrt{ns^*}\tilde G_n(\frac{i}{n},s^*)\tilde H(\frac{i}{n},s^*)I(i,t,d_r)/\Delta_{r,n}\right|\leq \mathcal Q_1+\mathcal Q_2,
				\end{align}
				where
				\begin{align}
					\mathcal Q_1=\left(\max_{1\leq r\leq m}\sup_{|t-d_r|\in (h_n,\bar s]}\left|\sum_{i\in K(t)}\sqrt{ns^*}\tilde G_n(\frac{i}{n},s^*)\tilde H(\frac{i}{n},s^*)/\Delta_{r,n}\right|\right),
					\\
					\mathcal Q_2=\left(\max_{1\leq r\leq m}\sup_{|t-d_r|\in (h_n,\bar s]}\left|\sum_{i\in K(d_r)}\sqrt{ns^*}\tilde G_n(\frac{i}{n},s^*)\tilde H(\frac{i}{n},s^*)/\Delta_{r,n}\right|\right).
				\end{align}
				To further study $\mathcal Q_1$ and $\mathcal Q_2$,  by similar arguments to \eqref{approxV1_1}, Proposition \ref{GaussianApprox}, Proposition \ref{techpropo-2019}, and Proposition B.2 of \cite{dette2015change} we have that $\sup_{t\in [\bar s, 1-\bar s]}|\tilde H(t,s^*)|=O_p(\log n)$.
				This fact and Proposition \ref{proposition-1-10-13} yield that $|\mathcal Q_1|=O_p(ns^*\sqrt{ns^*}\log n)$ and
				$|\mathcal Q_2|=o_p(ns^*\sqrt{ns^*}\log n)$.   Hence equation \eqref{April-10-233} is $O_p(ns^*\sqrt{ns^*}\log n)$. By using this fact together with expressions  \eqref{April-10-226}, \eqref{April-10-231} and \eqref{April-10-232} we 
				show equation \eqref{2019-04-07}. 
				Therefore the Theorem follows. \hfill $\Box$
				
				To prove Theorem \ref{Thm-Second-Stage}, we utilize th following proposition:
				\begin{proposition}\label{PropXiao}
					Suppose noises $\varepsilon_{i}$ satisfy (A1) and (A2) with $p>2$. Let $S^*_{a,b}=\max_{a \leq e \leq b}|\sum_{i=a}^e \varepsilon_{i}|$, $\nu=\sum_{j=1}^\infty \check{\mu}_j$, $\check{\mu}_j=(j^{\frac{p}{2}-1}\delta^p_p(L,j))^{\frac{1}{p+1}}$. Then we have
					\begin{align}
						&\p(S^*_{a,b}\geq x)\leq \frac{c_p}{x^p}((b-a)\nu^{p+1}+\sum_{i=a}^b\|\varepsilon_{i}\|_p^p)\notag\\ &+4\sum_{j=1}^\infty \exp\left(-\frac{c_p\check{\mu}_j^2x^2}{\nu^2(b-a)\delta^2_{p}(L,j)}\right)+2\exp\left(-\frac{c_px^2}{\sum_{i=a}^b\|\varepsilon_i\|_2^2}\right),
					\end{align}
					where $c_p$ is a constant only depending on $p$.
				\end{proposition}
				
				{\it Proof.} This follows from similar arguments to those in the proof of Theorem 2 (i) of \cite{liu2013probability}. \hfill $\Box$
				
				\ \\
				\noindent{\bf Proof of Theorem \ref{Thm-Second-Stage}.} 
					
					\noindent Consider the events 
					\begin{align}\label{April-19-2019}
						\mathcal E_n=\{\max_{1\leq r\leq m}|d_r-\hat d_r|\leq z_n\}
						\cap\{\hat m=m\}.
					\end{align}By Theorem \ref{thm2}, we have $\lim_{n\rightarrow \infty}\p(\mathcal E_n)= 1-\alpha$, where $\alpha$ is the significance level. 
					Define\begin{align}
						\tilde S_I=\sum_{i\in \lambda(I)}\varepsilon_{i}, ~~ \mu_r(t)=\E(V_r(t)|u_r,l_r),~~ \Delta_{r,n}=\beta(d_r-)-\beta(d_r+),\notag\\
						\tilde V_r(t)=\tilde S_{[l_r,t]}-\frac{\lambda([l_r,t])}{\lambda([l_r,u_r])}\tilde S_{[l_r,u_r]}.
					\end{align} 
					Elementary calculations show that uniformly for $1\leq r\leq m$,
					\begin{align}
						|\E(V_r(t)|u_r,l_r)-n\Delta_{r,n}\frac{(t\wedge d_r)u_r-td_r-u_rl_r+(t\vee d_r)l_r }{u_r-l_r}(1+O(\tfrac{1}{n}))|\leq M(n(u_r-l_r)z_n)
					\end{align}  
					for some sufficiently large constant $M$, where the term $(1+O(\tfrac{1}{n}))$ accounts for the error when $nl_r$ or $nu_r$ is not an integer.
					Without loss of generality, we consider the case that $t\geq d_r$. The case that $t<d_r$ follows from a similar argument.  When $t\geq d_r$, the above equation simplifies to 
					\begin{align}
						\mu_r(t)=\frac{n\Delta_{r,n}(u_r-t)(d_r-l_r)}{u_r-l_r}+O(nz_n^2).\label{March-6-40}
					\end{align}
					Furthermore, for $t\geq d_r$, \begin{align}\label{2019-9-8}
						V_r(t)-V_r(d_r)=S_{[d_r,t]}-\frac{\lambda([d_r,t])}{\lambda([l_r,u_r])}S_{[l_r,u_r]}.
					\end{align}
					As a result, \eqref{March-6-40} and \eqref{2019-9-8} lead to 
					\begin{align}
						\mu_r(t)-\mu_r(d_r)=\frac{n\Delta_{r,n}(d_r-t)(d_r-l_r)}{u_r-l_r}+O(n(d_r-t)z_n),\label{April-11-plus}\\
						\mu_r(t)+\mu_r(d_r)=\frac{n\Delta_{r,n}(2u_r-d_r-t)(d_r-l_r)}{u_r-l_r}+O(nz^2_n)\label{April-11-minus}.
					\end{align}
					Since $t\in [\tilde l_r,\tilde u_r]$, we have that \begin{align}\label{March-6-41}
						\min\{|t-u_r|,|t-l_r|\}\geq \frac{1+\tilde \alpha}{4+2\tilde \alpha}|u_r-l_r|.
					\end{align}
					It follows from equations \eqref{April-11-plus}, \eqref{April-11-minus}, \eqref{March-6-41} that on event $\mathcal E_n$,
					\begin{align}
						\frac{| d_r-u_r|}{|u_r-l_r|}\in \left[\frac{1+\tilde \alpha}{4+2 \tilde \alpha},\frac{3+\tilde \alpha}{4+2\tilde \alpha}\right],~~\frac{| d_r-l_r|}{|u_r-l_r|}\in \left[\frac{1+\tilde \alpha}{4+2\tilde \alpha},\frac{3+\tilde \alpha}{4+2\tilde \alpha}\right].
					\end{align}
					The above expression further yields that uniformly for $n(t-d_r)\geq 1$,
					\begin{align}\label{March-2019-45}
						\mu^2_r(t)-\mu^2_r(d_r)=\frac{n^2\Delta_{r,n}^2(2u_r-t-d_r)(d_r-l_r)^2(d_r-t)}{(u_r-l_r)^2}+O(n^2\Delta_{r,n}z_n^2(d_r-t))<0.
					\end{align}
					Here we point out that for sufficiently large $n$, the leading term of \eqref{March-2019-45} is negative. Now consider the following decomposition 
					\begin{align}\label{eq44-March2019}
						&V^2_r(t)-V^2_r(d_r)=I_r+II_r+III_r+\mu_r^2(t)-\mu_r^2(d_r),\\
						&I_r=\tilde V_r^2(t)-\tilde V_r^2(d_r),\ II_r=2(\tilde V_r(t)-\tilde V_r(d_r))\mu_r(t),III_r=2\tilde V_r(d_r)(\mu_r(t)-\mu_r(d_r))\label{eq45-March2019}
					\end{align}
					Let \begin{align}g_n=\frac{M\log n}{n\Delta_n^2}\label{gn}\end{align} for $\Delta_n=\min_{1\leq r\leq m}|\Delta_{r,n}|$. Since $ng_n\rightarrow \infty$, by equation \eqref{April-19-2019} it suffices to show that
					\begin{align}
						\p(\max_{1\leq r\leq m}\sup_{d_r+g_n\leq t\leq \tilde u_r}(V_r^2(t)-V_r^2(d_r))\geq 0|\mathcal E_n,u_r,l_r)\rightarrow 0.
					\end{align} 
					By equations \eqref{eq44-March2019} and \eqref{eq45-March2019}, showing the above equation amounts 
					to showing that 
					\begin{align}\label{eq47-March2019}
						\p(\max_{1\leq r\leq m}\sup_{d_r+g_n\leq t\leq \tilde u_r}(\Theta_r(t)+\frac{1}{3}[\mu_r^2(t)-\mu_r^2(d_r)])\geq 0|\mathcal E_n,u_r,l_r)\rightarrow 0
					\end{align} 
					for $\Theta_r(t)=I_r(t)$, $II_r(t)$ and $III_r(t)$, respectively. We shall show the situation that $\Theta_r(t)=II_r(t)$ in step 1,  that $\Theta_r(t)=III_r(t)$ in step 2 and that $\Theta_r(t)=I_r(t)$ in step 3. 
					\\\noindent {\bf Step 1 } Observe that $II_r(t)=2(II_{r,1}(t)+II_{r,2}(t))\mu_r(t)$, where \begin{align}
						II_{r,1}(t)=\tilde S_{(d_r,t]},~~II_{r,2}(t)=-\frac{\lambda([d_r,t])}{\lambda([l_r,u_r])}\tilde S_{[l_r,u_r]}.
					\end{align}Then by the triangle inequality, the LHS of \eqref{eq47-March2019} can be bounded by
					\begin{align}
						\p(\max_{1\leq r\leq m}\sup_{d_r+g_n\leq t\leq \tilde u_r}(2(\tilde V_r(t)-\tilde V_r(d_r))\mu_r(t)+\frac{1}{3}[\mu_r^2(t)-\mu_r^2(d_r)])\geq 0|\mathcal E_n,u_r,l_r)\leq \notag\\
						\sum_{r=1}^m\p(\sup_{d_r+g_n\leq t\leq \tilde u_r}(2(\tilde V_r(t)-\tilde V_r(d_r))\mu_r(t)+\frac{1}{3}[\mu_r^2(t)-\mu_r^2(d_r)])\geq 0|\mathcal E_n,u_r,l_r)\leq\notag \\
						\sum_{s=1}^2\sum_{r=1}^m\p(\sup_{d_r+g_n\leq t\leq \tilde u_r}(2II_{r,s}(t)\mu_r(t)+\frac{1}{6}[\mu_r^2(t)-\mu_r^2(d_r)])\geq 0|\mathcal E_n,u_r,l_r)\label{March18-52}
					\end{align} 
					
					For $II_{r,1}(t)$, notice that the last row of \eqref{March18-52} is bounded by
					\begin{align}\label{March18-53}
						&\p\left(\max_{1\leq u\leq\tilde g_{n,r}}\sup_{ \lceil n(d_r+ug_n)\rceil\leq k\leq  \lf n(d_r+(u+1)g_n)\rf}\left(II_{r,1}(\frac{k}{n})\mu_r(\frac{k}{n})+\frac{1}{6}(\mu_r^2(\frac{k}{n})-\mu_r^2(d_r))\right)\geq 0\bigg|\mathcal E_n,u_r,l_r\right)\notag\\
						=&\p\left(\max_{1\leq u\leq \tilde g_{n,r}}\sup_{ \lceil n(d_r+ug_n)\rceil\leq k\leq  \lf n(d_r+(u+1)g_n)\rf}|2\mu_r(\frac{k}{n})|\left(|II_{r,1}(\frac{k}{n})|+\frac{\mu_r^2(\frac{k}{n})-\mu_r^2(d_r)}{12|\mu_r(\frac{k}{n})|}\right)\geq 0\bigg|\mathcal E_n,u_r,l_r\right)\notag\\
						\leq &\p\left(\max_{1\leq u\leq \tilde g_{n,r}}\sup_{ \lceil n(d_r+ug_n)\rceil\leq k\leq  \lf n(d_r+(u+1)g_n)\rf}\left(|II_{r,1}(\frac{k}{n})|+\frac{\mu_r^2(\frac{k}{n})-\mu_r^2(d_r)}{12|\mu_r(\frac{k}{n})|}\right)\geq 0\bigg|\mathcal E_n,u_r,l_r\right)
					\end{align}
					for $k$ integers, and 
					\begin{align}
						\tilde g_{n,r}=\left\lceil\frac{1}{g_n}\left(\frac{\lf n\tilde u_r\rf+1}{n}-d_r\right)-1 \right\rceil.
					\end{align}
					Elementary calculations show that  \begin{align}\label{2019-9-8-01}
						u_r-d_r\in[(1+\tilde \alpha)z_n,(3+\tilde \alpha)z_n ], 
					\end{align}
					and uniformly for $d_r+g_n\leq t\leq \tilde u_r$
					\begin{align}\label{2019-9-8-02}
						u_r-t\in [(1+\tilde \alpha)z_n,(3+\tilde \alpha)z_n].
					\end{align}
					Expression \eqref{2019-9-8-01} and \eqref{2019-9-8-02} together with equations \eqref{March-6-40} and \eqref{March-2019-45} imply that for all  $1\leq u\leq \tilde g_{n,r}$,
					\begin{align}\label{March58-2019}
						\sup_{ \lceil n(d_r+ug_n)\rceil\leq k\leq  \lf n(d_r+(u+1)g_n)\rf}\left(\frac{\mu_r^2(\frac{k}{n})-\mu_r^2(d_r)}{12|\mu_r(\frac{k}{n})|}\right)\leq C(\tilde \alpha)|\Delta_{r,n}|(nd_r-\lceil n(d_r+ug_n)\rceil)<0,
					\end{align}
					where $C(\tilde \alpha)$ is a positive constant only depending on $\tilde \alpha$. \eqref{March58-2019} leads to the following bound for \eqref{March18-53}, which is 
					\begin{align}
						\p\left(\max_{1\leq u\leq \tilde g_{n,r}}\sup_{ k=\lceil n(d_r+ug_n)\rceil}^{  \lf n(d_r+(u+1)g_n)\rf}\left(|II_{r,1}(\frac{k}{n})|+C(\tilde \alpha)|\Delta_{r,n}|(nd_r-\lceil n(d_r+ug_n)\rceil)\right)\geq 0\bigg|\mathcal E_n,u_r,l_r\right).
					\end{align}
					Since $\tilde u_r=\hat d_r+z\leq d_r+2z$ and $|\mu_r(\frac{k}{n})|>0$, the above probability is further bounded by
					\begin{align}\label{March-55-2019}
						\p\left(\max_{1\leq u\leq  g_{r}}\sup_{k= \lceil n(d_r+ug_n)\rceil}^{  \lf n(d_r+(u+1)g_{n,r})\rf}\left(|II_{r,1}(\frac{k}{n})|+C(\tilde \alpha)|\Delta_{r,n}|(nd_r-\lceil n(d_r+ug_n)\rceil)\right)\geq 0\bigg|\mathcal E_n,u_r,l_r\right)
					\end{align}
					where 
					\begin{align}
						g_{n,r}=\left\lceil\frac{1}{g_n}\left(\frac{\lf n d_r+2nz_n\rf+1}{n}-d_r\right)-1 \right\rceil
					\end{align} which is non-random and is larger than $\tilde g_{n,r}$ on event $\mathcal{E}_n$.
					Therefore by using the triangle inequality we shall see that \eqref{March-55-2019} is further bounded by
					\begin{align}\label{March59-2019}
						&\sum_{u=1}^{ g_{n,r}}\p\left(\sup_{  \lceil n(d_r+ug_n)\rceil}^{  \lf n(d_r+(u+1)g_n)\rf}\left(|II_{r,1}(\frac{k}{n})|\right)\geq  C(\tilde \alpha)|\Delta_{r,n}|(\lceil n(d_r+ug_n)\rceil-nd_r)\bigg|\mathcal E_n,u_r,l_r\right)\notag\\
						=& 	\sum_{u=1}^{ g_{n,r}}\p\left(\sup_{  \lceil n(d_r+ug_n)\rceil}^{  \lf n(d_r+(u+1)g_n)\rf}\left(|II_{r,1}(\frac{k}{n})|\right)\geq  C(\tilde \alpha)|\Delta_{r,n}|(\lceil n(d_r+ug_n)\rceil-nd_r)\right).
					\end{align}
					By applying Proposition \ref{PropXiao} to equation \eqref{March59-2019} and using the definition of $II_{r,1}$, we have 
					\begin{align}\label{March61-2019}
						&\sum_{r=1}^m\p(\sup_{d_r+g_n\leq t\leq \tilde u_r}(2II_{r,1}(t)\mu_r(t)+\frac{1}{6}(\mu_r^2(t)-\mu_r^2(d_r)))\geq 0|\mathcal E_n,u_r,l_r)\notag\\
						&\leq C_p\sum_{r=1}^m\sum_{u=1}^{g_{n,r}}\left(\frac{ng_n}{(|\Delta_{r,n}|nug_n)^p}+\sum_{j=1}^\infty\exp\left(-\frac{c_p\check{\mu}_j^2(\Delta_{r,n} nug_n)^2}{\nu^2
							\delta_p^2(L,j)nug_n}\right)+\exp\left(-\frac{c_p(\Delta_{r,n}nug_n)^2}{nug_n}\right)\right),
					\end{align}
					where $\check{\mu}_j$ and  $\nu$ are defined in Proposition \ref{PropXiao}, and $C_p$, $c_p$ are constants only depending on $p$ and $\tilde \alpha$. 
					Since $g_{n}=\frac{M
						\log n}{n \Delta^2_n}$, we have that by assumptions on $m$ and $\Delta_n$,
					\begin{align}
						\sum_{r=1}^m\sum_{u=1}^{g_{n,r}}\left(\frac{ng_n}{(\Delta_{r,n}nug_n)^p}\right)\leq m|\Delta_{r,n}|^{-p}(M\log n)^{1-p}\Delta_n^{2p-2}\leq m\Delta_n^{p-2}(M\log n)^{1-p}\rightarrow 0.
					\end{align}
					Similarly, it follows that
					\begin{align}\label{March63-2019}
						\sum_{r=1}^m\sum_{u=1}^{g_{n,r}}\sum_{j=1}^\infty\exp\left(-\frac{c_p\check{\mu}_j^2(\Delta_{r,n} nug_n)^2}{\nu^2
							\delta_p^2(L,j)nug_n}\right)\leq m\sum_{u=1}^{g_{n,r}}\sum_{j=1}^\infty\exp\left(-C_p'j^{\frac{p-2}{p+1}}u\log n\right)\rightarrow 0,\\\label{March64-2019}
						\sum_{r=1}^m\sum_{u=1}^{g_{n,r}}\exp\left(-\frac{c_p(\Delta_{r,n} nug_n)^2}{nug_n}\right)\leq m\sum_{u=1}^{g_{n,r}}\exp\left(-C_p'u\log n\right)\rightarrow 0,
					\end{align}
					where $C_p'$ is some positive constant depending only on $p$ and $\tilde \alpha$, and the convergence to $0$ is guaranteed by sufficiently large choices of $M$ in equation \eqref{gn}. Furthermore, for $II_{r,2}(t)$, we have
					\begin{align}
						&\sum_{r=1}^m\p(\sup_{d_r+g_n\leq t\leq \tilde u_r}(2II_{r,2}(t)\mu_r(t)+\frac{1}{6}(\mu_r^2(t)-\mu_r^2(d_r)))\geq 0|\mathcal E_n,u_r,l_r)\notag\\
						& =\sum_{r=1}^m\p(\sup_{d_r+g_n\leq t\leq \tilde u_r}\left|\frac{2\mu_r(t)\lambda([d_r,t])}{\lambda([l_r,u_r])}\right|(-\tilde S_{[l_r,u_r]}\text{sgn}(\mu _r(t))\notag\\&~~~~~~~~~~~~~~~~~~~~~~\quad\quad\quad\quad+\frac{(\mu_r^2(t)-\mu_r^2(d_r))\lambda([l_r,u_r])}{12|\mu_r(t)\lambda([d_r,t])|})\geq 0\big|\mathcal E_n,u_r,l_r)\notag\\
						&\leq\sum_{r=1}^m\p\left(|\tilde S_{[l_r,u_r]}|\geq  C(\tilde \alpha)n|\Delta_{r,n}|z_n||\mathcal E_n,u_r,l_r\right)\label{A1792019}
					\end{align}
					for some constant $C(\tilde \alpha)$ only depends on $\tilde \alpha$, and ``sgn'' denotes the usual sign function. Notice that under $\mathcal E_n$, $v_r:=d_r-(3+\tilde \alpha)z_n\leq l_r$, then $\tilde S_{[l_r,u_r]}=\tilde S_{[v_r,u_r]}-\tilde S_{[v_r,l_r)}$. By the triangle inequality, the above expression is bounded by
					\begin{align}\label{March66-2019}
						2\sum_{r=1}^m\p&\left( \sup_{0<s\leq (6+2\alpha) z_n }|\tilde S_{[v_r,v_r+s]}|\geq  C(\tilde \alpha)nz_n|\Delta_{r,n}|\bigg|\mathcal E_n,u_r,l_r\right)\notag\\
						=	2\sum_{r=1}^m\p&\left( \sup_{0<s\leq (6+2\alpha) z_n }|\tilde S_{[v_r,v_r+s]}|\geq  C(\tilde \alpha)nz_n|\Delta_{r,n}|\right)\notag\\
						\leq &C_p \left((m\Delta_n^{p-2})(nz_n\Delta_n^2)^{1-p}+m\exp(-C_p'nz_n\Delta_n^2)\right)\rightarrow 0,
					\end{align}
					where $C_p$ and $C_p'$ are constants depending  only on $p$ and $\tilde \alpha$. By expressions \eqref{March18-52}, \eqref{March61-2019}--\eqref{March64-2019}, \eqref{A1792019}--\eqref{March66-2019}, we show \eqref{eq47-March2019} when $\Theta_r(t)$ is replaced by $II_r(t)$. \\
					\\\noindent {\bf Step 2} For $\Theta_r=III_r(t)$, we apply similar but simpler argument. Without loss of generality consider $\Delta_{r,n}>0$. By equation \eqref{April-11-minus} and the fact that $|\Delta_{r,n}|/z_n\rightarrow \infty$, for sufficiently large $n$ and $t\geq d_r+g_n$,
					\begin{align}
						\mu_r(t)-\,u_r(d_r)<0, ~~~\mu(d_r)>0.
					\end{align}
					Using the fact that $\mu^2_r(t)-\mu^2_r(d_r)=(\mu_r(t)+\mu_r(d_r)(\mu_r(t)-\mu_r(d_r))$ and equation \eqref{April-11-minus}, we obtain that LHS of \eqref{eq47-March2019} with $\Theta_r(t)=III_r(t)$ is bounded by 
					\begin{align}\label{March68-2019}
						&\sum_{1\leq r\leq m}\p(\inf_{d_r+g_n\leq t\leq \tilde u_r}(2\tilde V_r(d_r)+\mu_r(t)/3+ \mu_r(d_r)/3)\leq 0|\mathcal E_n,u_r,l_r)\notag\\
						\leq &\sum_{1\leq r\leq m}\p\left(|\tilde V_r(d_r)|\geq \frac{1}{3}\mu_r(d_r)\bigg|\mathcal E_n,u_r,l_r\right)\notag\\
						\leq &C_p \left((m\Delta_n^{p-2})(nz_n\Delta_n^2)^{1-p}+m\exp(-C_p'nz_n\Delta_n^2)\right)\rightarrow 0.
					\end{align}
					where $C_p$ and $C_p'$ are positive constants that only depend on $p$ and $\tilde \alpha$. The inequality in the second row of \eqref{March68-2019} is due to i): the positiveness of $\mu_r(t)$ when $t\in [d_r,\tilde \mu_r]$ and ii): the other terms except $\mu_r(t)$ are independent of $t$, and the inequality in the third row is due to Proposition \ref{PropXiao} and the fact that
					$\tilde S_{[l_r,u_r]}=\tilde S_{[v_r,u_r]}-\tilde S_{[v_r,l_r)}$ and $\tilde S_{[l_r,d_r]}=\tilde S_{[v_r,d_r]}-\tilde S_{[v_r,l_r)}$ for $v_r=(3+\tilde \alpha)z_n$. As a result, \eqref{eq47-March2019} holds when $\Theta_r(t)$ is replaced by $III_r(t)$. \\\ 
					\\\noindent {\bf Step 3} For $\Theta_r(t)=I_r(t)$, notice that 
					\begin{align}\label{March69-2019}
						\p(\max_{1\leq r\leq m}\sup_{d_r+g_n\leq t\leq \tilde u_r}(\tilde V_r^2(t)-\tilde V_r^2(d_r)+\mu_r^2(t)/3-\mu_r^2(d_r)/3)\geq 0|\mathcal E_n,u_r,l_r)\notag\\
						=\p(\cup_{1\leq r\leq m,\lceil n(d_r+g_n)\rceil\leq k\leq  \lf n\tilde u_r\rf}A_{k,r}|\mathcal E_n,u_r,l_r)
					\end{align}
					where \begin{align}\label{March70-2019}
						A_{k,r}=\{\tilde{V}_r(\frac{k}{n})^2-\tilde{V}_r(d_r)^2+\mu_r^2(\frac{k}{n})/3-\mu_r^2(d_r)/3\geq 0\}.
					\end{align}
					Since by \eqref{March-2019-45},  when $n$ is sufficiently large  we have $-\mu_r^2(\frac{k}{n})/3+\mu_r^2(d_r)/3>0$ for $\lceil n(d_r+g_n)\rceil\leq k\leq  \lf n\tilde u_r\rf,  1\leq r\leq m,$ which yields that
					\begin{align}
						A_{k,r} \in  A^\dag_{k,r}, ~~\text{where}~~	A^\dag_{k,r}=\{|\tilde{V}_r(\frac{k}{n})^2-\tilde{V}_r(d_r)^2|\geq -\mu_r^2(\frac{k}{n})/3+\mu_r^2(d_r)/3\}.
					\end{align}
					Furthermore, it is obvious that $A^\dag_{k,r}\in A^\dag_{k,r,1}\cup A^\dag_{k,r,2}$ where 
					\begin{align}\label{March73-2019}
						A^\dag_{k,r,1}=\{|\tilde{V}_r(\frac{k}{n})-\tilde{V}_r(d_r)|\geq 3^{-1/2}|\mu_r(\frac{k}{n})-\mu_r(d_r)|\},
						\\ A^\dag_{k,r,2}=\{|\tilde{V}_r(\frac{k}{n})+\tilde{V}_r(d_r)|\geq 3^{-1/2}|\mu_r(\frac{k}{n})+\mu_r(d_r)|\}.
					\end{align}
					Expressions \eqref{March70-2019}--\eqref{March73-2019} show that \eqref{March69-2019} is bounded by the sum of
					\begin{align}\label{March74-2019}
						\p(\max_{1\leq r\leq m}\sup_{d_r+g_n\leq t\leq \tilde u_r}(|\tilde{V}_r(t)-\tilde{V}_r(d_r)|-3^{-1/2}|\mu_r(t)-\mu_r(d_r)|)>0|\mathcal E_n,u_r,l_r)
					\end{align}
					and
					\begin{align}\label{March75-2019}
						\p(\max_{1\leq r\leq m}\sup_{d_r+g_n\leq t\leq \tilde u_r}(|\tilde{V}_r(t)+\tilde{V}_r(d_r)|-3^{-1/2}|\mu_r(t)+\mu_r(d_r)|)\geq 0|\mathcal E_n,u_r,l_r).
					\end{align}
					By similar arguments to the proof of steps 1 and 2, we shall see that both \eqref{March74-2019} and \eqref{March75-2019} converge to $0$. 
					As a result, \eqref{eq47-March2019}
					holds when $\Theta_r(t)$ is replaced by $I_r(t)$. Finally steps I, II and III prove \eqref{eq47-March2019} and complete the proof. \hfill $\Box$
					
					\noindent{\bf Proof of Theorem \ref{New.Thm4}.} For (i), it suffices to show \begin{align}
						\lim_{n\rightarrow \infty}\p\left(\max_{1\leq r\leq m}|\tilde d_r-d_r|\geq \frac{Mm^{\frac{1}{p-1}}\log n}{n}\bigg|\mathcal E_n,u_r,l_r\right)=0
					\end{align} and for (ii) it suffices to show \begin{align}
						\lim_{n\rightarrow \infty}\p\left(\max_{1\leq r\leq m}|\tilde d_r-d_r|\geq \frac{M\log^{\frac{2}{\beta}}n}{n\Delta_n^2}\bigg|\mathcal E_n,u_r,l_r\right)=0
					\end{align}
					for all $m<\frac{\bar s}{2}$, where the event $\mathcal E_n$ is defined in the Proof Theorem \ref{Thm-Second-Stage}. We first show (i), then show (ii).
					
					i). Let $g_n=\frac{Mm^{\frac{1}{p-1}}}{n}$ for some sufficiently large constant $M$.  By checking the proof of Theorem \ref{Thm-Second-Stage}, we shall see that under the conditions of this theorem, \eqref{March63-2019} and \eqref{March64-2019} hold. Equation \eqref{March66-2019} holds since our conditions guarantee that  $m=o(\bar s^{-1})$ and $z_n\geq C_1\bar s^2$ for some sufficiently large constant $C_1$. Therefore
					\begin{align}
						m(nz_n)^{1-p}\leq C_2n^{1-p}\bar s^{1-2p}\rightarrow 0
					\end{align}
					for some sufficiently large constant $C_2$. As a result, step 1 in the proof of Theorem \ref{Thm-Second-Stage} holds. Using similar arguments we find that step 2 and step 3 of the proof of Theorem \ref{Thm-Second-Stage} hold, which shows (i) holds. 
					

						ii) 
						For any integers $a<b$,  define $Z_{a,b}=\frac{\max_{a\leq q\leq b}\sum_{i=a}^q\varepsilon_{i}}{\sqrt{b-a}}$. Then by using Theorem  2 of \cite{wu2005nonlinear}, we get 
						\begin{align}
							\p(Z_{a,b}\geq u)\leq \exp(-tu^\beta)M(t)
						\end{align}
						for  $t<(\exp(1)\beta\gamma^\beta)^{-1}2^{-\frac{\beta}{2}}$, where $M(t)=\sup_{a,b}\exp(tZ_{a,b}^\beta)<\infty$. Using this fact, we shall see that
						\eqref{March-55-2019} is  bounded by $$\sum_{u=1}^{g_r}M_{\beta}\exp(-t_{\beta}(|\Delta_{r,n}|\sqrt{ung})^{\beta})$$
						for constants $t_\beta$ and $M_\beta$ depends only on $\beta$. This yields that equation \eqref{March18-53} is bounded by
						\begin{align}
							\sum_{u=1}^{g_r}M_{\beta}\exp(-t_{\beta}(|\Delta_r|\sqrt{ung_n})^{\beta}+\log m)\leq \sum_{u=1}^\infty\exp(-t_\beta'u^{\frac{\beta}{2}}\log n)\rightarrow 0
						\end{align}
						for $g_n=\frac{M_{\beta}\log^{\frac{2}{\beta}} n}{n\Delta_n^2}$ for some sufficiently small constant $t'_\beta$ and a large positive constant $M_{\beta}$. Both $t'_\beta$ and $M_{g,\beta}$  only depend on $\beta$. By using the condition that $\frac{nz_n\Delta_n^2}{\log ^{\frac{2}{\beta}}n}\rightarrow \infty$ we have that \eqref{March66-2019} holds, which shows that step 1 in the Proof of Theorem \ref{Thm-Second-Stage} holds. Similarly we shall see that  steps 2 and 3 are still valid in the proof of Theorem \ref{Thm-Second-Stage}, from which the ii) follows. \hfill $\Box$\

								\noindent{\bf Proof of Corollary \ref{Corol1}}
								
								To simplify the proof, noticing the condition that $g_n=o(m^{\frac{1}{p-1}-2}\log ^{\frac{2}{1-p}-2}n)$, we assume $n$ is sufficiently large such that $$g_n/n<m^{\frac{1}{p-1}}\log ^{\frac{2}{1-p}}n/n\in [0,1].$$ Let $C$ be a sufficiently small positive constant which varies from line to line, and $m^*=\frac{1}{n}m^{\frac{1}{p-1}}\log ^{\frac{2}{1-p}}n$ for short. Let $\iota_n=\frac{g_n\log^2n}{nm^*}$. It is easy to see that $\iota_n=o(1)$. Recall the quantity $V_r(\cdot)$ and event $\mathcal E_n$ in the proof of Theorem \ref{Thm-Second-Stage}. Define the notation
								\begin{align}
									\tilde M(n)=\big\{m:m\log^{-2} n\rightarrow \infty, m\bar s=o(1), m=o(n\underline s),\frac{nz_n}{m^{\frac{1}{p-1}}\log n}\rightarrow \infty\big\},\\
									\text{Linf}=\liminf_{n\rightarrow \infty, m\in \tilde M(n) },~~~
									\text{Lsup}=\limsup_{n\rightarrow \infty, m\in \tilde M(n) }
								\end{align}
								for short. Here $\tilde M(n)$ is a collection of sequence which satisfying the conditions for $m_n$ in Corollary \ref{Corol1}. Since the event 
								$$V_r^2\left(d_r-m^*\right)\vee V_r^2\left(d_r+m^*\right)>\sup_{t\in [d_r-g/n, d_r+g_n/n]}V_r^2(t)$$ implies that
								$|\tilde d_r-d_r|>g_n/n$, by Theorem \ref{thm2} it suffices to show that 
								\begin{align}
									\text{Linf}~	\p\left(\cup_{1\leq r\leq m}\left\{V_r^2\left(d_r+m^*\right)>\sup_{t\in [d_r,d_r+g_n/n]}V_r^2(t)\right\}\bigg|\mathcal E_n,u_r,l_r \right)\geq C>0,\label{case+}\\
									\text{Linf}~	\p\left(\cup_{1\leq r\leq m}\left\{V_r^2\left(d_r-m^*\right)>\sup_{t\in [d_r-g_n/n,d_r]}V_r^2(t)\right\}\bigg|\mathcal E_n,u_r,l_r  \right)\geq C>0. \label{case-}
								\end{align}In the following we shall show expression \eqref{case+}, and expression \eqref{case-} follows similarly.
								Similar to the proof of Theorem \ref{Thm-Second-Stage}, we have the following decomposition, which is 
								\begin{align}\label{April-20-2019}
									V^2_r(d_r+m^*)-V^2_r(t)=I_r(t)+II_r(t)+III_r(t)+\mu_r^2(d_r+m^*)-\mu_r^2(t),\\
									I_r(t)=\tilde V_r^2(d_r+m^*)-\tilde V_r^2(t),\ II_r(t)=2(\tilde V_r(d_r+m^*)-\tilde V_r(t))\mu_r(t),
									\\III_r(t)=2\tilde V_r(d_r+m^*)(\mu_r(d_r+m^*)-\mu_r(t)),
								\end{align}
								where the quantities $\tilde V_r(\cdot)$ and $\mu_r(\cdot)$ are defined in the proof of Theorem \ref{Thm-Second-Stage}. Recall the definition of $\Delta_{r,n}$ in the proof of Theorem \ref{Thm-Second-Stage}, we then have
								\begin{align}
									\mu_r(t)=\frac{n\Delta_{r,n}(\mu_r-t)(d_r-l_r)}{u_r-l_r}+O(nz_n^2),\label{eq55-2019}\\
									\mu_r(d_r+m^*)+\mu_r(t)=\frac{n\Delta_{r,n}(2\mu_r-t-d_r-m^*)(d_r-l_r)}{\mu_r-l_r}+O(nz_n^2),\\
									\mu_r(d_r+m^*)-\mu_r(t)=\frac{n\Delta_{r,n}(t-d_r-m^*)(d_r-l_r)}{\mu_r-l_r}+O(nz_n(t-d_r-m^*)).
								\end{align} As a result, we have that when $n$ is sufficiently large and $t\in [d_r,d_r+g_n/n]$, 
								\begin{align}\mu_r^2(d_r+m^*)-\mu^2_r(t)\leq Cn^2z_n(g_n/n-m^*)\leq -Cn^2z_nm^*/2<0,\\
									\mu^2(d_r+m^*)-\mu^2_r(t)\geq -Mn^2z_nm^*.\end{align} Hence by elementary calculations we have 
								\begin{align}
									&\p\left(V_r^2\left(d_r+m^*\right)>\sup_{t\in [d_r,d_r+g_n/n]}V_r^2(t) \bigg|\mathcal E_n,u_r,l_r \right)\notag\\
									=&\p\left(\inf_{t\in [d_r,d_r+g_n/n]}\left(V_r^2\left(d_r+m^*\right)-V_r^2(t)\right)>0\bigg|\mathcal E_n,u_r,l_r \right)\notag\\
									\geq& \p\bigg(\inf_{t\in [d_r,d_r+g_n/n]}\left(II_r(t)+\mu_r^2(d_r+m^*)-\mu_r^2(t)-C_{1,n}(t)-C_{3,n}(t)\right)>0,\notag\\&\sup_{t\in [d_r,d_r+g_n/n]} (|I_r(t)|-C_{1,n}(t))\leq 0,\sup_{t\in [d_r,d_r+g_n/n]} (|III_r(t)|-C_{3,n}(t))\leq 0\bigg|\mathcal E_n,u_r,l_r \bigg)\notag
								\end{align}
								for $C_{1,n}(t)=C_{3,n}(t)=-(\mu_r^2(d_r+m^*)-\mu_r^2(t))/4$. It remains to show that
								\begin{align}\label{new.eq59-2019}
									\lim_{n\rightarrow \infty, m\in \tilde M(n)}	m\max_{1\leq r\leq m}\p\bigg(&\sup_{t\in [d_r,d_r+g_n/n]} (|I_r(t)|- C_{1,n}(t))\geq 0\notag\\&\cup\sup_{t\in [d_r,d_r+g_n/n]} (|III_r|- C_{3,n}(t))\geq 0\bigg|\mathcal E_n,u_r,l_r \bigg)=0,\\
									\text{Linf}~	m\min_{1\leq r\leq m}\p\bigg(\inf_{t\in [d_r,d_r+g_n/n]}&\big(II_r(t)+\mu_r^2(d_r+m^*)\notag\\&-\mu_r^2(t)-C_{1,n}(t)-C_{3,n}(t)\big)>0\bigg|\mathcal E_n,u_r,l_r \bigg)>C.\label{new.eq60-2019}
								\end{align} This is  because   expressions \eqref{new.eq59-2019} and \eqref{new.eq60-2019} will lead to 
								\begin{align}
									\text{Linf}~ m\min_{1\leq r\leq m}\p\left(V_r^2\left(d_r+m^*\right)>\sup_{t\in [d_r,d_r+g_n/n]}V_r^2(t) \bigg|\mathcal E_n,u_r,l_r \right)\geq C.
								\end{align}
								Since $z_n\leq \bar s$, the quantities that $\{V_r^2\left(d_r+m^*\right)-\sup_{t\in [d_r,d_r+g_n/n]}V_r^2(t)\}_{r=1}^m$
								are independent of each other. As a consequence, it follows from $\log(1-x)\leq -x$ for $x\in(0,1)$ that 
								\begin{align}\label{April23-61-2019}
									\text{Linf}~\p\left(\cup_{1\leq r\leq m}\left\{V_r^2\left(d_r+m^*\right)>\sup_{t\in [d_r,d_r+g_n/n]}V_r^2(t)\right\} \bigg|\mathcal E_n,u_r,l_r \right)\geq 1-\exp(-C)>0,
								\end{align}
								hence \eqref{case+} holds and the corollary follows.
								In the following we shall investigate expression \eqref{new.eq60-2019} in detail, while expression \eqref{new.eq59-2019} follows from a similar argument to the proof of \eqref{new.eq60-2019}
								and the proof of Theorem \ref{New.Thm4}. Notice that the probability term of \eqref{new.eq60-2019} can be written as 
								\begin{align}
									&\p\left(\inf_{t\in [d_r,d_r+g_n/n]}\left(II_r+\frac{3}{2}(\mu_r^2(d_r+m^*)-\mu_r^2(t))\right)>0\bigg|\mathcal E_n,u_r,l_r \right)\notag\\\geq
									&\p\Big(\inf_{t\in [d_r,d_r+g_n/n]}\left(2(\tilde V_r(d_r+m^*)-\tilde V_r(t))\mu_r(t)\right)>\frac{3}{2}\sup_{t\in [d_r,d_r+g_n/n]}(-\mu_r^2(d_r+m^*)\notag\\&+\mu_r^2(t))\bigg|\mathcal E_n,u_r,l_r \Big)
									\notag\\\geq
									&\p\Big(\inf_{t\in [d_r,d_r+g_n/n]}\Big(2(\tilde V_r(d_r+m^*)-\tilde V_r(t))\mu_r(t)\Big)> Mn^2z_nm^*\bigg|\mathcal E_n,u_r,l_r \Big).\label{equation61-2019}
								\end{align}
								Without loss of generality consider $\Delta_{r,n}>0$. Then by expression \eqref{eq55-2019} we have that
								\begin{align}\label{April22-2019-1}
									\mu_r(t)\in [c_1nz_n,C_1nz_n], ~~~~ \forall t\in [d_r,d_r+g_n/n]
								\end{align}
								for some positive constants $0<c_1<C_1$.
								By definition, $\tilde V_r(t)=\tilde S_{[l_r,t]}-\frac{\lambda([l_r,t])}{\lambda([l_r,u_r])}\tilde S_{[l_r,u_r]}$. This expression leads to 
								\begin{align}\label{April22-2019-2}
									\tilde V_r(d_r+m^*)-\tilde V_r(t)=\tilde S_{[t,d_r+m^*]}-\frac{\lambda([t,d_r+m^*])}{\lambda([l_r,u_r])}\tilde S_{[l_r,u_r]}.
								\end{align}
								By expressions \eqref{April22-2019-1} and \eqref{April22-2019-2}, expression \eqref{equation61-2019} is larger than
								\begin{align}\label{912-0-2019}
									&\p\left(\inf_{t\in [d_r,d_r+g_n/n]}\big(\tilde V_r(d_r+m^*)-\tilde V_r(t))\big)> Mnm^*\bigg|\mathcal E_n,u_r,l_r \right)\notag\\
									&\geq \p\bigg(\inf_{t\in [d_r,d_r+g_n/n]}\big(\tilde S_{[t,d_r+m^*]}-D_{1,n} \big)>Mnm^*,\notag\\
									&\sup_{t\in [d_r,d_r+g_n/n]} \left|\frac{\lambda([t,d_r+m^*])}{\lambda([l_r,u_r])}\tilde S_{[l_r,u_r]}\right|\leq D_{1,n}
									\bigg|\mathcal E_n,u_r,l_r \bigg)
								\end{align}
								where $D_{1,n}=\frac{Mnm^*}{2}$.
								Notice that 
								\begin{align}
									&\p\left(\inf_{t\in [d_r,d_r+g_n/n]}\tilde S_{[t,d_r+m^*]}>Mnm^*\bigg|\mathcal E_n,u_r,l_r\right)\label{April23-67}\\=
									&\p\left(\inf_{t\in [d_r,\frac{\lf nd_r+g_n\rf-1}{n}]}\tilde S_{[t,\frac{\lf nd_r+g_n\rf-1}{n}]}+\tilde S_{[d_r+\frac{g_n}{n},d_r+m^*]}>Mnm^*\bigg|\mathcal E_n,u_r,l_r\right)\notag\\
									\geq & \p\left(\sup_{t\in [d_r,\frac{\lf nd_r+g_n\rf-1}{n}]}|\tilde S_{[t,\frac{\lf nd_r+g_n\rf-1}{n}]}|\leq \frac{Mnm^*}{2},\tilde S_{[d_r+\frac{g_n}{n},d_r+m^*]}>\frac{3}{2}Mnm^*\bigg|\mathcal E_n,u_r,l_r\right).\notag
								\end{align}
								Since $nm^*\rightarrow \infty$, by proposition 3.1 of \cite{mikosch1998large} and condition \eqref{condition48} in the main article, it follows that
								\begin{align}\label{April23-68}
									&\text{Linf}~m\min_{1\leq r\leq m}\p\left(\tilde S_{[d_r+\frac{g_n}{n},d_r+m^*]}>Mnm^*\bigg|\mathcal E_n,u_r,l_r\right)\notag\\=&\text{Linf}~m(Mnm^*-g_n)(1-F(Mnm^*))(1+o(1))\notag
									\\=&\text{Linf}~C^\dag m(Mnm^*-g_n)(Mnm^*)^{-p}\log^{-2}(Mnm^*)(1+o(1))\notag
									\\\geq &\text{Linf}~ C(mm^{-1})\log^2 n\log ^{-2}(m^{\frac{1}{p-1}}\log^{\frac{2}{1-p}} n)\geq C.
								\end{align}
								On the other hand, by proposition \ref{PropXiao}  we have
								\begin{align}\label{April-23-69}
									&\text{Lsup}~m\max_{1\leq r\leq m}\p\left( \sup_{t\in [d_r,\frac{\lf nd_r+g_n\rf-1}{n}]}|\tilde S_{[t,\frac{\lf nd_r+g_n\rf-1}{n}]}|>Mnm^*\bigg|\mathcal E_n,u_r,l_r\right)\notag\\
									\leq &\text{Lsup}~\frac{M_pmg_n}{(nm^*)^p}+2\exp(-\frac{M'_pm(nm^*)^2}{g_n})\leq \text{Lsup}~ M_p''(mm^{-1})(\iota_n\vee\log^{-1} n)=0,
								\end{align}
								where $M_p$, $M_p'$ and $M_p''$ are constants only depend on $p$. 
								
								\noindent Combining \eqref{April23-67} and \eqref{April-23-69} we show 
								\begin{align}\label{912-1-2019}
									\text{Linf}~m\min_{1\leq r\leq m}\p\left(\inf_{t\in [d_r,d_r+g_n/n]}\tilde S_{[t,d_r+m^*]}>Mnm^*\bigg|\mathcal E_n,u_r,l_r\right)\geq C.\end{align}
								By the proof of (i) of Theorem \ref{New.Thm4}, it follows that
								\begin{align}\label{912-2-2019}
									\lim_{n\rightarrow \infty, m\in \tilde M(n)}m\min_{1\leq r\leq m}\p\left(\sup_{t\in [d_r,\frac{\lf nd_r+g_n\rf-1}{n}]} \left|\frac{\lambda([t,d_r+m^*])}{\lambda([l_r,u_r])}\tilde S_{[l_r,u_r]}\right|\geq D_{1,n}\bigg|\mathcal E_n,u_r,l_r\right)=0.
								\end{align}
								Expressions \eqref{912-0-2019}, \eqref{912-1-2019} and \eqref{912-2-2019} show that \eqref{new.eq60-2019} holds, which completes the proof.\hfill $\Box$\\
								
								\noindent {\bf Proof of Example \ref{example1}. }
								
								\noindent Since
								\begin{align}\sum_{i=1}^\infty \max_{0\leq r\leq l}\sup_{t\in(c_r,c_{r+1}]}|a_{r,i}(t)|<\infty,
								\end{align}
								we have by property (A2) for $(\varepsilon_{i,n})_{i=1}^n$ and elementary calculations, it follows that
								\begin{align}
									\Omega_p=\left(\sum_{i=1}^\infty \max_{0\leq r\leq l}\sup_{t\in(c_r,c_{r+1}]}|a_{r,i}(t)|\right)\|\eta_0-\eta^*_0\|_p\leq 2\left(\sum_{i=1}^\infty \max_{0\leq r\leq l}\sup_{t\in(c_r,c_{r+1}]}|a_{r,i}(t)|\right)\|\eta_0\|_p,
								\end{align}
								where $\eta_0$ and $\eta_0^*$ are i.i.d. random variables. As a result, expression \eqref{eq80-March25} holds if
								\begin{align}\|\eta_0\|_p\leq M p^{\frac{1}{\beta}-\frac{1}{2}}\label{March26-90}
								\end{align}
								for $\beta \in (0,2)$ and some sufficiently large constant $M$. Then the lemma follows from $\frac{1}{\beta}-\frac{1}{2}>0$,  Proposition \ref{March-Propo4} and Jensen's inequality.\hfill $\Box$\\
								
								\begin{proposition}
									For any random variable $X$, the following two conditions are equivalent. \label{March-Propo4}
									\begin{description}
										\item (1) There exists a positive constant $t>0$ such that $\E(\exp(t|X|))<\infty$.
										\item (2) There exist constants $\sigma$ and $M$ such that $\E(|X|^p)\leq M\sigma^pp^p$ for $p>0$. 
									\end{description}
								\end{proposition}
								{\it Proof.} (1)$\Rightarrow$ (2). Notice that for $x>0$ and $p>0$, we have $x>p\log (x/p)$. This yields that $\exp(x)>(\frac{x}{p})^p$ for  $x\geq 0$ and $p>0$. Therefore
								\begin{align}
									\exp(t|X|)>\big(\frac{t|X|}{p}\big)^p
								\end{align}
								Take expectation on both sides of the above expression we have (2) follows.
								
								(2)$\Rightarrow$ (1). By Fubini's Theorem, we have that
								\begin{align}
									\exp(t|X|)=1+\sum_{k=1}^\infty\frac{t^k\E(|X|^k)}{k!}\leq1+M\sum_{k=1}^\infty\frac{t^k\sigma^kk^k}{k!}<1+M(\sum_{k=1}^\infty (te\sigma)^k)<\infty
								\end{align}
								for $t<(e\sigma)^{-1}$,
								where we have used the fact that $k!>(k/e)^k$ for positive integer $k$.\hfill $\Box$
								\\

									\noindent{\bf Proof of Theorem \ref{Save_b}.}
									By Lemma \ref{Lemma3} and Lemma \ref{Lemma4} and the bandwidth conditions, it suffices to show that 
									\begin{align}\label{May-2019-01}
										\sup_{t\in T_n^d} |H(t, \tilde s)-H^\dag(t, \tilde s)|=O_p(\log^{\frac{1}{2}-\epsilon} n\log \log n),
									\end{align}
									where $H(t,\tilde s)=\sup_{\underline s\leq s\leq \bar s}|H(t,s)|$ and  $H^\dag(t,\tilde s)=\max_{1\leq i\leq \delta_n}|H(t,s_i)|$. By 
									Proposition \ref{proposition-1-10-13} and Proposition \ref{GaussianApprox}, to show equation \eqref{May-2019-01} it is further equivalent to show that
									\begin{align}\label{May-2019-02}
										\sup_{t\in T_n^d} |\check H(t, \tilde s)-\check H^\dag(t, \tilde s)|=O_p(\log^{\frac{1}{2}-\epsilon} n\log \log n),
									\end{align}
									where $\check H(t,\tilde s)=\sup_{\underline s\leq s\leq \bar s}|\check H(t,s)|$, $\check H^\dag(t,\tilde s)=\max_{1\leq i\leq \delta_n}|\check H(t,s_i)|$,
									and 
									\begin{align}
										\check H(t,s)=\frac{1}{\sqrt{ns}}\sum_{i=1}^nW(\frac{i/n-t}{s})\sigma(\frac{i}{n})V_i
									\end{align}
									where $\{V_i,i\in \mathbb Z\}$ are $i.i.d.$ N(0,1) random variables in Proposition \ref{GaussianApprox}.
									Moreover, it follows from the the triangle inequality that
									\begin{align}
										\check H^\dag(t, \tilde s)\leq \check H(t,\tilde s)\leq \check H^\dag(t, \tilde s)+\max_{1\leq i\leq \delta_n-1}\sup_{s\in [s_{i},s_{i+1}]}|\check H(t,s)-\check H(t,s_i)|.
									\end{align}
									Therefore it suffices to show that
									\begin{align}\label{2019-9-13}
										\sup_{t\in T_n^d}\max_{1\leq i\leq \delta_n-1}\sup_{s\in [s_{i},s_{i+1}]}|\check H(t,s)-\check H(t,s_i)|=O_p(\log^{\frac{1}{2}-\epsilon} n\log \log n)
									\end{align}
									Write \begin{align}t_i=\frac{i\underline s}{n}\end{align} for the sake of brevity. By the triangle inequality, to show \eqref{2019-9-13} it suffices to prove
									\begin{align}
										\sup_{j: t_j\in T_n^d}\max_{1\leq i\leq \delta_n-1}&\sup_{s\in [s_{i},s_{i+1}]}|\check H(t_j,s)-\check H(t_j,s_i)|=O_p(\log^{\frac{1}{2}-\epsilon} n ),\label{2019-May-toshow1}\\
										\sup_{j: t_j\in T_n^d}\max_{1\leq i\leq \delta_n-1}&\sup_{s\in [s_{i},s_{i+1}],0\leq t^*\leq \frac{\underline s}{n}}\big|\big(\check H(t_j,s)-\check H(t_j,s_i)\big)\notag\\&-\big(\check H(t_j+t^*,s)-\check H(t_j+t^*,s_i)\big)\big|=O_p(\log^{\frac{1}{2}-\epsilon} n).\label{2019-May-toshow2}
									\end{align}
									Notice that
									\begin{align}
										\sup_{s\in [s_{i},s_{i+1}]}|\check H(t,s)-\check H(t,s_i)|\leq \int_{s_i}^{s_{i+1}}\left|\frac{\partial \check H(t,s)}{\partial s}\right|ds.
									\end{align}
									Then by the above inequality and the triangle inequality we have that for $j\in \{u\in \mathbb Z:t_u\in T_n^d\}$,
									\begin{align}\label{2019-May-03}
										\left\|\sup_{s\in [s_{i},s_{i+1}]}|\check H(t_j,s)-\check H(t_j,s_i)|
										\right\|_p\leq 
										\left\|\int_{s_i}^{s_{i+1}}\left|\frac{\partial \check H(t_j,s)}{\partial s}\right|ds\right\|_p\notag
										\\\leq \int_{s_i}^{s_{i+1}}\left\|\frac{\partial \check H(t_j,s)}{\partial s}\right\|_pds
									\end{align}
									for all $p\geq 1$. Since for $s\in [\underline s, \bar s] $, $\{\frac{\partial \check H(t_j,s)}{\partial s}\}_{j\in \{u\in \mathbb Z:t_u\in T_n^d\}}$ are centered Gaussian random variables with standard deviations bounded by $M/s$, the last term of inequality \eqref{2019-May-03} is bounded by
									\begin{align}
										M_0((p-1)!!)^{1/p}\log \left(\frac{s_{i+1}}{s_i}\right)\label{2019-May-04}
									\end{align}
									where $M_0$ is a constant independent of $p,j,n,i$ and integer $p\geq 1$. Moreover expressions \eqref{2019-May-03} and \eqref{2019-May-04} lead to that for $j\in \{u\in \mathbb Z:t_u\in T_n^d\}$, $1\leq i\leq \delta_n-1$,
									\begin{align}\label{2019-May-05}
										\E\left|\sup_{s\in [s_{i},s_{i+1}]}|\check H(t_j,s)-\check H(t_j,s_i)|\left(\log \left(\frac{s_{i+1}}{s_i}\right)\right)^{-1}
										\right|^{2 p}\leq  M_0^{2 p}(2p-1)!!.
									\end{align}
									for integer $p\geq 1$.
									Observe that
									\begin{align}\label{2019-May-06}
										\lim \sup_{p\rightarrow \infty} \frac{\nu^pM_0^{2 p}(2 p-1)!!}{p! }<1, p\in \mathbb Z
									\end{align}
									when $2\nu M_0^2<1$. Since $\exp(v)=1+\sum_{p=1}^\infty \frac{v^p}{p!}$, it follows from equations \eqref{2019-May-05} and \eqref{2019-May-06} that
									\begin{align}
										\E\left(\exp\left(\nu \sup_{s\in [s_{i},s_{i+1}]}\left|\frac{\check H(t_j,s)-\check H(t_j,s_i)}{\log s_{i+1}-\log s_i}\right|^2\right)\right)\leq M_1<\infty
									\end{align}
									for $j\in \{u\in \mathbb Z:t_u\in T_n^d\}$, $\nu\in (0,(2M_0^2)^{-1})$ and a constant $M_1$ independent of $n$, $i$ and $j$. By construction,
									\begin{align}\label{s_i-setting}
										\log s_{i+1}-\log s_i\asymp (\log n)^{-\epsilon}.
									\end{align} 
									Therefore, by equation \eqref{s_i-setting} and a similar argument to the Proof of Proposition \ref{max-gaussian} that applied to the term \begin{align}
										\E\left(\exp\left(\nu \max_{1\leq i\leq \delta_n-1, j: t_j\in T_n^d}\sup_{s\in [s_{i},s_{i+1}]}\left|\frac{\check H(t_j,s)-\check H(t_j,s_i)}{\log s_{i+1}-\log s_i}\right|^2\right)\right),
									\end{align} equation \eqref{2019-May-toshow1} follows. 
									On the other hand, using the triangle inequality, the LHS of equation \eqref{2019-May-toshow2} is bounded by 
									\begin{align}\label{May-7-2019}
										\sup_{j\in \{u\in \mathbb Z:t_u\in T_n^d\}}\max_{1\leq i\leq \delta_n-1}\sup_{0\leq t^*\leq \frac{\underline s}{n}}(\int_{s_i}^{s_{i+1}}\left|\frac{\partial \check H(t_j,s)}{\partial s}-\frac{\partial \check H(t_j+t^*,s)}{\partial s}\right|ds).
									\end{align}
									By Lipschitz continuity of $W'$, the fact that $W''$ exists except on a finite number of points, the triangle inequality and a similar  argument that applies to expressions \eqref{May-7-2019}, we have that expression \eqref{2019-May-toshow2} holds and the theorem follows. \hfill $\Box$
									\begin{theorem}\label{thm4-2021}
										Let $W(x)=A(x)-D(x)$ for $0\leq x\leq 1$, and $W(x)=-W(-x)$ for $-1\leq x<0$, where
										\begin{align}
											A(x)=\frac{1}{B(2,q+1)}x(1-x)^q
										\end{align} 
										for some $q>\max(2,k)$, $B(\cdot,\cdot)$ is the beta function
										and \begin{align}\label{eq:aaa}
											D(x)=(1-x)^2x^2(a_{v}x^{v}+a_{v-1}x^{v-1}+...+a_0),\end{align}
										where $v=\lceil \frac{k}{2}\rceil $ and $a_{v}$, $a_{v-1}$,...,$a_0$ are determined by solving $\int_0^1 x^uW(x)dx=0$ with $u=2g-1$, $g=1,2,...,v$ and
										$\int_0^1 W(x)dx=1$. Then if $q$ is sufficiently large, $W(\cdot)$ is an order $k$ filter satisfying both (W1) and (W2).
									\end{theorem}
									\noindent{\it Proof.} By definition it is easy to check (W1) holds. Since $W'(0)=\frac{1}{B(2,q+1)}>0$ and $W(\cdot)$ is an odd function, we have that
									$W'(0)F_w(0)<0$ which  leads to (W2) (ii). 
									It remains to show (W2) (i). Due to the fact that  $\int_0^1 A(x)dx=1$,
									straightforward calculations and the definition of beta function imply that the coefficients $\mathbf a=(a_{0},a_1 ...,a_v)^T$ are determined by
									\begin{align}
										A\mathbf a=\mathbf b
									\end{align}
									where $\mathbf b=(0,\frac{B(3,q+1)}{B(2,q+1)},\frac{B(5,q+1)}{B(2,q+1)},...,\frac{B(2v+1,q+1)}{B(2,q+1)})^T$ which is a $(v+1)\times 1$ vector, and the $(v+1)\times (v+1)$
									matrix $A$ is given by
									\begin{align}
										A=\begin{pmatrix} 
											B(3,3) &B(4,3) &... &B(v+3,3) \\
											B(4,3) &B(5,3) &... &B(v+4,3) \\
											B(6,3) &B(7,3) &... &B(v+6,3) \\
											B(8,3) &B(9,3) &... &B(v+8,3) \\
											...&...&...&...\\
											B(2v+2,3) &B(2v+3,3) &... &B(3v+2,3)
										\end{pmatrix}.
									\end{align} In the remaining of the proof, we let $M_v$ denote a generic 
									sufficiently large constant depending on $v$ which may vary from line to line.
									By Lemma \ref{lemma-5-24}, A is invertible. 
									Then the definition of $\mathbf b$, and the basic property of beta function lead to \begin{align}\label{new.125-5-22}
										|\mathbf a|\leq M_v/q.
									\end{align}
									To show $|F_w(x)|$ is maximized at $0$, it amounts to showing 
									\begin{align}
										\left(\int_{-1}^0W(s)ds+\int_0^xW(s)ds\right)^2\leq \left(\int_{-1}^0W(s)ds \right)^2, 
									\end{align}
									which is further equivalent to 
									\begin{align}
										0\leq \int_0^x(A(s)-D(s))ds\leq 2.
									\end{align}
									By expression \eqref{new.125-5-22} we have
									\begin{align}\label{boundforD}
										\sup_{x\in[0,1]}\left|\int_0^xD(s)ds\right|\leq \int_0^1|D(s)|ds=\sum_{i=0}^v|a_i|B(i+3,3)\leq q^{-1}M_v.
									\end{align}
									Since $A(s)>0$ for $s\in[0,1]$ and $\int_0^1 A(s)ds=1$, the above expression implies that when $q$ is sufficiently large, 
									\begin{align}
										\sup_{x\in [0,1]}\left(\int_0^x W(s)ds\right)=\sup_{x\in [0,1]}\left(\int_0^x(A(s)-D(s))ds\right)\leq 1+q^{-1}M_v<2. 
									\end{align}
									It now remains to show that for $x\in [0,1]$,
									\begin{align}
										\int_0^xW(s)ds=\int_0^x(A(s)-D(s))ds\geq 0.\label{LowerboundW}
									\end{align}
									Observe that for sufficiently large $q$, if $x\leq \frac{1}{q}$,
									\begin{align}\label{Filter-A1}
										\int _0^xA(s)ds&=\frac{1}{B(2,q+1)}\int_0^xs(1-s)^qds\notag\\
										&\geq  \frac{1}{B(2,q+1)}\int_0^xs(1-\frac{1}{q})^qds\geq \frac{\eta x^2}{2eB(2,q+1)}
									\end{align} 
									for a positive constant $\eta<1$, while straightforward calculations conclude that 
									\begin{align}\label{Filter-D1}
										\Big|\int _0^xD(s)ds\Big|\leq \int_0^x(\sum_{i=0}^v|a_i|)s^2ds\leq q^{-1}M_vx^3.
									\end{align}
									Equations \eqref{Filter-A1} and \eqref{Filter-D1} imply that when $q$ is sufficiently large, 
									\begin{align}
										\inf_{0\leq x\leq q^{-1}}\int_0^xW(s)ds=\inf_{0\leq x\leq q^{-1}}\int_0^x(A(s)-D(s))ds\geq 0.\label{LowerboundW1}
									\end{align}
									On the other hand, since for $x\in [\frac{1}{q},1]$,
									\begin{align}
										\int _0^xA(s)ds&=\frac{1}{B(2,q+1)}\int_0^xs(1-s)^qds\notag\\
										&\geq  \frac{1}{B(2,q+1)q}\int_0^{q^{-1}}(1-s)^qds\geq \frac{1-(1-q^{-1})^{q+1}}{B(2,q+1)q(q+1)},
									\end{align}
									and \begin{align}
										\lim_{q\rightarrow \infty}\frac{1-(1-q^{-1})^{q+1}}{B(2,q+1)q(q+1)}=1-e^{-1},
									\end{align}
									we have that when $q$ is sufficiently large,
									\begin{align}
										\inf_{x\in [q^{-1},1]}\int _0^xA(s)ds\geq \eta(1-e^{-1}), 
									\end{align}
									which together with equation \eqref{boundforD} imply
									\begin{align}
										\inf_{x\in [q^{-1},1]}\int _0^xW(s)ds\geq 0 \label{LowerboundW2}
									\end{align}
									for sufficiently large $q$. Now the expression \eqref{LowerboundW} holds in view of expressions \eqref{LowerboundW1} and \eqref{LowerboundW2}, which completes the proof. \hfill $\Box$
									\ \\\ \\
									
									\noindent {\bf Proof of Lemma \ref{LemmaSN}.} We first show the results for  $k=1,2,3,4$. Without loss of generality, we show the $k=4$ case. The results corresponding to filters with order $k=1,2,3$ follow similar arguments.
									Let $P_n(x)$ denote the $n_{th}$ order shifted Legender polynomials defined in $[0,1]$. Since $W(\cdot)\in\mathcal C^1(\mathbb R, C_{Lip})$, there is a sequence of functions $W_N(x)$ such that $W_N(x)\rightarrow W(x)$ uniformly in $[0,1]$, where $W_N(x)$ has the form of
									\begin{align}\label{WN(x)}
										W_N(x)=\sum_{i=0}^Na_iP_i(x)
									\end{align} 
									for $a_i\in \mathbb R,0\leq i\leq N$. Consider the following subset of $\mathcal W(4)$  $$\mathcal W_{4,N}=\{f:f(x)=\sum_{i=0}^Na_iP_i(x) \ \text{for some \ $N\in \mathbb Z$, $f\in \mathcal W(4)$} \}.$$ This subset has also been defined in Section \ref{Filter-Construct} of the main article. As a result, it suffices to show
									\begin{align}\label{Lexpand_0}
										\sup_{W\in
											\mathcal W_{4,N}}(SN(W))^2=\sup_{W\in
											\mathcal W_{4,N}}\frac{(\int_0^1 W(t)dt)^2}{\int_0^1 W^2(t)dt}\leq (3/8)^2,
									\end{align} and this bound is almost achievable, i.e., for any $\epsilon_0>0$, we can construct a $W\in \mathcal W_{4,N}$ such that $SN(W)\leq \frac{3}{8}-\epsilon_0.$
									Define \begin{align}
										A_N=\{(a_0,...,a_N)\in \mathbb R^{N+1}: \sum_{i=0}^Na_i=0; \sum_{i=0}^N(-1)^ia_i=0;\sum_{i=0}^Ni(i+1)a_i=0;\notag\\
										a_3+7a_2+21a_1+35a_0=0;a_0+a_1/3=0,a_0=1\}.
									\end{align} Then by the orthonormal properties of Legender  polynomials,
									to show \eqref{Lexpand_0} it is equivalent to show
									\begin{align}\label{Lexpand}
										(\max_{(a_0,...,a_N)\in A_N}\frac{a_0^2}{\sum_{i=0}^N\frac{a^2_i}{2i+1}})\leq (3/8)^2
									\end{align}
									for an $N\in \mathbb Z$, $N\geq 1$.
									\ Plug $a_0=1$, $a_1=-3$ into expression \eqref{Lexpand},  the LHS of this expression  is reduced to
									\begin{align}\label{reduced-obj}
										\min_{(a_2,...,a_N)\in S_N}{\sum_{i=2}^N\frac{a_i^2}{2i+1}}
									\end{align}
									where \begin{align}
										S_N=((a_2,...a_N)\in \mathbb R^{N-1}: \sum_{i=2}^Na_i=2; \sum_{i=2}^N(-1)^ia_i=-4;\notag\\
										\sum_{i=2}^Ni(i+1)a_i=6;a_3+7a_2=28).
									\end{align}
									Define \begin{align}
										B'_N=((a_2,...a_N)\in \mathbb R^{N-1}:a_3+7a_2=28).
									\end{align}
									By the fact that $S_N\subset S_N'$ we have
									\begin{align}
										\min_{(a_2,...,a_N)\in S_N}{\sum_{i=2}^N\frac{a_i^2}{2i+1}}&\geq \min_{(a_2,...,a_N)\in B'_N}{\sum_{i=2}^N\frac{a_i^2}{2i+1}}
										\\&= \min_{a_3+7a_2=28}\frac{a_2^2}{5}+\frac{a_3^2}{7}.
									\end{align}
									By Lagrange multiplier, we find that $\frac{a_2^2}{5}+\frac{a_3^2}{7}$ subject to $a_3+7a_2=28$ is minimised at  $a_2=\frac{35}{9}$and $a_3=\frac{7}{9}$.
								Together with $a_0=1$ and $a_1=-3$, the upper bound for the order $4$ filter can be obtained from \eqref{Lexpand}. 
								
								It  remain to show the bound is almost achievable. For this purpose, we argue that for any $\epsilon>0$, we can find $W^*\in \mathcal W_{4,N}$ with $N$ sufficiently large and even such that $SN(W^*)\leq 3/8-\epsilon$. Let $ N_1:=\lf N^{1/2}\rf+1$. We consider the case that $N_1<N$. 
								The coefficients  $a_0,a_1,...,a_N$ of shifted Legender polynomials for $W^*$ is set as
								$a_0=1$, $a_1=-3$, $a_2=\frac{35}{9}$,  $a_3=\frac{7}{9}$, $a_i=0$ for $
								i\neq 0,1,2,3, N_1-1,N_1,N$ where $a_{N_1-1}, a_{N_1}, a_N$ are determined by constraints $\sum_{i=2}^Na_i=2$, $\sum_{i=2}^N(-1)^ia_i=-4$ and
								$\sum_{i=2}^Ni(i+1)a_i=6$. Straightforward calculations show that $0\leq 3/8- SN(W^*)=O(N^{-1/2})$, which completes the proof. \hfill $\Box$
								

								\begin{proposition}(Zhou 2013)\label{GaussianApprox}
									Suppose conditions (A1)-(A3) hold. Then on a possibly richer probability space, there exist $i.i.d.$ standard normal random variables $V_1,\dots,V_n$ such that
									\begin{align}
										\max_{1\leq i\leq n}\left|\sum_{s=1}^i\varepsilon_s-\sum_{s=1}^i\sigma(s/n)V_s\right|=o_p(n^{1/4}\log^2n)
									\end{align}
								\end{proposition}

								\begin{proposition}\label{max-gaussian}
									Let $X_{1,n},\cdots X_{n^k,n}$ be an array of mean $0$ normal random variables such that $\max_{1\leq i\leq n^k}Var(X_{i,n})\leq U_n$, where $k$ is a finite number. Let $M_n=\max_{1\leq i\leq n^k}\{|X_{i,n}|\}$. Then $M_n=O_p(U_n^{1/2}\log ^{1/2}n)$.   
								\end{proposition} 
								Proof. Notice that \begin{align}\label{prop-max-gaussian}
									\mathbb {E}(e^{t |M_n |})= \mathbb E\left(\max_{1\leq i\leq n^k}e^{t |X_i|}\right)\leq \sum_{i=1}^{n^k} \mathbb E e^{t|X_i|}\leq 2n^k e^{\frac{t^2U_n}{2}}
								\end{align}
								The proposition follows by taking $\log$ in both sides of \eqref{prop-max-gaussian}, Jansen's inequality  and letting
								$t=(U_n^{-1}\log n)^{\frac{1}{2}}$.\hfill $\Box$
								\begin{proposition}\label{Prop-Burk}
									Let $X_i$ be a sequence of martingale difference with $\|X_i\|_p<\infty$. Then \begin{align}
										\left	\|\sum_{i=1}^nX_i\right\|_p^2\leq C p\sum_{i=1}^n\|X_i\|_p^2,
									\end{align}
									where $C\leq \left(\frac{p}{p-1}\right)^2$.
								\end{proposition}
								{\it Proof.} By Burkholder inequality, we have that 
								\begin{align}\label{Burk-1}
									\left	\|\sum_{i=1}^nX_i\right\|_p\leq C_p\left\|\sqrt{\sum_{i=1}^nX_i^2}\right\|_p=C_p\left\|\sum_{i=1}^nX_i^2\right\|_{p/2}^{1/2},
								\end{align}
								where $C_p\leq C\sqrt p$. Furthermore, straightforward calculations show that
								\begin{align}\label{Burk-2}
									\left\|\sum_{i=1}^nX_i^2\right\|_{p/2}\leq \sum_{i=1}^n\left\|X_i^2\right\|_{p/2}=\sum_{i=1}^n\|X_i\|_p^2
								\end{align}
								The proposition follows from \eqref{Burk-1} and \eqref{Burk-2}.

									\begin{lemma}\label{lemma-5-24}
											\begin{align}
												A=\begin{pmatrix} 
													B(3,3) &B(4,3) &... &B(v+3,3) \\
													B(4,3) &B(5,3) &... &B(v+4,3) \\
													B(6,3) &B(7,3) &... &B(v+6,3) \\
													B(8,3) &B(9,3) &... &B(v+8,3) \\
													...&...&...&...\\
													B(2v+2,3) &B(2v+3,3) &... &B(3v+2,3)
												\end{pmatrix}.
											\end{align} 
											where $v$ is a fixed number, $B(a,b)=\frac{\gamma(a)\gamma(b)}{\gamma(a+b)}$ is the usual beta function and $\gamma$ is the usual gamma function. Then $A$ is invertible.
										\end{lemma}
										{\it} Proof. 
											Define for $i\in \mathbb Z^+$,
											\begin{align}
												A=\begin{pmatrix} 
													B(3,i) &B(4,i) &... &B(v+3,i) \\
													B(4,i) &B(5,i) &... &B(v+4,i) \\
													B(6,i) &B(7,i) &... &B(v+6,i) \\
													B(8,i) &B(9,i) &... &B(v+8,i) \\
													...&...&...&...\\
													B(2v+2,i) &B(2v+3,i) &... &B(3v+2,i)
												\end{pmatrix}.
											\end{align} 
												Then $A'=B_3$. Notice that
												$B_1$ is a submatrix of Hilbert matrix which is invertible.
												By the definition of beta function, we see for $i\geq 2$, 
												\begin{align}
													B_{i}=B_{i-1}\circ H_{i-1}
												\end{align}
												where $\circ$ denotes the Hadamard product, where 
												\begin{align}
													H_{i-1}=(i-1)\begin{pmatrix} 
														&\frac{1}{i+2} &\frac{1}{i+3} &...&\frac{1}{i+v+2}\\
														&\frac{1}{i+3} &\frac{1}{i+4} &...&\frac{1}{i+v+3} \\
														&\frac{1}{i+5} &\frac{1}{i+6} &...&\frac{1}{i+v+5} \\
														&\frac{1}{i+7} &\frac{1}{i+8} &...&\frac{1}{i+v+7} \\
														&...&...&...&...                       \\
														&\frac{1}{i+2v+1} &\frac{1}{i+2v+2}&... &\frac{1}{i+3v+1}
													\end{pmatrix}_{(v+1)\times (v+1)}
												\end{align}
												Since $H_{i-1}/(i-1)$ is a submatrix of Hilbert matrix, $H_{i-1}$ is invertible for all $i\geq 2$. Now by Schur product theorem that
												\begin{align}
													\det(A\circ B)\geq \det(A)\det(B),
												\end{align}
												we have that $B_2$ is invertible, and also $B_3$ is invertible. This completes the proof.
												\hfill $\Box$
												
												\begin{proposition}\label{techpropo-2019}
													Let $Z(t)=\frac{1}{\sqrt{ns_n}}\sum_{i=1}^nV_iW(\frac{i/n-t}{s_n})$ for $V_i$ $i.i.d.$ standard normals. Then we have $\sup_{t\in [0,1]}\|Z(t)\|_p\leq Mp$ for some sufficiently large constant $M$ that does not depend on $n$ and $p$.
												\end{proposition}
												{\it Proof.} The proposition follows from Proposition \ref{Prop-Burk}
												and the fact that $\|V_1\|_p\leq M\sqrt{p}$ for some sufficiently large constant $M$.

\bibliographystyle{imsart-number}
	\begin{footnotesize}
		\setlength{\bibsep}{2pt}
		\bibliography{lit_BJ}

\begin{thebibliography}{55}

\bibitem{bai1997estimating}
\begin{barticle}[author]
\bauthor{\bsnm{Bai},~\bfnm{J.}\binits{J.}}
(\byear{1997}).
\btitle{Estimating multiple breaks one at a time}.
\bjournal{Econometric Theory}
\bvolume{13}
\bpages{315--352}.
\end{barticle}
\endbibitem

\bibitem{bai1998estimating}
\begin{barticle}[author]
\bauthor{\bsnm{Bai},~\bfnm{Jushan}\binits{J.}} \AND
  \bauthor{\bsnm{Perron},~\bfnm{Pierre}\binits{P.}}
(\byear{1998}).
\btitle{Estimating and testing linear models with multiple structural changes}.
\bjournal{Econometrica}
\bpages{47--78}.
\end{barticle}
\endbibitem

\bibitem{beibel1996note}
\begin{barticle}[author]
\bauthor{\bsnm{Beibel},~\bfnm{M.}\binits{M.}}
(\byear{1996}).
\btitle{A note on Ritov's Bayes approach to the minimax property of the cusum
  procedure}.
\bjournal{Ann. Statist.}
\bpages{1804--1812}.
\end{barticle}
\endbibitem

\bibitem{chen2012testing}
\begin{barticle}[author]
\bauthor{\bsnm{Chen},~\bfnm{Bin}\binits{B.}} \AND
  \bauthor{\bsnm{Hong},~\bfnm{Yongmiao}\binits{Y.}}
(\byear{2012}).
\btitle{Testing for smooth structural changes in time series models via
  nonparametric regression}.
\bjournal{Econometrica}
\bvolume{80}
\bpages{1157--1183}.
\end{barticle}
\endbibitem

\bibitem{chen2022inference}
\begin{barticle}[author]
\bauthor{\bsnm{Chen},~\bfnm{Likai}\binits{L.}},
  \bauthor{\bsnm{Wang},~\bfnm{Weining}\binits{W.}} \AND
  \bauthor{\bsnm{Wu},~\bfnm{Wei~Biao}\binits{W.~B.}}
(\byear{2022}).
\btitle{Inference of breakpoints in high-dimensional time series}.
\bjournal{J. Amer. Statist. Assoc.}
\bvolume{117}
\bpages{1951--1963}.
\end{barticle}
\endbibitem

\bibitem{dahlhaus1997fitting}
\begin{barticle}[author]
\bauthor{\bsnm{Dahlhaus},~\bfnm{R.}\binits{R.}}
(\byear{1997}).
\btitle{Fitting time series models to nonstationary processes}.
\bjournal{Ann. Statist.}
\bvolume{25}
\bpages{1--37}.
\end{barticle}
\endbibitem

\bibitem{dahlhaus2006statistical}
\begin{barticle}[author]
\bauthor{\bsnm{Dahlhaus},~\bfnm{R.}\binits{R.}} \AND
  \bauthor{\bsnm{Subba~Rao},~\bfnm{S.}\binits{S.}}
(\byear{2006}).
\btitle{Statistical inference for time-varying ARCH processes}.
\bjournal{Ann. Statist.}
\bvolume{34}
\bpages{1075--1114}.
\end{barticle}
\endbibitem

\bibitem{daubechies2011synchrosqueezed}
\begin{barticle}[author]
\bauthor{\bsnm{Daubechies},~\bfnm{I.}\binits{I.}},
  \bauthor{\bsnm{Lu},~\bfnm{J.}\binits{J.}} \AND
  \bauthor{\bsnm{Wu},~\bfnm{H-T.}\binits{H.-T.}}
(\byear{2011}).
\btitle{Synchrosqueezed wavelet transforms: An empirical mode
  decomposition-like tool}.
\bjournal{Appl. Comput. Harmon. Anal.}
\bvolume{30}
\bpages{243--261}.
\end{barticle}
\endbibitem

\bibitem{dette2020multiscale}
\begin{barticle}[author]
\bauthor{\bsnm{Dette},~\bfnm{Holger}\binits{H.}},
  \bauthor{\bsnm{Eckle},~\bfnm{Theresa}\binits{T.}} \AND
  \bauthor{\bsnm{Vetter},~\bfnm{Mathias}\binits{M.}}
(\byear{2020}).
\btitle{Multiscale change point detection for dependent data}.
\bjournal{Scand. J. Stat.}
\bvolume{47}
\bpages{1243--1274}.
\end{barticle}
\endbibitem

\bibitem{dette2018multiscale}
\begin{barticle}[author]
\bauthor{\bsnm{Dette},~\bfnm{Holger}\binits{H.}},
  \bauthor{\bsnm{Eckle},~\bfnm{Theresa}\binits{T.}} \AND
  \bauthor{\bsnm{Vetter},~\bfnm{Mathias}\binits{M.}}
(\byear{2020}).
\btitle{Multiscale change point detection for dependent data}.
\bjournal{Scand. J. Stat.}
\bvolume{47}
\bpages{1243-1274}.
\end{barticle}
\endbibitem

\bibitem{dette2016detecting}
\begin{barticle}[author]
\bauthor{\bsnm{Dette},~\bfnm{H.}\binits{H.}} \AND
  \bauthor{\bsnm{Wied},~\bfnm{D.}\binits{D.}}
(\byear{2016}).
\btitle{Detecting relevant changes in time series models}.
\bjournal{J. R. Stat. Soc. Ser. B. Stat. Methodol.}
\bvolume{78}
\bpages{371--394}.
\end{barticle}
\endbibitem

\bibitem{dette2015change}
\begin{barticle}[author]
\bauthor{\bsnm{Dette},~\bfnm{H.}\binits{H.}},
  \bauthor{\bsnm{Wu},~\bfnm{W.}\binits{W.}} \AND
  \bauthor{\bsnm{Zhou},~\bfnm{Z.}\binits{Z.}}
(\byear{2019}).
\btitle{Change point analysis of second order characteristics in non-stationary
  time series}.
\bjournal{Statist. Sinica}
\bvolume{29}
\bpages{611--643}.
\end{barticle}
\endbibitem

\bibitem{dumbgen1991asymptotic}
\begin{barticle}[author]
\bauthor{\bsnm{D{\"u}mbgen},~\bfnm{L.}\binits{L.}}
(\byear{1991}).
\btitle{The asymptotic behavior of some nonparametric change-point estimators}.
\bjournal{Ann. Statist.}
\bvolume{19}
\bpages{1471--1495}.
\end{barticle}
\endbibitem

\bibitem{eubank1994nonparametric}
\begin{barticle}[author]
\bauthor{\bsnm{Eubank},~\bfnm{R.~L.}\binits{R.~L.}} \AND
  \bauthor{\bsnm{Speckman},~\bfnm{P.~L}\binits{P.~L.}}
(\byear{1994}).
\btitle{Nonparametric estimation of functions with jump discontinuities}.
\bjournal{Lect. Notes. Monogr. Ser.}
\bpages{130--144}.
\end{barticle}
\endbibitem

\bibitem{fan1994fast}
\begin{barticle}[author]
\bauthor{\bsnm{Fan},~\bfnm{J.}\binits{J.}} \AND
  \bauthor{\bsnm{Marron},~\bfnm{J.~S.}\binits{J.~S.}}
(\byear{1994}).
\btitle{Fast implementations of nonparametric curve estimators}.
\bjournal{J. Comput. Graph. Statist.}
\bvolume{3}
\bpages{35--56}.
\end{barticle}
\endbibitem

\bibitem{frick2014multiscale}
\begin{barticle}[author]
\bauthor{\bsnm{Frick},~\bfnm{K.}\binits{K.}},
  \bauthor{\bsnm{Munk},~\bfnm{A.}\binits{A.}} \AND
  \bauthor{\bsnm{Sieling},~\bfnm{H.}\binits{H.}}
(\byear{2014}).
\btitle{Multiscale change point inference}.
\bjournal{J. R. Stat. Soc. Ser. B. Stat. Methodol.}
\bvolume{76}
\bpages{495--580}.
\end{barticle}
\endbibitem

\bibitem{gao2008nonparametric}
\begin{barticle}[author]
\bauthor{\bsnm{Gao},~\bfnm{J.}\binits{J.}},
  \bauthor{\bsnm{Gijbels},~\bfnm{I.}\binits{I.}} \AND
  \bauthor{\bsnm{Van~Bellegem},~\bfnm{S.}\binits{S.}}
(\byear{2008}).
\btitle{Nonparametric simultaneous testing for structural breaks}.
\bjournal{J. Econometrics}
\bvolume{143}
\bpages{123--142}.
\end{barticle}
\endbibitem

\bibitem{gijbels1999estimation}
\begin{barticle}[author]
\bauthor{\bsnm{Gijbels},~\bfnm{I.}\binits{I.}},
  \bauthor{\bsnm{Hall},~\bfnm{P.}\binits{P.}} \AND
  \bauthor{\bsnm{Kneip},~\bfnm{A.}\binits{A.}}
(\byear{1999}).
\btitle{On the estimation of jump points in smooth curves}.
\bjournal{Ann. Inst. Statist. Math.}
\bvolume{51}
\bpages{231--251}.
\end{barticle}
\endbibitem

\bibitem{hajek2010smooth}
\begin{barticle}[author]
\bauthor{\bsnm{H{\'a}jek},~\bfnm{P.}\binits{P.}} \AND
  \bauthor{\bsnm{Johanis},~\bfnm{M.}\binits{M.}}
(\byear{2010}).
\btitle{Smooth approximations}.
\bjournal{J. Funct. Anal.}
\bvolume{259}
\bpages{561--582}.
\end{barticle}
\endbibitem

\bibitem{horowitz2001adaptive}
\begin{barticle}[author]
\bauthor{\bsnm{Horowitz},~\bfnm{J.~L.}\binits{J.~L.}} \AND
  \bauthor{\bsnm{Spokoiny},~\bfnm{V.~G.}\binits{V.~G.}}
(\byear{2001}).
\btitle{An adaptive, rate-optimal test of a parametric mean-regression model
  against a nonparametric alternative}.
\bjournal{Econometrica}
\bvolume{69}
\bpages{599--631}.
\end{barticle}
\endbibitem

\bibitem{huang1998empirical}
\begin{barticle}[author]
\bauthor{\bsnm{Huang},~\bfnm{N.~E.}\binits{N.~E.}},
  \bauthor{\bsnm{Shen},~\bfnm{Z.}\binits{Z.}},
  \bauthor{\bsnm{Long},~\bfnm{S.~R.}\binits{S.~R.}},
  \bauthor{\bsnm{Wu},~\bfnm{M.~C.}\binits{M.~C.}},
  \bauthor{\bsnm{Shih},~\bfnm{H.~H.}\binits{H.~H.}},
  \bauthor{\bsnm{Zheng},~\bfnm{Q.}\binits{Q.}},
  \bauthor{\bsnm{Yen},~\bfnm{N.}\binits{N.}},
  \bauthor{\bsnm{Tung},~\bfnm{C.~C.}\binits{C.~C.}} \AND
  \bauthor{\bsnm{Liu},~\bfnm{H.~H.}\binits{H.~H.}}
(\byear{1998}).
\btitle{The empirical mode decomposition and the Hilbert spectrum for nonlinear
  and non-stationary time series analysis}.
\bjournal{Proc. R. Soc. Lond. A.}
\bvolume{454}
\bpages{903--995}.
\end{barticle}
\endbibitem

\bibitem{khismatullina2018multiscale}
\begin{barticle}[author]
\bauthor{\bsnm{Khismatullina},~\bfnm{Marina}\binits{M.}} \AND
  \bauthor{\bsnm{Vogt},~\bfnm{Michael}\binits{M.}}
(\byear{2018}).
\btitle{Multiscale inference and long-run variance estimation in non-parametric
  regression with time series errors}.
\bjournal{J. R. Stat. Soc. Ser. B. Stat. Methodol.}
\end{barticle}
\endbibitem

\bibitem{killick2012optimal}
\begin{barticle}[author]
\bauthor{\bsnm{Killick},~\bfnm{Rebecca}\binits{R.}},
  \bauthor{\bsnm{Fearnhead},~\bfnm{Paul}\binits{P.}} \AND
  \bauthor{\bsnm{Eckley},~\bfnm{Idris~A}\binits{I.~A.}}
(\byear{2012}).
\btitle{Optimal detection of changepoints with a linear computational cost}.
\bjournal{J. Amer. Statist. Assoc.}
\bvolume{107}
\bpages{1590--1598}.
\end{barticle}
\endbibitem

\bibitem{langrene2019fast}
\begin{barticle}[author]
\bauthor{\bsnm{Langren{\'e}},~\bfnm{N.}\binits{N.}} \AND
  \bauthor{\bsnm{Warin},~\bfnm{X.}\binits{X.}}
(\byear{2019}).
\btitle{Fast and stable multivariate kernel density estimation by fast sum
  updating}.
\bjournal{J. Comput. Graph. Statist.}
\bpages{1--27}.
\end{barticle}
\endbibitem

\bibitem{liu2013probability}
\begin{barticle}[author]
\bauthor{\bsnm{Liu},~\bfnm{W.}\binits{W.}},
  \bauthor{\bsnm{Xiao},~\bfnm{H.}\binits{H.}} \AND
  \bauthor{\bsnm{Wu},~\bfnm{W.~B.}\binits{W.~B.}}
(\byear{2013}).
\btitle{Probability and moment inequalities under dependence}.
\bjournal{Statist. Sinica}
\bvolume{23}
\bpages{1257--1272}.
\end{barticle}
\endbibitem

\bibitem{loader1996change}
\begin{barticle}[author]
\bauthor{\bsnm{Loader},~\bfnm{C.~R.}\binits{C.~R.}}
(\byear{1996}).
\btitle{Change point estimation using nonparametric regression}.
\bjournal{Ann. Statist.}
\bvolume{24}
\bpages{1667--1678}.
\end{barticle}
\endbibitem

\bibitem{mikosch1998large}
\begin{barticle}[author]
\bauthor{\bsnm{Mikosch},~\bfnm{T.}\binits{T.}} \AND
  \bauthor{\bsnm{Nagaev},~\bfnm{A.~V.}\binits{A.~V.}}
(\byear{1998}).
\btitle{Large deviations of heavy-tailed sums with applications in insurance}.
\bjournal{Extremes}
\bvolume{1}
\bpages{81--110}.
\end{barticle}
\endbibitem

\bibitem{muller1992change}
\begin{barticle}[author]
\bauthor{\bsnm{M{\"u}ller},~\bfnm{H-G}\binits{H.-G.}}
(\byear{1992}).
\btitle{Change-points in nonparametric regression analysis}.
\bjournal{Ann. Statist.}
\bpages{737--761}.
\end{barticle}
\endbibitem

\bibitem{muller1997two}
\begin{barticle}[author]
\bauthor{\bsnm{M{\"u}ller},~\bfnm{H-G.}\binits{H.-G.}} \AND
  \bauthor{\bsnm{Song},~\bfnm{K-S.}\binits{K.-S.}}
(\byear{1997}).
\btitle{Two-stage change-point estimators in smooth regression models}.
\bjournal{Statist. Probab. Lett.}
\bvolume{34}
\bpages{323--335}.
\end{barticle}
\endbibitem

\bibitem{politis1999subsampling}
\begin{bbook}[author]
\bauthor{\bsnm{Politis},~\bfnm{D.~N.}\binits{D.~N.}},
  \bauthor{\bsnm{Romano},~\bfnm{J.~P.}\binits{J.~P.}} \AND
  \bauthor{\bsnm{Wolf},~\bfnm{M.}\binits{M.}}
(\byear{1999}).
\btitle{Subsampling}.
\bpublisher{Springer Science \& Business Media}.
\end{bbook}
\endbibitem

\bibitem{qiu2003jump}
\begin{barticle}[author]
\bauthor{\bsnm{Qiu},~\bfnm{P.}\binits{P.}}
(\byear{2003}).
\btitle{A jump-preserving curve fitting procedure based on local
  piecewise-linear kernel estimation}.
\bjournal{J. Nonparametr. Stat.}
\bvolume{15}
\bpages{437--453}.
\end{barticle}
\endbibitem

\bibitem{qu2008testing}
\begin{barticle}[author]
\bauthor{\bsnm{Qu},~\bfnm{Zhongjun}\binits{Z.}}
(\byear{2008}).
\btitle{Testing for structural change in regression quantiles}.
\bjournal{J. Econometrics}
\bvolume{146}
\bpages{170--184}.
\end{barticle}
\endbibitem

\bibitem{rho2019bootstrap}
\begin{barticle}[author]
\bauthor{\bsnm{Rho},~\bfnm{Y.}\binits{Y.}} \AND
  \bauthor{\bsnm{Shao},~\bfnm{X.}\binits{X.}}
(\byear{2019}).
\btitle{Bootstrap-Assisted Unit Root Testing With Piecewise Locally Stationary
  Errors}.
\bjournal{Econometric Theory}
\bvolume{35}
\bpages{142--166}.
\end{barticle}
\endbibitem

\bibitem{ritov1990decision}
\begin{barticle}[author]
\bauthor{\bsnm{Ritov},~\bfnm{Y.}\binits{Y.}}
(\byear{1990}).
\btitle{Decision theoretic optimality of the CUSUM procedure}.
\bjournal{Ann. Statist.}
\bpages{1464--1469}.
\end{barticle}
\endbibitem

\bibitem{schmidt2013multiscale}
\begin{barticle}[author]
\bauthor{\bsnm{Schmidt-Hieber},~\bfnm{J.}\binits{J.}},
  \bauthor{\bsnm{Munk},~\bfnm{A.}\binits{A.}} \AND
  \bauthor{\bsnm{D{\"u}mbgen},~\bfnm{L.}\binits{L.}}
(\byear{2013}).
\btitle{Multiscale methods for shape constraints in deconvolution: confidence
  statements for qualitative features}.
\bjournal{Ann. Statist.}
\bvolume{41}
\bpages{1299--1328}.
\end{barticle}
\endbibitem

\bibitem{seifert1994fast}
\begin{barticle}[author]
\bauthor{\bsnm{Seifert},~\bfnm{B.}\binits{B.}},
  \bauthor{\bsnm{Brockmann},~\bfnm{M.}\binits{M.}},
  \bauthor{\bsnm{Engel},~\bfnm{J.}\binits{J.}} \AND
  \bauthor{\bsnm{Gasser},~\bfnm{T.}\binits{T.}}
(\byear{1994}).
\btitle{Fast algorithms for nonparametric curve estimation}.
\bjournal{J. Comput. Graph. Statist.}
\bvolume{3}
\bpages{192--213}.
\end{barticle}
\endbibitem

\bibitem{shao2010self}
\begin{barticle}[author]
\bauthor{\bsnm{Shao},~\bfnm{Xiaofeng}\binits{X.}}
(\byear{2010}).
\btitle{A self-normalized approach to confidence interval construction in time
  series}.
\bjournal{J. R. Stat. Soc. Ser. B. Stat. Methodol.}
\bvolume{72}
\bpages{343--366}.
\end{barticle}
\endbibitem

\bibitem{shao2010}
\begin{barticle}[author]
\bauthor{\bsnm{Shao},~\bfnm{X.}\binits{X.}} \AND
  \bauthor{\bsnm{Zhang},~\bfnm{X.}\binits{X.}}
(\byear{2010}).
\btitle{Testing for change points in time series}.
\bjournal{J. Amer. Statist. Assoc.}
\bvolume{105}
\bpages{1228--1240}.
\end{barticle}
\endbibitem

\bibitem{siegmund1988}
\begin{barticle}[author]
\bauthor{\bsnm{Siegmund},~\bfnm{D.}\binits{D.}}
(\byear{1988}).
\btitle{Confidence sets in change-point problems}.
\bjournal{Int. Stat. Rev.}
\bpages{31--48}.
\end{barticle}
\endbibitem

\bibitem{stuaricua2005nonstationarities}
\begin{barticle}[author]
\bauthor{\bsnm{St{\u{a}}ric{\u{a}}},~\bfnm{C.}\binits{C.}} \AND
  \bauthor{\bsnm{Granger},~\bfnm{C.}\binits{C.}}
(\byear{2005}).
\btitle{Nonstationarities in stock returns}.
\bjournal{Rev. Econ. Stat.}
\bvolume{87}
\bpages{503--522}.
\end{barticle}
\endbibitem

\bibitem{sun1993tail}
\begin{barticle}[author]
\bauthor{\bsnm{Sun},~\bfnm{J.}\binits{J.}}
(\byear{1993}).
\btitle{Tail probabilities of the maxima of Gaussian random fields}.
\bjournal{Ann. Probab.}
\bpages{34--71}.
\end{barticle}
\endbibitem

\bibitem{sun1994simultaneous}
\begin{barticle}[author]
\bauthor{\bsnm{Sun},~\bfnm{J.}\binits{J.}} \AND
  \bauthor{\bsnm{Loader},~\bfnm{C.~R.}\binits{C.~R.}}
(\byear{1994}).
\btitle{Simultaneous confidence bands for linear regression and smoothing}.
\bjournal{Ann. Statist.}
\bvolume{22}
\bpages{1328--1345}.
\end{barticle}
\endbibitem

\bibitem{weyl1939volume}
\begin{barticle}[author]
\bauthor{\bsnm{Weyl},~\bfnm{H.}\binits{H.}}
(\byear{1939}).
\btitle{On the volume of tubes}.
\bjournal{Amer. J. Math.}
\bvolume{61}
\bpages{461--472}.
\end{barticle}
\endbibitem

\bibitem{wu2005nonlinear}
\begin{barticle}[author]
\bauthor{\bsnm{Wu},~\bfnm{W.~B.}\binits{W.~B.}}
(\byear{2005}).
\btitle{Nonlinear system theory: Another look at dependence}.
\bjournal{Proc. Natl. Acad. Sci. U.S.A.}
\bvolume{102}
\bpages{14150--14154}.
\end{barticle}
\endbibitem

\bibitem{wu2018gradient}
\begin{barticle}[author]
\bauthor{\bsnm{Wu},~\bfnm{W.}\binits{W.}} \AND
  \bauthor{\bsnm{Zhou},~\bfnm{Z.}\binits{Z.}}
(\byear{2018}).
\btitle{Gradient-based structural change detection for nonstationary time
  series M-estimation}.
\bjournal{Ann. Statist.}
\bvolume{46}
\bpages{1197--1224}.
\end{barticle}
\endbibitem

\bibitem{zhang2003adaptive}
\begin{barticle}[author]
\bauthor{\bsnm{Zhang},~\bfnm{C.~M.}\binits{C.~M.}}
(\byear{2003}).
\btitle{Adaptive tests of regression functions via multiscale generalized
  likelihood ratios}.
\bjournal{Canad. J. Statist.}
\bvolume{31}
\bpages{151--171}.
\end{barticle}
\endbibitem

\bibitem{zhang2000minimax}
\begin{barticle}[author]
\bauthor{\bsnm{Zhang},~\bfnm{J.}\binits{J.}} \AND
  \bauthor{\bsnm{Fan},~\bfnm{J.}\binits{J.}}
(\byear{2000}).
\btitle{Minimax kernels for nonparametric curve estimation}.
\bjournal{Int. J. Comput. Math.}
\bvolume{12}
\bpages{417--445}.
\end{barticle}
\endbibitem

\bibitem{zhang2016testing}
\begin{barticle}[author]
\bauthor{\bsnm{Zhang},~\bfnm{T.}\binits{T.}}
(\byear{2016}).
\btitle{Testing for jumps in the presence of smooth changes in trends of
  nonstationary time series}.
\bjournal{Electron. J. Stat.}
\bvolume{10}
\bpages{706--735}.
\end{barticle}
\endbibitem

\bibitem{zhao2021segmenting}
\begin{barticle}[author]
\bauthor{\bsnm{Zhao},~\bfnm{Zifeng}\binits{Z.}},
  \bauthor{\bsnm{Jiang},~\bfnm{Feiyu}\binits{F.}} \AND
  \bauthor{\bsnm{Shao},~\bfnm{Xiaofeng}\binits{X.}}
(\byear{2022}).
\btitle{{Segmenting Time Series via Self-Normalisation}}.
\bjournal{J. R. Stat. Soc. Ser. B. Stat. Methodol.}
\bvolume{84}
\bpages{1699-1725}.
\end{barticle}
\endbibitem

\bibitem{zhou2010nonparametric}
\begin{barticle}[author]
\bauthor{\bsnm{Zhou},~\bfnm{Z.}\binits{Z.}}
(\byear{2010}).
\btitle{Nonparametric inference of quantile curves for nonstationary time
  series}.
\bjournal{Ann. Statist.}
\bvolume{38}
\bpages{2187--2217}.
\end{barticle}
\endbibitem

\bibitem{zhou2013heteroscedasticity}
\begin{barticle}[author]
\bauthor{\bsnm{Zhou},~\bfnm{Z.}\binits{Z.}}
(\byear{2013}).
\btitle{Heteroscedasticity and autocorrelation robust structural change
  detection}.
\bjournal{J. Amer. Statist. Assoc.}
\bvolume{108}
\bpages{726--740}.
\end{barticle}
\endbibitem

\bibitem{zhou2014inference}
\begin{barticle}[author]
\bauthor{\bsnm{Zhou},~\bfnm{Z.}\binits{Z.}}
(\byear{2014}).
\btitle{Inference of weighted $ V $-statistics for nonstationary time series
  and its applications}.
\bjournal{Ann. Statist.}
\bvolume{42}
\bpages{87--114}.
\end{barticle}
\endbibitem

\bibitem{zhou2009local}
\begin{barticle}[author]
\bauthor{\bsnm{Zhou},~\bfnm{Z.}\binits{Z.}} \AND
  \bauthor{\bsnm{Wu},~\bfnm{W.}\binits{W.}}
(\byear{2009}).
\btitle{Local linear quantile estimation for nonstationary time series}.
\bjournal{Ann. Statist.}
\bvolume{37}
\bpages{2696--2729}.
\end{barticle}
\endbibitem

\bibitem{zhou2010simultaneous}
\begin{barticle}[author]
\bauthor{\bsnm{Zhou},~\bfnm{Z.}\binits{Z.}} \AND
  \bauthor{\bsnm{Wu},~\bfnm{W.~B.}\binits{W.~B.}}
(\byear{2010}).
\btitle{Simultaneous inference of linear models with time varying
  coefficients}.
\bjournal{J. R. Stat. Soc. Ser. B. Stat. Methodol.}
\bvolume{72}
\bpages{513--531}.
\end{barticle}
\endbibitem

\bibitem{zhu2010algorithm}
\begin{barticle}[author]
\bauthor{\bsnm{Zhu},~\bfnm{Y-K.}\binits{Y.-K.}} \AND
  \bauthor{\bsnm{Hayes},~\bfnm{W.~B.}\binits{W.~B.}}
(\byear{2010}).
\btitle{Algorithm 908: Online exact summation of floating-point streams}.
\bjournal{ACM Trans. Math. Software}
\bvolume{37}
\bpages{1–-13}.
\end{barticle}
\endbibitem

\end{thebibliography}


\begin{thebibliography}{10}

\bibitem{dette2018multiscale}
\begin{barticle}[author]
\bauthor{\bsnm{Dette},~\bfnm{Holger}\binits{H.}},
  \bauthor{\bsnm{Eckle},~\bfnm{Theresa}\binits{T.}} \AND
  \bauthor{\bsnm{Vetter},~\bfnm{Mathias}\binits{M.}}
(\byear{2020}).
\btitle{Multiscale change point detection for dependent data}.
\bjournal{Scand. J. Stat.}
\bvolume{47}
\bpages{1243-1274}.
\end{barticle}
\endbibitem

\bibitem{dette2015change}
\begin{barticle}[author]
\bauthor{\bsnm{Dette},~\bfnm{H.}\binits{H.}},
  \bauthor{\bsnm{Wu},~\bfnm{W.}\binits{W.}} \AND
  \bauthor{\bsnm{Zhou},~\bfnm{Z.}\binits{Z.}}
(\byear{2019}).
\btitle{Change point analysis of second order characteristics in non-stationary
  time series}.
\bjournal{Statist. Sinica}
\bvolume{29}
\bpages{611--643}.
\end{barticle}
\endbibitem

\bibitem{frick2014multiscale}
\begin{barticle}[author]
\bauthor{\bsnm{Frick},~\bfnm{K.}\binits{K.}},
  \bauthor{\bsnm{Munk},~\bfnm{A.}\binits{A.}} \AND
  \bauthor{\bsnm{Sieling},~\bfnm{H.}\binits{H.}}
(\byear{2014}).
\btitle{Multiscale change point inference}.
\bjournal{J. R. Stat. Soc. Ser. B. Stat. Methodol.}
\bvolume{76}
\bpages{495--580}.
\end{barticle}
\endbibitem

\bibitem{liu2013probability}
\begin{barticle}[author]
\bauthor{\bsnm{Liu},~\bfnm{W.}\binits{W.}},
  \bauthor{\bsnm{Xiao},~\bfnm{H.}\binits{H.}} \AND
  \bauthor{\bsnm{Wu},~\bfnm{W.~B.}\binits{W.~B.}}
(\byear{2013}).
\btitle{Probability and moment inequalities under dependence}.
\bjournal{Statist. Sinica}
\bvolume{23}
\bpages{1257--1272}.
\end{barticle}
\endbibitem

\bibitem{mikosch1998large}
\begin{barticle}[author]
\bauthor{\bsnm{Mikosch},~\bfnm{T.}\binits{T.}} \AND
  \bauthor{\bsnm{Nagaev},~\bfnm{A.~V.}\binits{A.~V.}}
(\byear{1998}).
\btitle{Large deviations of heavy-tailed sums with applications in insurance}.
\bjournal{Extremes}
\bvolume{1}
\bpages{81--110}.
\end{barticle}
\endbibitem

\bibitem{sun1994simultaneous}
\begin{barticle}[author]
\bauthor{\bsnm{Sun},~\bfnm{J.}\binits{J.}} \AND
  \bauthor{\bsnm{Loader},~\bfnm{C.~R.}\binits{C.~R.}}
(\byear{1994}).
\btitle{Simultaneous confidence bands for linear regression and smoothing}.
\bjournal{Ann. Statist.}
\bvolume{22}
\bpages{1328--1345}.
\end{barticle}
\endbibitem

\bibitem{wu2005nonlinear}
\begin{barticle}[author]
\bauthor{\bsnm{Wu},~\bfnm{W.~B.}\binits{W.~B.}}
(\byear{2005}).
\btitle{Nonlinear system theory: Another look at dependence}.
\bjournal{Proc. Natl. Acad. Sci. U.S.A.}
\bvolume{102}
\bpages{14150--14154}.
\end{barticle}
\endbibitem

\bibitem{zhou2010nonparametric}
\begin{barticle}[author]
\bauthor{\bsnm{Zhou},~\bfnm{Z.}\binits{Z.}}
(\byear{2010}).
\btitle{Nonparametric inference of quantile curves for nonstationary time
  series}.
\bjournal{Ann. Statist.}
\bvolume{38}
\bpages{2187--2217}.
\end{barticle}
\endbibitem

\bibitem{zhou2014inference}
\begin{barticle}[author]
\bauthor{\bsnm{Zhou},~\bfnm{Z.}\binits{Z.}}
(\byear{2014}).
\btitle{Inference of weighted $ V $-statistics for nonstationary time series
  and its applications}.
\bjournal{Ann. Statist.}
\bvolume{42}
\bpages{87--114}.
\end{barticle}
\endbibitem

\bibitem{zhou2010simultaneous}
\begin{barticle}[author]
\bauthor{\bsnm{Zhou},~\bfnm{Z.}\binits{Z.}} \AND
  \bauthor{\bsnm{Wu},~\bfnm{W.~B.}\binits{W.~B.}}
(\byear{2010}).
\btitle{Simultaneous inference of linear models with time varying
  coefficients}.
\bjournal{J. R. Stat. Soc. Ser. B. Stat. Methodol.}
\bvolume{72}
\bpages{513--531}.
\end{barticle}
\endbibitem

\end{thebibliography}
	\end{footnotesize}
\end{document}